%% file: Henry.tex
\documentclass[review,3p,12pt]{elsarticle}
\usepackage{setspace} 

\usepackage[perpage]{footmisc}

\usepackage{graphicx,color,float}
\usepackage{amsmath}
\usepackage[normalem]{ulem}
\usepackage{subcaption}
\usepackage{hyphenat}
\usepackage{hyperref}
\usepackage[nottoc]{tocbibind}

\numberwithin{equation}{section}

\usepackage{lineno,amssymb}
\journal{Nucl. Inst. and Meth. A}

\begin{document}
\newcommand {\T} {{\cal T}}
\newcommand {\DC} {{\delta \chi^2}}
\newcommand {\A} {{A}}
\newcommand {\al} {c} 
\newcommand {\bl} {b}
\newcommand {\erf} {\text{erf}}
\newcommand {\ttheta} {\theta} 
\newcommand {\x}  {{\theta_{\text{true}}}}
\newcommand {\y}  {{\theta_{\min}}}
\newcommand {\Tm} {{\Theta_{\min}}}
\newcommand {\Tt} {{\Theta_{\text{true}}}}
\newcommand {\Eta} {{\cal H}}

\newcommand {\hmu} {\hat{\mu}}
\newcommand {\hnu} {\hat{\nu}}
\newcommand {\hpi} {\hat{\pi}}
\newcommand {\htau} {\hat{\tau}}
\newcommand {\hth} {\hat{\theta}} 
\newcommand {\dg} {\text{diag}}
\newcommand {\p} {p}
\newcommand {\s} {s}
\newcommand {\ba} {{\bf a}}
\newcommand {\bb} {{\bf b}}
\newcommand {\bc} {{\bf c}}
\newcommand {\0} {{\bf 0}}
\newcommand {\f} {{\bf f}}
\newcommand {\e} {{\bf e}}
\newcommand {\B} {{B^*}}
\newcommand {\itau} {M^{-1}}
\newcommand {\mtau} {{M}}

\newcommand {\rf} {\text{ref}}
\newcommand {\tr} {\text{true}}
\newcommand {\NH} {{H_0}}
\newcommand {\IH} {{H_1}}

\newcommand {\Prob} {\text{Prob}}
\newcommand {\E} {\text{E}}
\newcommand {\Var} {\text{Var}}
\newcommand{\ind}{\stackrel{\text{indep}}{\sim}}
\newcommand{\cas}{\buildrel \text{a.s.} \over \longrightarrow}
\newcommand{\casleft}{\buildrel \text{a.s.} \over \longleftarrow}
\newcommand{\cd}{\buildrel d \over \rightarrow}
\newcommand{\cp}{\buildrel P \over \longrightarrow}
\newcommand{\apsim}{\buildrel \text{approx.} \over \sim}

\newtheorem{lemma}{Lemma}

\newcommand {\Tbarn} {\overline{\Delta T_{H_0}}}
\newcommand {\Tbara} {\overline{\Delta T_{H_1}}}
\newcommand {\n} {{n}}
\newcommand {\Ni} {{N_i}}
\newcommand {\nutn} {\eta} 
\newcommand {\nuta} {\zeta} 
\newcommand {\nutnT} {\eta^0} 
\newcommand {\nutaT} {\zeta^0} 
\newcommand {\muT} {\mu^0} 
\newcommand {\nuT} {\nu^0} 
\newcommand {\piT} {{\pi^{0}}} 
\newcommand {\tauT} {\tau^0} 
\newcommand {\tmu} {\tilde{\mu}}
\newcommand {\tnu} {\tilde{\nu}}

\newcommand {\hnutn} {\hat{\nutn}}
\newcommand {\hnuta} {\hat{\nuta}}

\newcommand {\paratheta} {\beta} 
\newcommand {\PT} {B}

\newcommand {\D} {\overline{D}}

\def\Journal#1#2#3#4{{#1} {\bf #2}, #3 (#4)}
\def\NCA{\rm Nuovo Cimento}
\def\NPA{{\rm Nucl. Phys.} A}
\def\NIM{\rm Nucl. Instrum. Methods}
\def\NIMA{{\rm Nucl. Instrum. Methods} A}
\def\NPB{{\rm Nucl. Phys.} B}
\def\PLB{{\rm Phys. Lett.}  B}
\def\PRL{\rm Phys. Rev. Lett.}
\def\PRD{{\rm Phys. Rev.} D}
\def\PRC{{\rm Phys. Rev.} C}
\def\ZPC{{\rm Z. Phys.} C}
\def\JPG{{\rm J. Phys.} G}


\title{Modeling Impurity Concentrations in Liquid Argon Detectors}

\author[BNL]{Aiwu Zhang\corref{cor1}\textsuperscript{}{\footnote{Now with Stony Brook University.}}}\ead{aiwu.zhang@stonybrook.edu}
\author[BNL]{Yichen Li\corref{cor1}}\ead{yichen@bnl.gov}
\author[BNL]{Craig Thorn\corref{cor1}}\ead{thorn@bnl.gov}
\author[MSU]{Carl Bromberg}
\author[BNL]{Milind V. Diwan}
\author[BNL]{Steve Kettell}
\author[Pitt]{Vittorio Paolone}
\author[BNL]{Xin Qian}
\author[BNL]{James Stewart}
\author[BNL]{Wei Tang\textsuperscript{}{\footnote{Now with University of Tennessee.}}}
\author[BNL]{Chao Zhang}

\cortext[cor1]{Corresponding authors}

\address[BNL]{Physics Department, Brookhaven National Laboratory, Upton, NY, USA}
\address[MSU]{Department of Physics and Astronomy, Michigan State University, East Lansing, MI, USA}
\address[Pitt]{Department of Physics and Astronomy, University of Pittsburgh, Pittsburgh, PA, USA}


\begin{abstract}
Impurities in noble liquid detectors used for neutrino and dark matter experiments can significantly impact the quality of data. 
We present an experimentally verified model for describing the dynamics of impurity distributions in liquid argon (LAr) detectors.
The model considers sources, sinks, and transport of impurities within and between the gas and liquid argon phases. 
Measurements of oxygen concentrations in a 20-L LAr multi-purpose test stand are compared to calculations made with this model to show that an accurate description of the concentrations under various operational conditions can be obtained. A result of this analysis is a determination of Henry's coefficient for oxygen in LAr. 
These calculations also show that some processes have small effects on the impurity dynamics and excluding them yields a solution as a sum of two exponential terms. 
This solution provides a simple way to extract Henry's coefficient with negligible approximation error.
It is applied to the data and the Henry's coefficient for oxygen in LAr is obtained as 0.84$^{+0.09}_{-0.05}$, consistent with literature results. Based on the analysis of the data with the model, we further suggest that, for a large liquid argon detector, barriers to flow (``baffles") installed in the gas phase to restrict flow can help reduce the ultimate impurity concentration in the LAr.
\end{abstract}


\maketitle

\thispagestyle{plain}


\setstretch{1.25}
\tableofcontents
\input{intro_CET.tex}
\input{process_CET.tex}
\input{allEquations_4thOrder.tex}
\input{discuss_steady_state}
\input{time_CET.tex}
\input{measurement_CET.tex}
\input{numerical_calc.tex}

\input{Henry_Coeff.tex}

\input{uncertainty_discuss.tex}
\input{baffle_CET.tex}

\input{summary.tex}

\section*{Acknowledgments}
\addcontentsline{toc}{section}{Acknowledgments}
This work is supported by Laboratory Directed Research and Development (LDRD) 
of Brookhaven National Laboratory and U.S. Department of Energy, Office of Science, 
Office of High Energy Physics and Early Career Research program under contract number
DE-SC0012704.

\input{bibliography.tex}



\end{document}

%% file: intro_CET.tex
\counterwithin*{footnote}{section}

\section{Introduction}

Liquid argon time projection chambers (LArTPCs) and calorimeter detectors~\cite{Chen:1976pp,rubbia77,willis74,Nygren:1976fe} have been constructed and operated in several neutrino~\cite{Amerio:2004ze,ArgoNeuT2012,Berns:2013usa,Hahn:2016tia,Cavanna:2014iqa} and dark matter~\cite{Alexander:2013hia,badertscher2013ardm,Zani:2014lea} experiments.
These detectors range in size from hundreds of liters to hundreds of cubic meters.
A few even larger LArTPCs~\cite{Antonello:2015lea,Abi:2017aow} are under construction.
Moreover, the Deep Underground Neutrino Experiment (DUNE)~\cite{Acciarri:2016crz}
is proposing $\sim$1$\times 10^4$~m$^3$ LAr detector modules to precisely examine neutrino oscillation physics~\cite{Diwan:2004, Diwan:2016gmz,Qian:2015waa}.
LAr satisfies three essential requirements for a detector medium: it is dense (1.4~g/cm$^3$), the ionization charge and scintillation light can propagate over many meters, and it is commercially available in large quantities at relatively low cost.  The long propagation distance allows one to make very large and relatively cheap detectors, with electronic readout devices at the periphery of the active detector volume.

Impurities in LAr (such as oxygen, water, and nitrogen) can significantly attenuate the charge or light signals. Charge attenuation will lead to a decrease in energy resolution as well as a potential loss of efficiency for short and minimum ionizing tracks; light signal attenuation can lead to reduction of the detector efficiency as well. The loss of charge signal occurs as a result of the process of electron attachment to impurities~\cite{Bakale1976}. The negative ions formed by electron attachment drift so slowly that they do not produce a significant signal within the electronic readout time window~\cite{Antonello:2014eha}. The loss of light signal is primarily the result of the processes of optical absorption~\cite{Jones2013N2} and scintillation quenching (non-radiative de-excitation)~\cite{Acciarri:2008kv}.
For example, oxygen, with an electron affinity of either 0.45 or 0.9~eV, depending on the final state~\cite{Chen2002_affinities,Rienstra2002_affinities}, is particularly detrimental to the charge propagation. 
At one part per billion (ppb) oxygen contamination the mean lifetime for electrons in LAr is $\sim$0.3~ms in a 500~V/cm electric field, which corresponds to a mean drift distance of $\sim$0.5~m. The mean drift distance and lifetime are inversely proportional to the oxygen contamination ~\cite{microboone2017, microbooneNote}. 
Water, whose properties in LAr are essentially unknown, also appears to be a significant contributor in limiting electron lifetime, although its electron affinity is zero~\cite{Chipman1978}. 
Nitrogen, another common impurity, has a negative electron affinity~\cite{GILMORE1965}, and shows relatively little attenuation to charge signals~\cite{Hofmann:1976de,Biller:1989yq}, but it is particularly effective at quenching scintillation signals~\cite{Acciarri:2008kv}.

Since commercial LAr with typical impurity levels of one part per million (ppm) is not satisfactory as a detector medium, considerable care must be taken to purify the argon and to minimize the introduction of impurities through leakage and surface desorption from materials inside the detector. 
The necessity to remove and control impurities to extremely low levels ($<$1~ppb) in very large LArTPCs contributes significantly to the high costs of their cryogenic systems (as much as 25\% of the total detector cost).
It is thus desired to have a verified realistic engineering model of the introduction, transport, and removal of possible impurities. Such a model can help understand the contamination and purification systems, which can further lead to cost reductions in the construction and operation of large LAr detectors.
In this paper, we develop a mathematical model to describe the dynamics of impurity sources, sinks, and distribution in LAr detectors and apply it to predict the purity performance of a typical LAr detector.

The rest of this paper is organized as follows. 
In Sec.~\ref{sec:model}, the significant processes governing the behavior of an impurity in a LAr detector are introduced and the differential equations that characterize each process are presented.
A set of coupled differential equations that describe the evolution of the concentrations of an impurity over time in different phases, which constitute the full mathematical model, are presented and discussed.
In Sec.~\ref{sec:time}, we simplify the model by focusing on the most important processes, in order to obtain expressions for impurity concentrations in closed forms as a function of time and the physical parameters of the model.
This simplified model is used to illustrate 
aspects of the time dependence of impurity concentrations under 
common conditions, and to 
demonstrate some important factors in obtaining ultra pure LAr.
In Sec.~\ref{sec:measurement}, we describe measurements of the time dependence of oxygen concentration in LAr obtained with a 20-L LAr multi-purpose test stand with a gas argon (GAr) recirculation and purification system.  We show that the model can provide a good description of the data under many operating conditions, and that a value of Henry's coefficient (the ratio of the oxygen concentration in GAr to that in LAr at equilibrium) can be obtained which agrees with past literature; in addition, oxygen leak rates are determined and limits on sorption parameters are discussed.  Further, comparing the model to the data demonstrates that the amount of oxygen entering the LAr from leaks at the top of the cryostat is inversely proportional to the evaporation rate.  Based on this observation, we suggest that a properly designed passive baffle installed inside and near the top of the gas phase in a large LAr detector could further suppress the impurities that leak into the detector.  Finally, in Sec.~\ref{sec:summary}, we discuss improvements to remove the current limitations of the model, and suggest further measurements that could improve our understanding of impurities in LAr.

%% file: process_CET.tex
\section{A Model of the Processes Governing the Distribution of an Impurity in a LAr Detector}\label{sec:model}

A comprehensive understanding of the distribution of an impurity in a LArTPC requires modeling its time and spatial dependence throughout the detector volume. 
Here we consider only the time dependence.
Specifically, we assume that the spatial distribution of an impurity is uniform in each of the four phases in the cryostat: the gas, the liquid, and the surfaces in the liquid and gas.  This assumption can be justified most easily in small systems, where the large surface area to volume leads to large convection from heat leaking through the imperfect insulation. However, in very large systems, like actual LAr TPCs, the spatial distribution may well be non-uniform and, if so, it must be separately simulated, e.g., through finite element analysis~\cite{Voirin2012}.  The uniformity assumption is further discussed in Sections~\ref{subsec:process_1} and  \ref{sec:baffle}.

\begin{figure}[!htbp]
\centering
\vspace{-1.0 cm}
\begin{minipage}[]{0.32\textwidth}
 \includegraphics[width=1.1\textwidth]{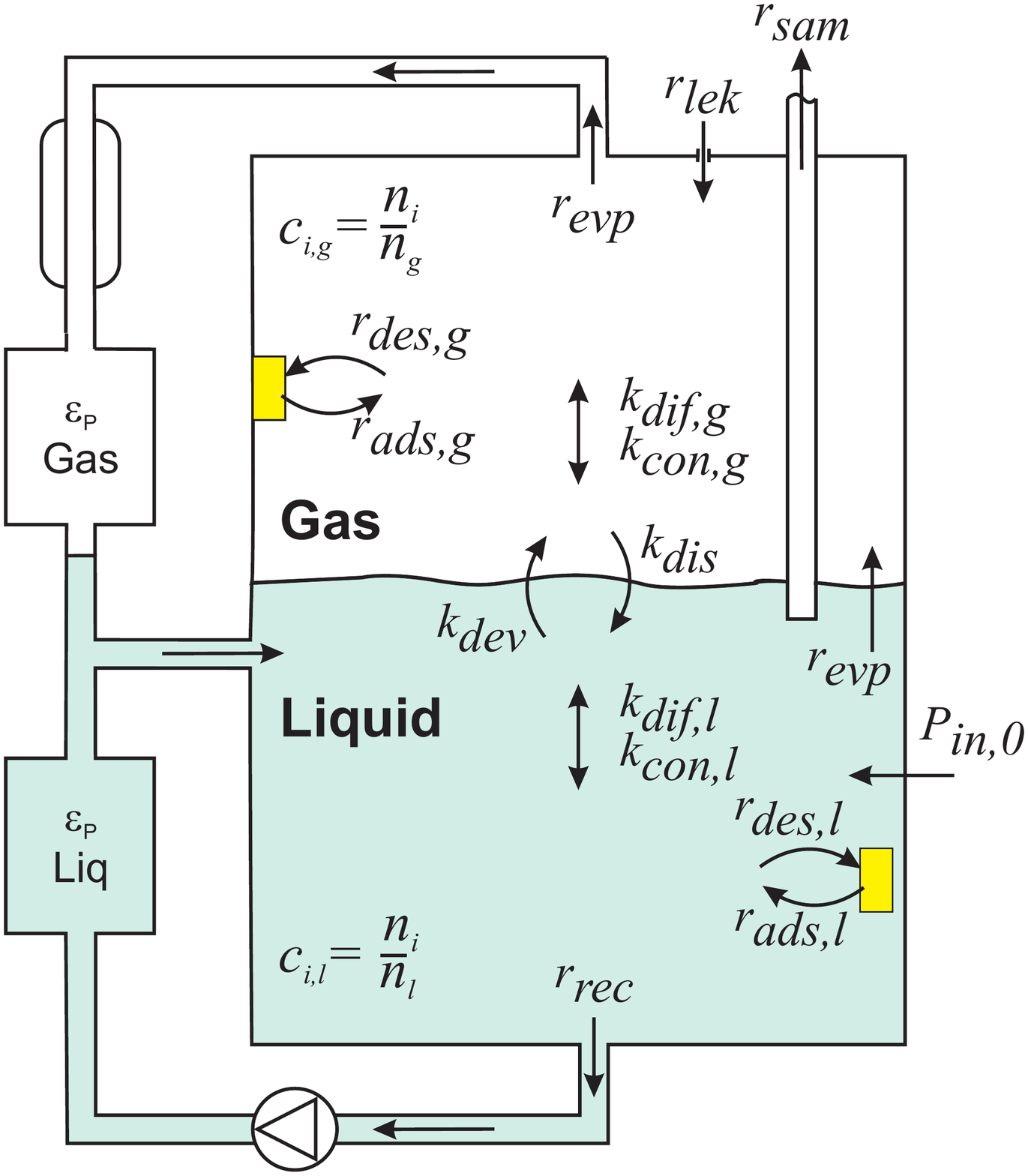}
  \end{minipage}
 \begin{minipage}[]{0.55\textwidth}
\includegraphics[width=1.1\textwidth]{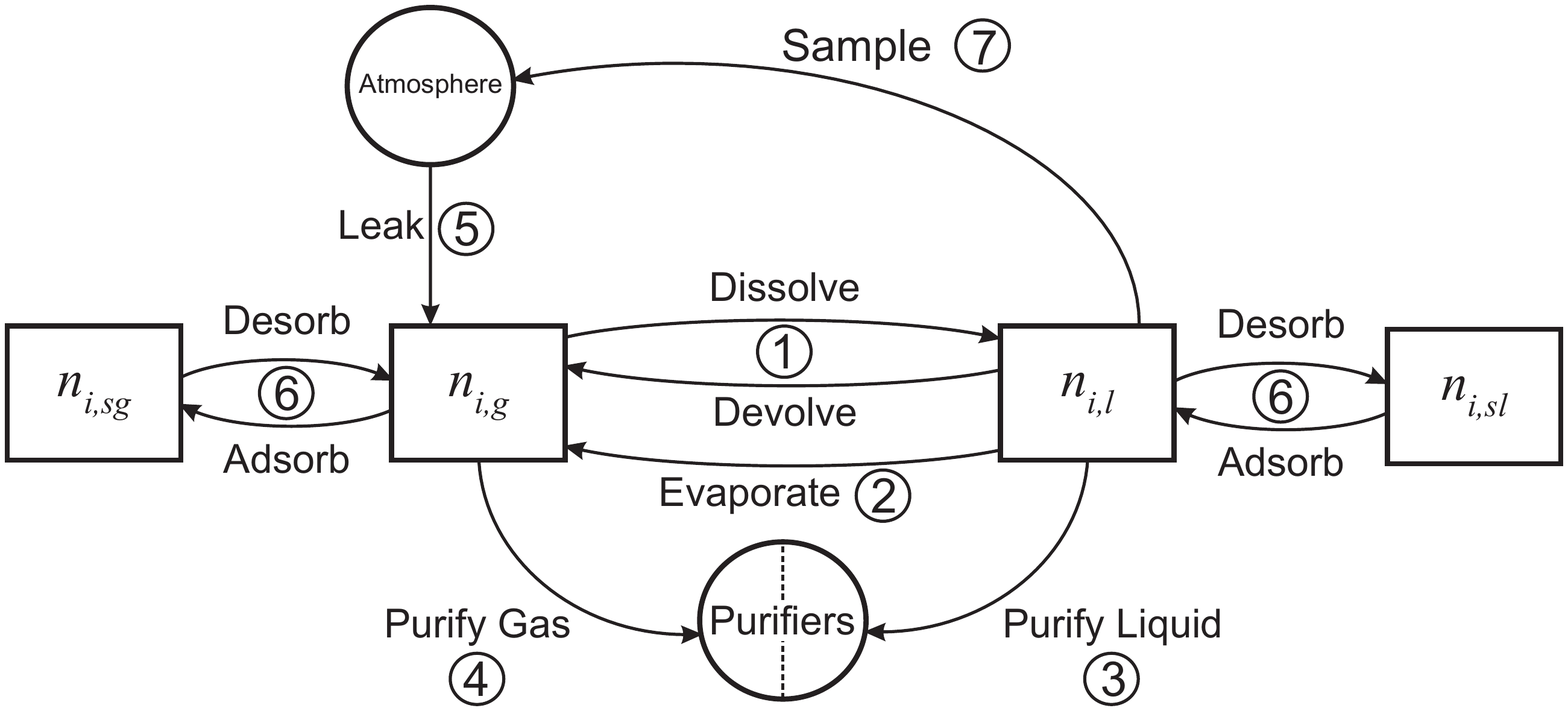}
  \end{minipage}
 \vspace{-0.5 cm}
 \caption{Left: A schematic showing an idealized LAr detector geometry discussed here. Right: A diagram of the model processes.  The rectangles represent the places where argon and impurities reside and the symbols are the molar amounts of the impurities in the gas and liquid.  The circles are infinite sinks (the purifiers) and sources (the atmosphere) of impurities. Each process in the model is identified by a number, corresponding to the sequence in Sec.~\ref{subsec:allprocess}. The arrows indicate the direction in which the impurity is transported; each line is characterized by a rate parameter indicated in the schematic.  A description of all the symbols appears in Table~\ref{tab:eq_def}.}
\label{fig:state}
\end{figure}

The model presented here includes the dominant sources, sinks, and transport processes of an impurity in a cryostat containing both LAr and GAr near thermal equilibrium. 
A diagrammatic representation of the processes is displayed in Fig.~\ref{fig:state}. The processes are: (1) impurity exchange (dissolution and devolution) at the gas-liquid interface, (2) evaporation of the liquid, (3) purification of the liquid, (4) purification of the gas and condensation of the purified gas into the liquid, (5) impurity leakage from the atmosphere, (6) sorption (i.e. adsorption and desorption) of an impurity on the surfaces in contact with the gas and liquid in the cryostat, and (7) liquid loss as a result of continuously removing a sample  to measure impurities (with the sample exhausted to the atmosphere).

In the following subsections, we present the equations governing the time evolution of eight molar amounts: two species (argon and impurity) in four places (in the gas, in the liquid, and on surfaces in contact with the gas and the liquid).  We assume that the amount of argon adsorbed on the surfaces is constant (or negligibly small), so that equations for the amounts of argon on the two surfaces do not appear, reducing the number of quantities from eight to six. Table~\ref{tab:eq_def} defines these quantities as well as the physical parameters used in the model. 

\begin{table}[!htbp]
\vspace{-0.5cm}
\caption{List of symbols used in the differential equations in the model. The subscript $p$ (for phase) is to be replaced by $g$ or $l$, to indicate either GAr or LAr, while the subscript $xx$ in $H_{xx}$ denotes it is defined as a mole fraction. }
\vspace{-0.5cm}
\begin{center}
\begin{tabular}{lll}
\hline
\hline
Symbol & Meaning   & Unit\\
\hline
$n_{p}$    & quantity of GAr (LAr)  & mol\\
$n_{i,p}$  & quantity of impurity in GAr (LAr)& mol \\
$n_{i,sp}$  & quantity of impurity on surfaces in GAr (LAr)& mol\\
$n_{i,sp}^{sat}$  & quantity of impurity on surface at saturation & mol \\
$c_{i,p}=n_{i,p}/n_{p}$  & concentration of impurity in GAr (LAr) & mol/mol\\
$c_{i,sp}$  & concentration of impurity on surface in GAr (LAr) & mol/m$^{2}$\\
$c_{i,sp}^{sat}$  & saturation concentration of impurity on surface & mol/$m^{2}$\\
$\theta_p=c_{i,sp} / c_{i,sp}^{sat}$    & surface concentration as a fraction of saturation &-\\
$H_{xx}$      & Henry's coefficient for impurity in Ar & mol/mol \\
$K_{ad,p}$      & sorption equilibrium constant  for surface in GAr (LAr)& mol/mol \\
$k_{dis}$  & rate constant for dissolving impurity into liquid from gas  & s$^{-1}$ \\
$k_{dev}$  & rate constant for devolving impurity into gas from liquid & s$^{-1}$  \\
$k_{des,p}$ & desorption rate constant of impurity from surface in GAr (LAr) & s$^{-1}$ \\
$k_{ads,p}$ & adsorption rate constant of impurity onto surface in GAr (LAr) & s$^{-1}$ \\
$r_{evp}$  & evaporation rate of liquid to gas   & mol/s\\
$r_{rec}$  & circulation rate of liquid through purifier   & mol/s\\
$r_{lek}$  & leak rate of impurity from atmosphere into gas & mol/s\\
$r_{srp,p}$  & net sorption rate from surfaces into GAr (LAr) & mol/s\\
$r_{sam}$  & sampling rate of liquid & mol/s\\
$\epsilon_P$   & fractional efficiency of the gas purifier & -\\
$\rho_{p}$    & molar density of GAr (LAr )& mol/m$^{3}$\\
$\delta\rho$ &  ratio of $\rho_{g}$ to ($\rho_{l}$ - $\rho_{g}$)  & -\\
$A_{surf}$ & surface area of liquid-gas interface& m$^2$\\
$A_{ads,p}$ & surface area of cryostat in contact with GAr (LAr)& m$^2$\\
$P_{0}$     & heat ``leakage" into LAr through imperfect cryostat insulation & W\\
$P_{in,H}$ & heat power provided by a heater immersed in LAr & W \\
$P_{in}$      & total heat power absorbed by LAr & W\\
$\Delta{}H_{evp}$  & molar enthalpy of vaporization of Ar &  J/mol  \\
$\Delta{}H_{sol}$  & molar enthalpy of solution of impurity in LAr &  J/mol  \\
\hline
\end{tabular}
\label{tab:eq_def}
\end{center}
\end{table}

The full model is described by a set of coupled ordinary differential equations describing the seven processes that govern the impurity concentrations in the liquid, the gas, and the surfaces.
Solving these differential equations with proper initial conditions and values of the physical parameters gives the impurity concentrations as a function of time.
Note that impurity concentrations in the liquid and gas are quantified by mole fractions, whereas impurities adsorbed on surfaces are quantified by moles per unit area. 
We make the following additional assumptions: (1) the impurity concentrations are very small, so that we can write mole fractions as the ratio of impurity to argon; (2) the extensive properties (density, thermal properties, etc.) of the impure mixtures in gas and liquid are identical to those of pure argon; (3) the temperatures in LAr and GAr in the cryostat are constant; and (4) there is no limit to the solubility of the impurity in either the gas or liquid.
Because both gas and liquid are always present, the pressure in the system is also constant, at the vapor pressure of argon at the operating temperature.  Small differences in temperature and pressure throughout the system are necessary to cause evaporation and condensation at the appropriate locations and to drive the flow of gas, but we assume that these differences are small enough that they do not significantly alter any of the parameters of the model. Aside from these approximations, there are no other assumptions of linearity in the model.

\subsection{The Processes}\label{subsec:allprocess}

\subsubsection{Impurity Exchange at the Gas-to-Liquid Interface}\label{subsec:process_1}

Transfer of an impurity from the gas phase into liquid is called dissolution and the opposite is called devolution. At each encounter of an impurity molecule with the surface, in each direction, there is a fixed probability of the molecule passing into the other phase.  The probability for an impurity molecule that strikes the surface to pass from gas into liquid (dissolution) is generally substantially different than the probability to pass from liquid into gas (devolution).

The number of molecules passing the surface out of one phase is proportional to the concentration of the solute, the rate of encounters with the surface, and the ``transmission" probability per encounter.
Therefore, the rate of passage (in mol/s) from phase 1 to phase 2 is
\begin{equation}
 \frac{dn_{i, 1 \rightarrow 2}}{dt} = \rho_{i, 1} \cdot p_{1 \rightarrow 2} \cdot A_{surf} \cdot \bar {s}_{i, 1}/4,\\
\end{equation}
which is Eq.~2 of Ref.~\cite{Nathanson1996}.
In this expression, $\rho_{i, 1}$ is the molar volumetric concentration (mol/m$^3$) of impurity $i$ in phase 1, $p_{1 \rightarrow 2}$ is the probability per encounter per unit surface area of a molecule passing from phase 1 into phase 2 ($0 \leq p_{1 \rightarrow 2} \leq 1$), $A_{surf}$ is the contact surface area (m$^2$), and $\bar {s}_{i, 1}$ is the mean molecular speed (m/s) of impurity molecules in phase 1. 
The number 4 comes from computing the integral of a differential element of flux, $\rho \cdot \cos(\theta) ds \, d\theta \, d\phi$, over the range $0\leq \phi \leq 2\pi$ and $0\leq \theta \leq \pi /2$ (i.e., all velocity vectors pointing towards the surface of the liquid) for a Maxwell velocity distribution, which gives a value of $\rho \bar {s}/4$.
The probability passing from the gas to the liquid, $p_{g \rightarrow l}$, is conventionally called the mass accommodation coefficient~\cite{Davidovits2011}.

The net rate of change of impurity molecules in one phase is the difference between the rates of those entering and those leaving. For the gas phase, we can write
\begin{equation}  \label{eq:mole rate in gas}
\begin{aligned}
  \frac{dn_{i, g}}{dt} = & - \frac{dn_{i, g \rightarrow l}}{dt} + \frac{dn_{i, l \rightarrow g}}{dt} \\
   = &  (-c_{i, g} \cdot \rho_{g} \cdot s_{i, g} \cdot p_{g \rightarrow l}+c_{i, l} \cdot \rho_{l} \cdot s_{i, l} \cdot p_{l \rightarrow g} ) \cdot A_{surf}/4. \\
 \end{aligned}
\end{equation}
Here we have used the molar densities of gas and liquid argon, $\rho_{g}$ and $\rho_{l}$, to replace the molar volumetric concentrations with the mole fraction concentrations $c_{i, g}$ and $c_{i, l}$, respectively:
\begin{equation}
\begin{aligned}
 c_{i, g} &=   \rho_{i,g} /  \rho_{g} , \\
 c_{i, l}  &=  \rho_{i,l} /  \rho_{l} . \label{eq:concentration definitions}
\end{aligned}
\end{equation}
Only impurity molecules traveling toward the surface and within a short distance from the surface (approximately one mean free path) in either phase have a substantial probability of passing into the other phase. 
The quantity of argon within a distance, $\lambda_g$, of the gas surface is
\begin{equation}
  n_{g} = \rho_{g} \cdot \lambda_g \cdot A_{surf} .
\label{eq:argon in gas}
\end{equation}
Using this relationship to eliminate $A_{surf}$ in Eq.~\eqref{eq:mole rate in gas}, the rate of change in the amount of impurity in the gas becomes
\begin{equation}
  \frac{dn_{i, g}}{dt} = n_{g} (-c_{i, g} \cdot k_{dis} + c_{i, l} \cdot k_{dev} ) ,
  \label{eq:concentration in gas} 
\end{equation}
with the transport rate constants expressed (in s$^{-1}$) as
\begin{equation}
\begin{aligned}
  k_{dis} & = \frac {1}{4} \cdot \frac {s_{i, g}}  {\lambda_{ g} } \cdot p_{g \rightarrow l}   \label{eq:kDis definition} ,\\
  k_{dev} & = \frac {1}{4} \cdot \frac {\rho_{l}} {\rho_{g}} \cdot \frac {s_{i, l} } {\lambda_{ g }} \cdot p_{l \rightarrow g} .
\end{aligned}
\end{equation}

The value of $k_{dis}$ can be estimated from the values of the parameters in this equation.
The mean speed of a Maxwell distribution is $\sqrt{8RT/\pi M}$ (R is the ideal gas constant and $M$ is molar mass), which is $\sim$23~m/s at 90~K.
The mean free path can be estimated from the diffusion coefficient as $\lambda = 3D/s_{i,g}$, where the diffusion coefficient ($D$) for oxygen in argon is about~ 0.02~cm$^2$/s at 90~K~\cite{Marrero1972}. 
No measurements of mass accommodation coefficients exist for LAr; assuming a very small value\footnote{Even for relatively insoluble molecules in liquid water near room temperature, the mass accommodation coefficients are larger than 0.0002~\cite{Davidovits2011, Nathanson1996}, and they increase rapidly with decreasing temperature.} of $10^{-5}$ gives a lower limit of 100~s$^{-1}$ for $k_{dis}$. We will use this value in our calculations (Sec.~\ref{sec:model_calc}).  

In order to maintain the rates of transport from one phase to the other, as given in Eq.~\eqref{eq:kDis definition}, either diffusion or convection in the two phases must be large enough to move impurities into and out of the two near-surface regions at a higher rate than they are transported through the surface, ensuring a uniform distribution throughout each phase. 
Since the diffusion rates in gas and liquid argon (typically $\sim$$10^{-2}$~cm$^2$/s and $\sim$$10^{-6}$~cm$^2$/s, respectively) are too small to have any effect in mixing even for small volumes,
to satisfy the assumption that spatial non-uniformities are negligible, we require large convection rates in the gas and liquid.  However, even if this is not the case, an approximate result may still be obtained for the average concentrations by using effective rates.  With little heat input into the system to produce thermal convection and without mechanical agitation, the impurity distributions become highly nonuniform~\cite{Bettini1991}.

From the above discussion, the differential equations for the amount of impurity in argon in each phase for the dissolution/devolution process are written as
\begin{equation}\label{eq:process1}
\begin{aligned}
  \left(\frac{dn_{i,g}}{dt}\right)_{1} &= n_{g} (-c_{i,g} \cdot k_{dis} + c_{i,l} \cdot k_{dev}),  \\
  \left(\frac{dn_{i,l}}{dt}\right)_{1} &=  - \left(\frac{dn_{i,g}}{dt}\right)_{1}, \\ 
  \left(\frac{dn_{l}}{dt}\right)_{1} &= 0, \\
  \left(\frac{dn_{g}}{dt}\right)_{1} &= 0, \\
  \left(\frac{dn_{i,sl}}{dt}\right)_{1} &= 0,\\
  \left(\frac{dn_{i,sg}}{dt}\right)_{1} &= 0.
\end{aligned}
\end{equation}
Here we have used conservation of mass to obtain the second equation. The parentheses with the subscript ``1'' on the derivatives in Eq.~\eqref{eq:process1} indicate that the rates apply only for process discussed in this subsection. Similarly, this format will be applied to the expressions for other processes, and the rates for quantities that do not change with time (i.e. are zero) will not be shown.

The competition between dissolution and devolution results in an equilibrium between the gas and liquid characterised by an equilibrium ``constant":
\begin{equation}\label{eq:henry_def}
K_{H,xx}(T) = \frac{k_{dev}(T)}{k_{dis}(T)},
\end{equation}
where the subscript $xx$ denotes that the ``constant" is defined in terms of mole fractions; note that these quantities are also temperature ($T$) dependent.

This equilibrium was first studied by the chemist William Henry for various gases dissolved in water~\cite{Henry:1803}. He observed that the concentration of a dissolved gas in water is proportional to its concentration in water vapor.
In steady state, all time derivatives are zero and from the first equation in Eq.~\eqref{eq:process1} (provided $n_g > 0$) we obtain 
\begin{equation}\label{eq:henry_law}
K_{H,xx} = \frac{c_{i,g}(T)}{c_{i,l}(T)}|_\text{equilibrium} \equiv H_{xx},
\end{equation}
which is the statement of Henry's Law.
The proportionality constant $H_{xx}$ is known as Henry's coefficient.  We can use Eqs.~\eqref{eq:henry_def} and~\eqref{eq:henry_law} to replace $k_{dev}$ in all expressions:
\begin{equation}
    k_{dev}=H_{xx} \cdot k_{dis}.
\end{equation}
In the following development only $H_{xx}$ and $k_{dis}$ will appear.

Since Henry's coefficient is an equilibrium constant, its temperature dependence is described by the van't Hoff equation~\cite{Sander:acp:2015}:
\begin{equation}\label{eq:vant_Hoff}
 \frac{d \log H_{xx}}{ d(1/T)} =  \frac{-\Delta{}H_{sol}}{R},
\end{equation}
where $R$ is the ideal gas constant and $\Delta{}H_{sol}$ is the enthalpy change of solution.
In the approximation that $\Delta H_{sol}$ is constant, this can be integrated to give the expression for the temperature dependence:
\begin{equation}\label{eq:H_T_relation}
H_{xx}(T) = H_{xx}(T_0) \, e^ {- \Delta H_{sol} \left(\frac {1} {T} - \frac {1} {T_0}\right) /R }.
\end{equation}

Henry's coefficient has been reported in many different forms in literature: as the ratio defined here and its inverse, and by using various other measures for the gas and liquid concentrations. These conventions, and conversions between them, are summarized in Ref.~\cite{Sander:acp:2015}.
The ratio used here is often called ``Henry's law volatility". For brevity we will simply use ``Henry's coefficient". 
Henry's coefficient spans in a very large range for different common impurities in LAr, for example, it is $\sim $10$^{-5}$ for Xe~\cite{yunker} and $\sim$4150  for He~\cite{karasz}.

Henry's coefficient for oxygen in LAr can be extracted from measurements reported in the literature. We have not found any direct statement of Henry's coefficient itself; instead what is commonly reported is the relative volatility of the two components of the solution as a function of the mole fraction of one of them~\cite{Clark517,Burn1962,Wilson1964,Wang1960}. 
The relative volatility is the ``double" ratio of mole fractions
\begin{equation}
\alpha_{1-2}(x_1) \equiv  \frac{K_{H}(1 \,in\, 2) }{K_{H}(2 \,in\, 1) } = \frac {y_1 / x_1} {y_2 / x_2} = \frac{x_2 \cdot (1-x_1)}{x_1 \cdot (1-x_2)},
\end{equation}
where $x$ and $y$ represent the equilibrium concentration of the components in liquid and gas phases, and the indices $1$ and $2$ are for the two components in a binary mixture (argon and the impurity), respectively, with $x_i+y_i=1, i=1$ or $2$. Note that $\alpha_{1-2}=\alpha_{2-1}^{-1}$. This quantity is often reported in the literature of mixtures because it directly indicates the difficulty of separating the two components by distillation. Henry's coefficient for oxygen in argon can be expressed in terms of the relative volatility by
\begin{equation}
H_{xx}(\text O_2 \:in\: \text {Ar}) = 1/\alpha_{Ar-O_2} (x_{Ar}=1).
\end{equation}
There is one paper reporting activity coefficient data for binary mixtures~\cite{Pool1962} from which we can calculate the Henry's coefficient for oxygen in LAr. In addition, we have found literature reporting calculations based on the Scatchard-Hildebrand theory of solutions~\cite{bazua} and on equation of states~\cite{BENDER1973, Kunz:REFPROP}, which can be used to derive Henry's coefficients for oxygen in LAr.
Table~\ref{table:HenryValues} lists eight values of Henry's coefficient for oxygen in LAr deduced from the literature; the mean value is $0.910\pm0.006$. 
Five of the papers~\cite{Clark517, Burn1962, Wang1960, bazua, Kunz:REFPROP} in this table provide data at more than one temperature. 
From the reported temperature variation of Henry's coefficient with temperature, the enthalpy change of solution ($\Delta H_{sol}$) can be determined using Eq.~\eqref{eq:vant_Hoff} as $0.20\pm0.03$~kJ/mol. This temperature dependence has been used to correct all the measurements to a common temperature of 88.9~K.
 
\begin{table}[!htbp]
\caption{\label{table:HenryValues}Values of Henry's coefficient for oxygen in argon at a temperature of 88.9~$K$, determined from the literature.  The type of data in the reference is indicated by $\alpha$ for relative volatility, $\gamma$ for activity coefficient, S-H for Scatchard-Hildebrand theory, and EOS for an equation of state of mixtures.}
\vspace{-0.5cm}
\begin{center}
\begin{tabular}{|c|c|c|c|}
\hline
Value            & Ref. & Type &Year \\\hline
0.907           & \cite{Clark517} & $\alpha$ & 1954 \\\hline 
0.913           & \cite{Wang1960} & $\alpha$ & 1960 \\\hline 
0.901           & \cite{Burn1962} & $\alpha$ & 1962 \\\hline 
0.885           & \cite{Pool1962} & $\gamma$ & 1962 \\\hline
0.903           & \cite{Wilson1964} & $\alpha$ & 1964 \\\hline
0.910           & \cite{bazua} & S-H &1971 \\\hline
0.940           & \cite{BENDER1973} & EOS & 1973 \\\hline
0.889           & \cite{Kunz:REFPROP} & EOS & 2012 \\\hline
\end{tabular}
\end{center}
\end{table}

\subsubsection{Evaporation of the Liquid}\label{process_2}
Heat input to the liquid, for example from imperfect cryostat insulation and electrical dissipation in electronic components in the liquid, causes the impure liquid to evaporate and both impurity and argon will be transferred from the liquid into the gas.
Heat can also be removed by a heat exchanger in the liquid or by cold surfaces in the gas ullage which deposit condensed gas into the liquid (a so-called ``raining condenser" ~\cite{andrews}), and these heat sinks will decrease the rate of evaporation.
The rates of change for the amount of impurity and argon in gas and liquid are given by 
\begin{equation}\label{eq:evaporation_eqs}
\begin{aligned}
  \left(\frac{dn_{i,l}}{dt}\right)_{2} &= - c_{i,l} \cdot r_{evp},\\
  \left(\frac{dn_{i,g}}{dt}\right)_{2} &= - \left(\frac{dn_{i,l}}{dt}\right)_{2}, \\
  \left(\frac{dn_{l}}{dt}\right)_{2} &= -r_{evp},\\
  \left(\frac{dn_{g}}{dt}\right)_{2} &= -\left(\frac{dn_{l}}{dt}\right)_{2},
\end{aligned}
\end{equation}
with $r_{evp}$ being the evaporation rate defined as 
\begin{equation}\label{eq:evaporation_rate}
r_{evp} \equiv \frac{P_{in}}{\Delta{}H_{evp}}= \frac{P_{0}+P_{in,H}+P_{in,P}-P_{out,exr}-P_{out,cdr}}{\Delta{}H_{evp}}.
\end{equation}
Here $P_{in}$ is the total heating power into the LAr, which is expressed as the sum of the cryostat heat ``leakage" entering the liquid ($P_{0}$), the electrical heater power ($P_{in,H}$), and the liquid pump power ($P_{in,P}$), minus the power removed by exchangers in the liquid ($P_{out,exr}$) and condensers in the ullage ($P_{out,cdr}$). $\Delta H_{evp}$ is the enthalpy of vaporization of LAr which is 6445.6~J/mol at 90~K as calculated from the equation of state of fluid argon from Ref.~\cite{Tegeler1999}.
For all comparisons of calculations to data in this paper we will set $P_{in,P}$ and $P_{out,exr}+P_{out,cdr}$ to 0, since we have neither a liquid pump nor any cold, condensing surfaces in the gas.
   
\subsubsection{Liquid Purification}\label{process_3}

An impurity can be removed from the liquid by pumping the impure liquid through a material (a ``scrubber'' or ``purifier'') which reacts with or absorbs the impurity. Once the liquid is purified, it is returned into the bulk.
In practice it is found that the purifier removes impurities (oxygen and water) efficiently,  therefore we assume the returned liquid is pure, and obtain the following equation for this process:
\begin{equation}\label{eq:purification}
  \left(\frac{dn_{i,l}}{dt}\right)_{3} = - c_{i,l} \cdot r_{rec},
\end{equation}
where $r_{rec}$ (mol/s) is the flow rate of liquid through the purifier.

\subsubsection{Gas Purification and Condensation}\label{process_4}
In a similar manner, an impurity in the gas can be removed by flowing the gas through a purifier. The purified gas then needs to be condensed into liquid and returned into the LAr bulk. 
The reaction of oxygen with the getter (GetterMax) is irreversible under all reasonable operating conditions, so it is 100\% efficient at removing oxygen until it is saturated. On the other hand, the sorption mechanism of the molecular sieve is thermally reversible, so depending on temperature, the degree of impurity loading, flow velocity, and flow path length, the efficiency can be less than 100\%. Especially for a gas purifier that is not at the liquid temperature, as is in our case, it may not remove all impurities from the argon.  To allow for this possibility, we introduce the purifier’s efficiency $\epsilon_{P}$ ($0 \leq \epsilon_{P} \leq 1$).
An additional advantage of introducing $\epsilon_P$ is that we can also use  $\epsilon_P=0$ to represent the case when the purifier is bypassed, and $\epsilon_P=1$ for a perfect purifier.
The equations for the purification and condensation processes are written as
\begin{equation}\label{eq:gaspurification_eqs}
\begin{aligned}
  \left(\frac{dn_{i,l}}{dt}\right)_{4} &= (1-\epsilon_{P}) \cdot c_{i,g} \cdot r_{evp},\\
  \left(\frac{dn_{i,g}}{dt}\right)_{4} &= - c_{i,g} \cdot r_{evp},\\
  \left(\frac{dn_{l}}{dt}\right)_{4} &=r_{evp},\\
  \left(\frac{dn_{g}}{dt}\right)_{4} &= - \left(\frac{dn_{l}}{dt}\right)_{4},
\end{aligned}
\end{equation}
where $r_{evp}$ is the evaporation rate expressed in Eq.~\eqref{eq:evaporation_rate}.

The evaporated gas can also be purified after condensing it to a liquid.  This can be done either by flowing the condensed impure liquid through a dedicated purifier, or, as is more often done for large detectors, by injecting it into the liquid purification stream before the liquid purifier~\cite{microboone2017}.  In the sense of this model, this is gas purification; the important requirement is that impure gas be removed from the ullage and purified before being returned as liquid to the bulk LAr.

\subsubsection{Leak from Outside of the System}\label{process_5}

In actual LAr detectors, leaks of air are almost always significant; for large detectors with many square meters of welded container surface and many feedthroughs, the leak rate can be large.
Of course these larger rates mix with larger masses of LAr, so all else being equal, the impurity concentrations due to leaks will scale inversely with detector size as the 2/3 power of the volume. 
In any case, it is important to include their effect.  For leakage into the volume of the gas phase 
the differential equation is 
\begin{equation}\label{eq:air_leak_eqs}
 \begin{aligned}
  \left(\frac{dn_{i,g}}{dt}\right)_{5} &= r_{lek},
 \end{aligned}
\end{equation}
where $r_{lek}$ is the leak rate of the impurity (in mol/s).  We assume mixing of the leaking gas with the bulk of the gas in the ullage. 
Impurity leakage into the liquid volume can be included by introducing an equation analogous to Eq.~\eqref{eq:air_leak_eqs} for the liquid volume, with an appropriate leak rate. 
Because our cryostat (Sec.~\ref{sec:setup}) is a simple welded stainless steel container that has been helium leak checked to $10^{-9}$~torr$\cdot$L/s we do not consider leakage into to the liquid volume here. Even in large cryostats the leakage into the liquid may be minimal, since impurities must enter the liquid by diffusion against the outward flow of argon through small holes in the walls. The leak rate will then be proportional to the diffusion constant which is $\sim$500 times smaller for the liquid than the gas~\cite{Medvedev2017}. 

We do not consider virtual leaks (volumes of trapped gas) in our model.  the rate of impurity leakage from  virtual leak has a complicated, model dependent history.  As the cryostat is cooled from room temperature the pressure of the trapped gas, presumably at atmospheric pressure, will be reduced to about 1/3 atmosphere. 
Argon, either gas or liquid will flow into the volume to equalize the pressure with the LAr.  During this time the impurity must diffuse out against the inward flow of argon, and after this the impurity will dissolve in the LAr and diffuse out through the LAr, which is about 100 times smaller than diffusion through gas.  In addition, the amount of impurity is limited. For reasonably sized trapped volumes, say less than 0.1~cm$^3$, the initial amount of impurity trapped as gas is less than 4 micro mole.  Even if all this is deposited instantly into the bulk LAr of a small system like ours ($\sim$700~moles) it results in an instantaneous increase of only 6 ppb; if it leaks over one hour period it contributes a leak rate of only $2\times10^{-12}$~mol/s.  Thus, unless the detector is exceptionally free of ``real" leaks, virtual leaks should at most be noticeable only for a short time after filling with LAr. If virtual leaks are to be considered, they could be included with a model similar to that suggested in the next subsection.

\subsubsection{Sorption on Surfaces}\label{process_6} 
The second source of impurity we consider is the adsorption and desorption (collectively sorption) of an impurity on the surfaces of the cryostat and any solid objects within it that are in contact with the gas or liquid.  Desorption rates are the analog of outgassing rates for a surface in a partial vacuum. Since there are only solid metal surfaces in our cryostat, we do not consider absorption (the binding of impurity in the bulk of the solid with diffusion to and from the surface). However, as for virtual leaks, a model similar to the one used here might be a reasonable first approximation.  Sorption is a thermally activated, reversible process describing the removal of a ``volatile'' impurity adsorbed on an ``inert" surface, but not chemically bound to the atoms of the surface material (i.e., physisorption as opposed to chemisorption). Since the surfaces in contact with GAr are relatively the warmer areas of the cryostat compared to surfaces in the liquid, one might expect the desorption rate into the gas volume from the upper most surfaces of the cryostat to dominate.  However, as the temperature increases the ability of surfaces to bind molecules (the ``saturation" of a surface) decreases, so colder surfaces in the LAr are able to hold and release relatively more impurity.  Therefore we consider the sorption process on surfaces in contact with the gas and liquid volumes separately.

The sorption process, like dissolution and devolution, is characterized by a pair of inverse reactions, desorption and adsorption, described by a pair of rate constants, $k_{des}$ and $k_{ads}$, respectively, whose ratio is an equilibrium constant, $K_{ad}\equiv k_{ads}/k_{des}$. 
However, for sorption the amount of material that can be bound on the surface is generally limited - the surface concentration saturates when the impurity concentration in the bulk phase increases. This alters the kinetics; we follow the pseudo-first order kinetics described in Ref.~\cite{Azizian2008} (see Eq.~1 there with $n=1$) and write the equations relating gas concentration to the surface concentration for the impurity as follows:
\begin{equation}\label{eq:outgassing_langmuir}
 \begin{aligned}
 \frac{d\theta_{g}}{dt} & = c_{i,g} \cdot k_{ads,g} \cdot (1-\theta_{g}) - k_{des,g} \cdot \theta_{g}, \\
 c_{i,sg} & \equiv \frac{n_{i,sg}}{A_{ads,g}}, \\
 c_{i,sg}^{sat} & \equiv \frac{n_{i,sg}^{sat}}{A_{ads,g}}, \\
 \theta_{g} & \equiv \frac{c_{i,sg}}{c_{i,sg}^{sat}}. \\
 \end{aligned}
\end{equation}
Here subscript ``g" is used to denote the quantities in the gas phase. $n_{i,sg}$ is the amount of impurity (in mol) absorbed on the surface in contact with GAr, $A_{ads,g}$ is the surface area, $c_{i,sg}$ (in mol/m$^{2}$) is the impurity concentration on the surface, $n_{i,sg}^{sat}$ and $c_{i,sg}^{sat}$ are the saturation (maximum) impurity amount and concentration, respectively, that can be bound to the surface, and $\theta_{g}$ is the fraction of the maximum possible (saturation) surface coverage ($0\leq \theta_{g} \leq1$). 

From Eq.~\eqref{eq:outgassing_langmuir} and the conservation of impurity between the gas and surface, we write the following equations for this process:
\begin{equation}\label{eq:outgassing_1}
\begin{aligned}
\left(\frac{dn_{i,sg}}{dt} \right)_{6} & = n_{i,g}^{sat} \cdot k_{des,g}\left ( c_{i,g}  \cdot K_{ad,g} \cdot (1-\theta_g) -   \theta_g \right ), \\
\left( \frac{dn_{i,g}}{dt} \right)_{6} &= -\left(  \frac{dn_{i,sg}}{dt} \right)_{6},
\end{aligned}
\end{equation}
where we have used the definition of the equilibrium constant, $K_{ad,g} \equiv k_{ads,g}/k_{des,g}$,  to replace $k_{ads}$ with $k_{des}$.  Note that the molar rate of change into the gas can be either positive or negative, depending on whether the first term on the right is smaller or larger than the second, respectively.  The maximum positive (negative) rate into the gas occurs when $\theta_g=1$ ($\theta_g=0$).

In steady state, the rate in Eq.~\eqref{eq:outgassing_1} is zero, and solving for $\theta_g$ yields the well-known Langmuir isotherm equation (see Ref.~\cite{Langmuir1918}):
\begin{equation}\label{eq:langmuirKinetics}
\begin{aligned}
\theta_{g} & = \frac{c_{i,g}\cdot K_{ad,g}}{1+ c_{i,g}\cdot K_{ad,g}}.
\end{aligned}
\end{equation}
 In the limit of zero impurity concentration in the gas, the ratio $\theta_{g}/c_{i,g}$ approaches $K_{ad,g}$. This is the Henry's coefficient for the sorption process on surface in contact with gas.  Since $\theta_g$ is dimensionless, if concentrations are expressed in ppm, $K_{ad,g}$ has dimension of ppm$^{-1}$. 

Sorption on the surface in contact with the liquid is also included in the model, by introducing a set of equations identical to  Eq.~\eqref{eq:outgassing_1}, except that the subscript ``$g$" is replaced by ``$l$" everywhere:
\begin{equation}\label{eq:sorption_in_liquid}
\begin{aligned}
\left(\frac{dn_{i,sl}}{dt} \right)_{6} & = n_{i,l}^{sat} \cdot k_{des,l}\left ( c_{i,l}  \cdot K_{ad,l} \cdot (1-\theta_l) -   \theta_l \right ), \\
\left( \frac{dn_{i,l}}{dt} \right)_{6} &= -\left(  \frac{dn_{i,sl}}{dt} \right)_{6}.
\end{aligned}
\end{equation}
The dependence of the impurity coverage on the surface in the liquid on the liquid impurity concentration is similarly described by a Langmuir isotherm characterized by $K_{ad,l}$, the Henry's sorption coefficient for the liquid-surface contact.

\begin{figure}[!htbp]
\vspace{-0.1cm}
\centering
\includegraphics[width=0.75\textwidth]{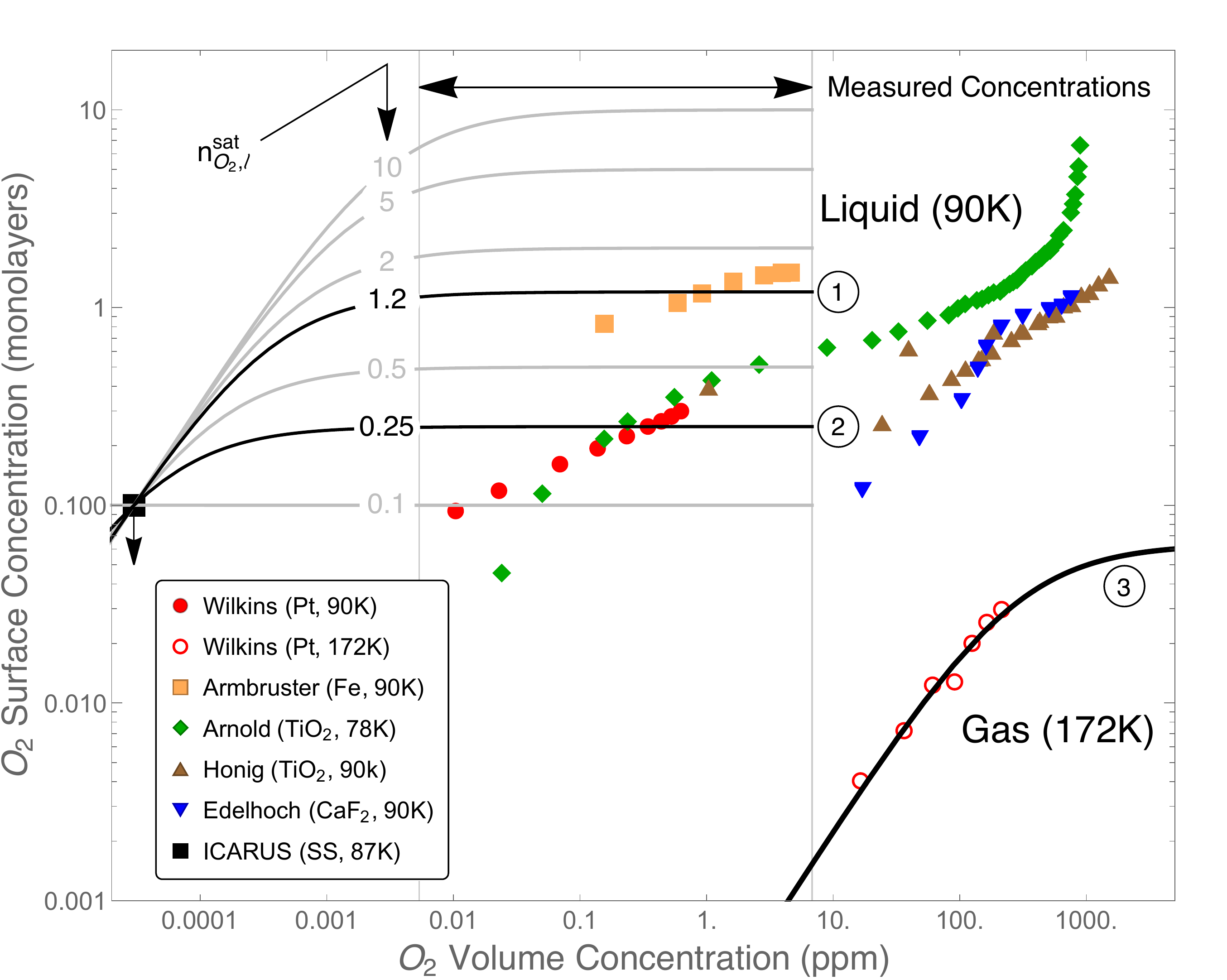}
\caption{Isotherms for sorption of oxygen on several non-porous surfaces as reported in the literature.  The data are taken from  Wilkins~\cite{Wilkins_Pt}, Armbruster~\cite{Armbruster_FeO2}, Arnold~\cite{Arnold_TiO2}, Honig~\cite{Honig_TiO2}, Edelhoch~\cite{Edelhoch_CaF2}, and ICARUS~\cite{Bettini1991}.   The three lines labeled \textcircled{1} to \textcircled{3} are the best fits of a Langmuir isotherm to: (1) the data of Armbruster constrained to pass through the single point from ICARUS~\cite{Bettini1991}; (2) the data of Wilkins at 90~K, Arnold, Honig, and Edelhoch also constrained to the point from ICARUS; and (3) the data of Wilkins at 172~K.} 
\label{fig:sorption_data}
\end{figure}

There is very little quantitative information about the sorption of oxygen on surfaces at cryogenic temperatures. 
We have found data on sorption of oxygen on platinum~\cite{Wilkins_Pt}, iron~\cite{Armbruster_FeO2}, TiO$_2$~\cite{Arnold_TiO2,Honig_TiO2}, and CaF$_2$~\cite{Edelhoch_CaF2}, which are shown in Fig.~\ref{fig:sorption_data}.
We have not found any literature on oxygen sorption on stainless steel.  It is important to note that argon is also adsorbed on these materials, although to a lesser degree~\cite{Armbruster_FeAr}, and will compete with oxygen for surface coverage: we ignore this, so the coverage in these isotherms should be considered upper limits for oxygen sorption in the presence of argon.  To generate Fig.~\ref{fig:sorption_data}, the reported amounts adsorbed on the surface have been converted to surface coverage in monolayers, using a value for the adsorbed area occupied by a single oxygen molecule of 0.136~nm$^2$~\cite{Adsorp_Xsection}.
The reported pressures of oxygen have been converted to concentrations by assuming a mixture of ideal gases. 
Note that the surface saturation is much larger in the liquid (90~K) than in the gas (172~K), which might be expected since oxygen at normal pressure is a liquid at the LAr temperature.

Although the data shown in Fig.~\ref{fig:sorption_data} are for sorption of oxygen on materials other than stainless steel (the material of choice for LAr cryostats) and there is inconsistency among them, they probably provide useful estimates for our purposes. 
This is the case because for non-porous materials (those for which the geometrical area of a sample is equal to the effective surface available for absorption) the isotherms are generally similar, as is illustrated by the isotherms below 10 ppm for Pt at 90~K and TiO$_2$ at 90~K and 78~K in the figure. The exception is the isotherm for Fe at 90~K, which has a saturation coverage about three times higher than the others.  This is probably due to the fact that the adsorption of oxygen on iron is primarily chemical in nature (i.e. not thermally reversible).   Indeed, other sources report ``no firm evidence of physical adsorption"~\cite{Gulbransen1942,McClellan_SS_Sorp} for steel, stainless steel, and chromium at room temperature.  For this reason, we will consider the data for Fe at 90~K to be an upper limit on saturated oxygen surface concentration.

A single value for adsorption of oxygen on stainless steel at cryogenic temperatures can be deduced from measurements of the ICARUS Collaboration~\cite{Bettini1991}. 
They observe an increase in the surface coverage of less than 0.1 monolayers in equilibrium with an oxygen concentration of 0.03~ppb in the LAr. 
With the reasonable assumption that the initial coverage obtained after evacuating the cryostat at room temperature is negligible, we conclude an upper limit of 0.1 monolayers at 0.03~ppb. This value is shown as the point labeled ICARUS in Fig.~\ref{fig:sorption_data}. 
Since this measurement, which includes the effect of competition with Ar sorption, is much larger than the coverage implied by extrapolating any of the other isotherm data sets measured at 90~K to 0.03~ppb, the assumption that sorption of Ar does not significantly alter oxygen sorption may well be justified.

To obtain the sorption parameters to be used in numerical calculations with our model, we have required the Langmuir function, Eq.~\eqref{eq:langmuirKinetics}, to pass through the ICARUS data point.   That provides a constraint on the the two parameters of the Langmuir function.  The  remaining parameter to be determined is the saturation oxygen concentration on the surface, which we take to be the mean of the literature data within the range of concentrations observed in the measurements to be discussed below. 
The line labeled \textcircled{1} uses the mean of the data for Fe at 90~K, and should be taken as a ``high estimate" isotherm for 90~K. 
The line labeled \textcircled{2} uses the mean of the data for Pt and CaF$_2$ at 90~K and TiO$_2$ at both 78 and 90~K, and should be considered a ``low estimate" for sorption to surfaces in the LAr. The figure also shows the Langmuir functions for other values of the saturation oxygen concentration on the surface as gray lines labeled with the saturation concentration in monolayers.
The line labeled \textcircled{3} is a fit for two parameters to the data for Pt at 172~K and will be taken as the ``best estimate" for sorption for surfaces in the GAr. The sorption parameters for each of the three sets we use in the model are summarized in Table~\ref{tab:sorption_params}.

\begin{table}[!htbp]
\caption{Values of sorption parameters for O$_2$ on smooth surfaces obtained by fitting the data shown in Fig.~\ref{fig:sorption_data}.  Each set number in the table represents the curve with that number in Fig.~\ref{fig:sorption_data}. These values will be used, without adjustment, in the numerical calculations presented in Sec.~\ref{sec:model_calc}.  Note that $K_{ad} \equiv k_{ads}/k_{des}$}
\vspace{-0.6cm}
\begin{center}
\begin{tabular}{|c|c|c|c|c|c|}
\hline
Set  &  T(K)   & $c_{i,s}^{sat}$~(monolayers) & $K_{ad}$~(ppm$^{-1}$)  &    $k_{des}$~(s$^{-1}$)       & Source\\
\hline
  S1   &  90  &           1.2         & 3200         &  $1\times10^{-6}$   & ~\cite{Armbruster_FeO2, Bettini1991}\\\hline
  S2   &  90  &         0.25         &     8100        &  $1\times10^{-6}$   & ~\cite{Wilkins_Pt, Arnold_TiO2, Honig_TiO2, Edelhoch_CaF2, Bettini1991}\\\hline
  S3   & 195 &          0.063      &     0.0036  &  $1\times10^{-3}$   & ~\cite{Wilkins_Pt}\\
\hline
\end{tabular}
\label{tab:sorption_params}
\end{center}
\end{table}

The Langmuir isotherm assumes that each adsorbed molecule has the same interaction energy with the surface and that there is only one molecule adsorbed per site. Thus it is appropriate for describing adsorption of less than one monolayer, and it can be used only to describe the low coverage portion of the isotherms.
Fortunately this is the region most relevant for oxygen in LArTPCs.  At high oxygen concentrations in LAr, multiple layers can be adsorbed (see the region above 100~ppm in Fig.~\ref{fig:sorption_data}) and other isotherms such as the Generalized Statistical Thermodynamic Adsorption (GSTA)~\cite{GST_Model_First,Ladshaw_GSTA} or the Brunauer-Emmett-Teller (BET)~\cite{BET_Isotherm} are required to describe the high concentration range.  The BET model does not give a better representation of the data in this concentration range; it is intended to represent multi layer surface concentrations. 
The GSTA can provide a better description of the shape of the isotherms over the entire range, but requires many more parameters to describe the concentration range. Considering the quality of the existing sorption data, this extra complexity is not justified. 

There are very few data in the literature on the rates of sorption for oxygen in LAr. The ICARUS Collaboration has observed rapid adsorption of oxygen in LAr on clean stainless steel walls of a small cryostat~\cite{Bettini1991}, and conclude that sorption from surfaces in the liquid can significantly contribute to impurity concentrations.  These measurements were made under isothermal conditions, with no heat transfer into the LAr, so that there is no convection.  They show a slow decrease in the concentration over time as the oxygen diffuses to the walls and is adsorbed; stirring the LAr causes the concentration to decrease rapidly, as the oxygen is moved rapidly to the walls.
During the periods of stirring, the change in oxygen concentration for the data shown in Fig.~9 of Ref.~\cite{Bettini1991} can  be used to estimate a desorption rate constant of $k_{des} \simeq 10^{-6}$~s$^{-1}$.  During the periods with no stirring only diffusion mixes the liquid, and the effective sorption rates are then about 20 times slower. Armbruster~\cite{Armbruster_FeO2} shows data implying a value of $k_{des}$ at 195~K of about $10^{-3}$~s$^{-1}$.  We take  this as the desorption rate constant at room temperature. The desorption rate constants used in the calculations are shown in Table~\ref{tab:sorption_params}. 

\subsubsection{Loss of Ar Due to Sampling}\label{process_7}

Samples of GAr or LAr can be withdrawn from LArTPCs as required to monitor impurity concentrations. The instruments used for this purpose require a sufficient flow rate of gas to obtain results with a reasonable accuracy and response time. However, with continuous sampling for a long period in small systems like the one described in Sec.~\ref{sec:measurement}, the amount of Ar lost can be significant compared to the remaining amount. In that case the sampling process should be taken into account.

Here we consider sampling from both LAr or GAr, although the measurements described here use only liquid sampling. The effect of sampling can be described as the sum of two steps.
In the first step, an amount of (impure) argon (either liquid or gas) is removed instantaneously from the system (with no gas-liquid contact). In the second step, some additional LAr must be evaporated to increase the amount of gas in order to keep the total volume and the gas pressure constant.  The sum of the amounts of gas and liquid for the two steps is equal to the amount of liquid removed by sampling.
In the case of sampling from LAr, writing the equations for the total amount and total volume of the two phases in terms of a sampling rate, $r_{sam}$ (in mol/s), and solving for the rate of change in moles in the two phases gives the following differential equations:
\begin{equation}\label{eq:sampling_eqs_liq}
 \begin{aligned}
  \left(\frac{dn_{i,l}}{dt}\right)_{7} &= - c_{i,l} \cdot r_{sam}\cdot(1+  \delta\rho),\\
  \left(\frac{dn_{i,g}}{dt}\right)_{7} &= c_{i,l} \cdot r_{sam} \cdot \delta\rho,\\
  \left(\frac{dn_{l}}{dt}\right)_{7} &= - r_{sam}\cdot{}(1+\delta\rho),\\
  \left(\frac{dn_{g}}{dt}\right)_{7} &= r_{sam} \cdot \delta\rho,
 \end{aligned}
\end{equation}
where $\delta\rho$ is a density ratio defined as
\begin{equation}\label{eq:delta_rho}
 \delta \rho \equiv \frac{\rho_{g}}{\rho_{l}-\rho_g},
\end{equation}
with $\rho_g$ ($\rho_l$) denoting the molar density of GAr (LAr). For argon at 90~K, $\delta\rho=0.005$. The sampling rate is not required to be constant.  For convenience, for the analysis to be discussed, we take the sampling rate to be a step function of time. 

If the gas is sampled instead, the first two equations in Eq.~\eqref{eq:sampling_eqs_liq} for the rate of change of impurity amounts are replaced by:
\begin{equation}\label{eq:sampling_eqs_gas}
 \begin{aligned}
  \left(\frac{dn_{i,l}}{dt}\right)_{7} &= - c_{i,g} \cdot r_{sam}\cdot(1+  \delta\rho),\\
  \left(\frac{dn_{i,g}}{dt}\right)_{7} &= c_{i,g} \cdot r_{sam} \cdot \delta\rho,\\
  \end{aligned}
\end{equation}
while the other two equations remain the same.
Since only the liquid is sampled in our measurements, we do not include Eq.~\eqref{eq:sampling_eqs_gas} in  the following discussion.

%% file: allEquations_4thOrder.tex
\subsection{Full Model: Coupled Differential Equations with All Processes}\label{summation_of_eq} 

We obtain the differential equations for the entire system by summing over all seven processes. The resulting equations for the rate of change of the amounts of GAr and LAr are
\begin{equation}\label{eq:argonquant_diffeqs}
 \begin{aligned}
  \frac{dn_{l}(t)}{dt} &= - r_{sam}\cdot{}(1+\delta\rho),\\
  \frac{dn_{g}(t)}{dt} &= r_{sam} \cdot \delta\rho.
  \end{aligned}
\end{equation}
If we restrict the sampling rate to be constant, the solution of these equations produces a linear decrease in the liquid and linear increase in the gas with time:
\begin{equation}\label{eq:argonquant}
 \begin{aligned}
  n_{l}(t)&=n_{0,l}-r_{sam} \cdot  t \cdot (1+\delta\rho),\\
  n_{g}(t)&=n_{0,g}+r_{sam} \cdot  t \cdot \delta\rho,
 \end{aligned}
\end{equation}
where $n_{0,l}$ and $n_{0,g}$ are the amounts of Ar in the liquid and gas phases at $t=0$, respectively. 
These expressions for the total amounts of GAr and LAr can be substituted into the summed differential equations for impurity concentrations in the LAr, GAr, and on the surfaces to obtain four differential equations, for the time evolution of the impurity concentrations in each of these four phases:
\begin{equation}\label{eq:dcigdt}
  \begin{aligned}
 & a_1 \cdot c_{i,g}(t) + a_2 \cdot c_{i,l}(t) + a_3 \cdot \theta_{l}(t) + a_4 \cdot \theta_{l} (t)\cdot c_{i,l}(t) + a_5 \cdot \frac {dc_{i,l}(t)} {dt} = 0,\\
 & a_6 + a_7 \cdot c_{i,g}(t) + a_8 \cdot c_{i,l}(t) + a_9 \cdot \theta_{g}(t) + a_{10} \cdot \theta_{g}(t) \cdot c_{i,g} (t) + a_{11} \cdot \frac {dc_{i,g}(t)} {dt} = 0, \\
 & a_{12} \cdot c_{i,l}(t) + a_{13} \cdot \theta_{l} (t) + a_{14} \cdot c_{i,l}(t) \cdot \theta_{l}(t) + a_{15} \cdot \frac {d\theta_{l}(t)} {dt} = 0,  \\
 & a_{16} \cdot c_{i,g}(t) + a_{17} \cdot \theta_{g} (t) + a_{18} \cdot c_{i,g}(t) \cdot \theta_{g}(t) + a_{19} \cdot \frac {d\theta_{g}(t)} {dt} = 0,
  \end{aligned}
\end{equation}
The coefficients in these equations are expressed in terms of the physical parameters as follows:
\begin{equation}\label{eq:basic_coefficients}
\begin{aligned}
a_1 &= -k_{dis} \cdot (n_{0,g}+\delta \rho \cdot r_{sam}\cdot t) - r_{evp} \cdot (1-\epsilon_P),\\
a_2 &=  H_{xx}\cdot k_{dis} \cdot (n_{0,g}+\delta \rho \cdot r_{sam}\cdot t) + K_{ad,l} \cdot k_{des,l} \cdot n_{i,sl}^{sat}  + r_{evp} + r_{rec},\\
a_3 &=  -k_{des,l} \cdot n_{i,sl}^{sat},\\
a_4 &= - K_{ad,l} \cdot k_{des,l} \cdot n_{i,sl}^{sat} ,\\
a_5&=  n_{0,l}-(\delta \rho +1) \cdot r_{sam}\cdot t,\\
a_6 &=  r_{lek},\\
a_7 &=  -K_{ad,g}\cdot k_{des,g} \cdot n_{i,sg}^{sat} - k_{dis} \cdot (n_{0,g}+\delta \rho  \cdot r_{sam} \cdot t) - r_{evp} - \delta \rho \cdot r_{sam},\\
a_8 &=  H_{xx}\cdot k_{dis}\cdot (n_{0,g}+\delta \rho \cdot r_{sam}\cdot t)+\delta \rho \cdot r_{sam}+r_{evp},\\
a_9 &= k_{des,g} \cdot n_{i,sg}^{sat},\\
a_{10} &=  K_{ad,g} \cdot k_{des,g} \cdot n_{i,sg}^{sat},\\
a_{11} &=  -n_{0,g}-\delta \rho \cdot r_{sam}\cdot t,\\
a_{12} &=  -K_{ad,l} \cdot k_{des,l},\\
a_{13} &=  k_{des,l},\\
a_{14} &=  K_{ad,l} \cdot k_{des,l},\\
a_{15} &=  1, \\
a_{16} &=  -K_{ad,g} \cdot k_{des,g},\\
a_{17} &=  k_{des,g},\\
a_{18} &=  K_{ad,g} \cdot k_{des,g},\\
a_{19} &=  1.
\end{aligned}
\end{equation}

Note that even if all the parameters in these expressions are assumed to be constants, an explicit time dependence is introduced by the sampling process into six of the coefficients ($a_1$, $a_2$, $a_5$, $a_7$, $a_8$, and $a_{11}$).  However, it is not necessary for any of the other parameters to be constants if the differential equations are to be solved numerically.

%% file: discuss_steady_state.tex
\subsection{Impurity  Concentrations at Steady State}\label{discuss_full_model}

The set of four coupled first order differential equations in Eq.~\eqref{eq:dcigdt} can be reduced to a single fourth order differential equation for each of the four impurity concentrations.  However, since this differential equation is nonlinear, and cannot be solved to obtain a closed form function, it offers little insight into the time dependence of the impurity concentrations.  Note that the non-linearity is introduced only by the sorption processes, which appear as the quadratic terms, $c_{i,l}(t) \cdot \theta_{l}(t)$ and $c_{i,g}(t) \cdot \theta_{g}(t)$ in the four differential equations of Eq.~\eqref{eq:dcigdt}.
If we remove the sorption processes, we are left with a set of two coupled linear first order equations with time dependent coefficients (introduced by the sampling process),  which are still not solvable in terms of the elementary functions of analysis.   However, if we further neglect the sampling processes, the resulting linear, second order differential equation with constant coefficients can be solved easily, and in Sec.~\ref{sec:time} we will present and discuss this solution.

Before solving the simplified differential equation, we find the steady state concentrations, i.e., the impurity concentrations achieved after a sufficiently long time that all time derivatives are zero.  
Such a steady state cannot be achieved with the sampling process: if continued long enough, sampling will remove all argon and impurities, at which time all concentrations are undefined.  Therefore we assume no sampling, and find the following  expressions for the concentrations  during extended stable operation of a LArTPC:
\begin{equation}\label{eq:coefficients1}
\begin{aligned}
c_{i,l}^{ss} &=  \frac{r_{lek} \cdot (k_{dis} \cdot n_{0,g} + r_{evp} \cdot (1-\epsilon_P))} {k_{dis} \cdot n_{0,g} \cdot (H_{xx} \cdot r_{evp} \cdot \epsilon_P  + r_{rec}) + r_{evp} \cdot (r_{evp} \cdot \epsilon_P+ r_{rec})}, \\
c_{i,g}^{ss} &= \frac{r_{lek} \cdot (H_{xx} \cdot k_{dis} \cdot n_{0,g}+r_{evp}+r_{rec})} {k_{dis} \cdot n_{0,g} \cdot (H_{xx} \cdot r_{evp} \cdot \epsilon_P+r_{rec})+r_{evp} \cdot (r_{evp} \cdot \epsilon_P+r_{rec})}, \\
\theta_{i,l}^{ss} &= \frac{c_{i,l}^{ss}\cdot K_{ad,l}}{1+c_{i,l}^{ss}\cdot K_{ad,l}},\\
\theta_{i,g}^{ss} &= \frac{c_{i,g}^{ss}\cdot K_{ad,g}}{1+c_{i,g}^{ss}\cdot K_{ad,g}}.
\end{aligned}
\end{equation}
The superscript ``ss" in the expressions denotes the steady state.  The third and fourth expressions reduce to the Langmuir expressions, Eq.~\eqref{eq:langmuirKinetics}, evaluated at the steady state concentrations in the gas and liquid. 

Note that if there is no leak, and with any purification, all of the impurity concentrations go to zero. With a non-zero leak rate, and no purification, the impurity concentrations in gas and liquid approach infinity and the surface coverages go to one.  For a finite, non-zero, leak rate the steady state concentrations can be reduced toward zero by increasing the liquid purification rate and/or the evaporation rate (assuming that the gas is purified, $\epsilon_P>0$).  These expressions also show that the steady state ratio of gas to liquid concentration does not give Henry's coefficient unless $k_{dis}$ is sufficiently large, or unless there is no purification.  In simple language, leaks, and rapid impurity dissolution prevent achieving highest ultimate purity; whereas large liquid re-circulation rate, high evaporation rate, efficient purification of the gas, and large Henry's coefficient improve the ultimate purity of the liquid.

Purifying large amounts (moles) of liquid is much easier than purifying the same amount of gas, simply because the molar volume is so much larger for gas. Assuming a Henry's coefficient of one, achieving the same purity by replacing liquid purification with gas purification requires a volumetric flow rate of gas $\sim$200 ($=\rho_l/\rho_g$) times that of liquid, and a power input of 223~kW ($=\Delta H_{evp} \cdot \rho_l \cdot 1$~L/s) into the liquid to replace 1~L/s of liquid flow.
Adding this power to the liquid and then removing it to condense the gas would incur a large operating cost.  Of course, with an increased capital expense, this cost for power could be reduced by condensing the gas in an efficient heat exchanger and using the heat output to evaporate the bulk liquid.  Gas purification alone is clearly not a practical solution to achieve ultimate high purity in large LArTPCs. However, large gas flow rates can provide a significant reduction in the transport of impurities to the liquid/gas surface, reducing the effective impurity source rate from leaks and desorption, thereby increasing the ultimate purity of the LAr. This is further discussed in Sec.~\ref{sec:baffle}.

Another special case without liquid purification is worth discussion, since we will present measurements in Sec.~\ref{sec:measurement} made with gas purification only.  Setting $r_{rec}=0$ in Eq.~\eqref{eq:coefficients1} we obtain the following steady state concentrations:
\begin{equation}\label{eq:steadystate_gen}
\begin{aligned}
c_{i,l}^{ss} &=  \frac{r_{lek} \cdot \left (k_{dis} \cdot n_{0,g} + r_{evp} \cdot (1 -\epsilon_P) \right)} {r_{evp} \cdot  \epsilon_P \cdot (k_{dis} \cdot n_{0,g} \cdot H_{xx} + r_{evp})}, \\
c_{i,g}^{ss} &= \frac {r_{lek}} {r_{evp} \cdot \epsilon_P}. \\
\end{aligned}
\end{equation}
The impurity concentration in the gas is just the ratio of the leak rate to the purification rate (the ratio of source to sink rates) as would be expected.
Recall from the discussion following Eq.~\eqref{eq:kDis definition} that $k_{dis}$ depends on the mean speed, mean free path and mass accommodation coefficient of the mixture, and is estimated to be 100~s$^{-1}$. 
If the dissolution rate is large enough (which means $k_{dis} \cdot n_{0,g} \gg r_{evp}$), the impurity concentration in the liquid becomes
\begin{equation}\label{eq:steadystate_liq}
c_{i,l}^{ss}\ =  \frac{r_{lek}}  {H_{xx}\cdot  r_{evp}\cdot \epsilon_P} ,\\
\end{equation}
and computing the ratio $c_{i,g}^{ss}/c_{i,l}^{ss}$ we recover the Henry's Law (Eq.~\eqref{eq:henry_law}). 

At the other extreme, as $k_{dis} \to 0$ we have
\begin{equation}\
c_{i,l}^{ss}\ =  \frac{r_{lek}\cdot (1-\epsilon_P)}  {r_{evp}\cdot \epsilon_P} ,
\end{equation}
and if $r_{lek} \neq 0$ and $\epsilon_{P} \neq 0$ then the ratio of the gas to liquid concentrations for the impurity becomes
\begin{equation}\label{eq:2.37}\
  \frac {c_{i,g}^{ss}} {c_{i,l}^{ss}}\ =  \frac{1}  {1-\epsilon_P} ,
\end{equation}
which is obviously not  Henry's law.  There is no contradiction, because in this case no impurity passes between the gas and the liquid, and an equilibrium between the gas and liquid impurity concentrations can not be achieved, as is required by Henry's law.  The two limits with respect to $\epsilon_P$ are easily understood: with perfect purification ($\epsilon_P\rightarrow 1$) the condensed gas introduces no impurity to the liquid, but impurity leaves the liquid by evaporation, so the concentration in the liquid goes to zero; with no purification ($\epsilon_P\rightarrow 0$) the gas and liquid are continually mixed by evaporation and re-condensation, and the gas and liquid concentrations approach equality.

%% file: time_CET.tex

\section{Reduced Model and Its Closed Form Solution}\label{sec:time}

\subsection{Solution of a Reduced Model}
In order to obtain an insight into the effects of the various processes on the time evolution of impurities in a LAr detector, it is useful to solve the differential equations describing the system.
As was pointed out in the previous section, a solution in simple closed form can only be obtained if the sampling and sorption processes are neglected. 
This is not only a convenient way to produce an easily understood solution, but as we will show for oxygen impurity in LAr in Sec.~\ref{sec:measurement}, it is also a reasonable approximation to the full solution. However, it is not a reasonable approximation for water impurity in LAr.

Making this approximation reduces the four equations of Eq.~\eqref{eq:dcigdt} to the following two coupled equations \begin{equation}
\begin{aligned}
 & a_1^{(0)} \cdot c_{i,g}(t) + a_2^{(0)} \cdot c_{i,l}(t) + a_5^{(0)} \cdot \frac {dc_{i,l}(t)} {dt} =0,\\
 &  a_6^{(0)} + a_7^{(0)} \cdot c_{i,g}(t) + a_8^{(0)} \cdot c_{i,l}(t)  + a_{11}^{(0)} \cdot \frac {dc_{i,g}(t)} {dt} =0,
  \end{aligned}
\end{equation}
where the coefficients $a_n^{(0)}$ are the corresponding $a_n$ coefficients of Eq.~\eqref{eq:basic_coefficients} with $r_{sam}$, $K_{ad,l}$, and $k_{des,l}$, $K_{ad,g}$, and $k_{des,g}$ all set to zero. With $r_{sam}=0$, the time dependence of the coefficients disappears and all coefficients are constants.
These two coupled first order equations can easily be converted into two uncoupled second order equations for the impurity concentrations in the gas and liquid:
\begin{equation}
\label{eq_reduced2ndorder}
\begin{aligned}
 & b_{0,g} +b_1 \cdot c_{i,g}(t) + b_2 \cdot  \frac {dc_{i,g}(t)} {dt} + \frac {dc_{i,g}^2(t)} {dt^2} =0,\\
 & b_{0,l} +b_1 \cdot c_{i,l}(t) + b_2 \cdot  \frac {dc_{i,l}(t)} {dt} + \frac {dc_{i,l}^2(t)} {dt^2} =0,\\
  \end{aligned}
\end{equation}
where the $b$ coefficients are related to the $a_{n}^{(0)}$ coefficients by 
\begin{equation}\label{eq:reduced_coeffs}
b_{0,l} = - \frac {a_1^{(0)} \cdot a_6^{(0)}} {a_5^{(0)} \cdot a_{11}^{(0)}}, \ b_{0,g} = \frac {a_2^{(0)} \cdot a_6^{(0)}} {a_5^{(0)} \cdot a_{11}^{(0)}}, \ b_1 = \frac {a_2^{(0)}\cdot a_7^{(0)} - a_1^{(0)} \cdot a_8^{(0)}} {a_5^{(0)} \cdot a_{11}^{(0)}}, \ b_2 = \frac {a_5^{(0)} \cdot a_7 ^{(0)}+ a_2^{(0)} \cdot a_{11}^{(0)}} {a_5^{(0)} \cdot a_{11}^{(0)}}.
\end{equation}
The solutions to Eq.~\eqref{eq_reduced2ndorder} are 
\begin{equation}\label{eq:two_eq_solution}
\begin{aligned}
c_{i,l}(t) & =  c_{i,l}^{ss} + C_1 \cdot e^{-k_F t} + C_2 \cdot e^{-k_S t},\\
c_{i,g}(t) & = c_{i,g}^{ss} + C_3 \cdot e^{-k_F t} + C_4 \cdot e^{-k_S t}.
\end{aligned}
\end{equation}
The coefficients $C_1$-$C_4$ are determined by the initial conditions and the steady state concentrations ($c_{i,l}^{ss}$ and $c_{i,g}^{ss}$) are given by Eq.~\eqref{eq:coefficients1}.  The two rate constants characterizing the time dependence are
\begin{equation}\label{eq:rate_constants}
\begin{aligned}
k_F = & \frac{1}{2} \left(b_2 + \sqrt{b_2^2 - 4 b_1}\right), \\
k_S = & \frac{1}{2} \left(b_2 - \sqrt{b_2^2 - 4 b_1}\right), \\
\end{aligned}
\end{equation}
where $k_F$  and $k_S$ are fast (large) and slow (small) rate constants describing the transient behavior of the impurity concentrations.  The solutions show that after a time much greater than $k_F^{-1}$, the impurity concentrations in the liquid and gas both decrease exponentially at a rate $k_S$, until after a time greater than $k_S^{-1}$ constant values of $c_{i,l}^{ss}$ and $c_{i,g}^{ss}$ are obtained.

\subsection{General Results of the Reduced Model} \label{discuss_kSkF}

The solutions of the reduced model for the liquid and gas concentrations for the impurity can be explicitly produced as two complicated algebraic expressions by substituting Eqs.~\eqref{eq:rate_constants}, \eqref{eq:reduced_coeffs}, and \eqref{eq:basic_coefficients} ($r_{sam}=0, K_{ad}=0$) into Eq.~\eqref{eq:two_eq_solution}. 
In order to understand the time dependence of the model, we expand the two rate constants in Eq.~\eqref{eq:rate_constants} in infinite series using two small dimensionless quantities: the ratio of gas to liquid mass ($\delta n$) and the evaporation rate constant to the dissolution rate constant ($\delta \mathcal{E}$). 
\begin{equation}\label{eq:more_liq_than_gas}
\begin{aligned}
\delta n &\equiv \frac{n_{0,g}} {n_{0,l}} \ll 1 ,\\
\delta \mathcal{E} &\equiv  \frac{r_{evp} } {k_{dis} \cdot n_{0,l}} \ll 1 .\\
\end{aligned}
\end{equation}
For a large LAr detector the ullage (gas volume) is generally maintained at $\sim5$\% of the cryostat volume to help maintain pressure stability.
This gives a value of $\delta n\sim2\times 10^{-4}$. For the much smaller system described in Sec.~\ref{sec:measurement}, $\delta n$ is $\sim2\times 10^{-3}$. The value of $\delta \mathcal{E}$ is not well defined because the value of $k_{dis}$ is not known,  but even assuming a small value for $k_{dis}$ of 0.1~s$^{-1}$, for our measurements $\delta \mathcal{E} \sim 4\times 10^{-5}$.  In addition, $\delta\mathcal{E}$ will be further reduced if the detector volume is larger. This is because $n_{0,l}$ increases proportionally with volume, while the heat load (and hence $r_{evp}$) increases proportionally with surface area,  so as a fixed geometry is scaled, $\delta\mathcal{E}$ decreases as the inverse of the scale.

Expressing $k_F$ in power series of $\delta n$ and $\delta \mathcal{E}$, we obtain
\begin{equation}\label{eq:kFexpansion}
\begin{aligned}
k_F = &k_{dis} \cdot (1+H_{xx} \cdot \delta n) \cdot \\
&\left( 1+\frac {\delta \mathcal{E} \cdot \left(1+\delta n \cdot (1+H_{xx} \cdot \delta n)\right)} {\delta n \cdot (1+H_{xx} \cdot \delta n)^2} - \frac {\delta \mathcal{E}^2 \cdot \left(1-H_{xx} \cdot (1-\delta n )\right)} {\delta n \cdot (1+H_{xx} \cdot \delta n)^4}+O(\delta \mathcal{E}^3)\right).
\end{aligned}
\end{equation}
For the values of $\delta n$ and $\delta \mathcal{E}$ given above, the series converges rapidly. The leading term is $>$98\% of the true $k_F$ value, and it is essentially independent of Henry's coefficient.
For large $k_{dis}$ the value of $k_F$ is very close to $k_{dis}$; $k_F$ therefore describes the rapid equilibration of the gas and liquid impurity concentrations toward the Henry's law ratio.  It is difficult to measure because it is not significantly larger than the response time (a few minutes) of our impurity measurement system.

The slow time constant, which characterizes the slow process of removing  impurity from the liquid by evaporation, can similarly be expanded in series
\begin{equation}\label{eq:kSexpansion}
 k_S  = \frac {H_{xx} \cdot r_{evp}} {n_{0,l} \cdot (1+H_{xx} \cdot \delta n)} \left(1 + \frac {\delta \mathcal{E} \cdot (1-H_{xx} \cdot (1 -  \delta n))} {H_{xx} \cdot \delta n  \cdot (1+H_{xx} \cdot \delta n)^2} - O(\delta \mathcal{E}^2 ) \right).
 \end{equation}
This series also converges rapidly.
The second term in the parentheses is about $4 \times10^{-6}$, and each of the following terms are about $-0.00$2 times the previous one.   Therefore, the slow time constant is closely proportional to Henry's coefficient and is essentially independent of  $k_{dis}$ if $k_{dis} > k_{evp}$.   For the system to be described here, if $H_{xx}=1$, $r_{evp}=0.02$~mol/s and $n_{0,l}=500$~mol, we find $1/k_S\sim7$~hours. 

It is clear from the leading term in Eq.~\eqref{eq:kSexpansion} that Henry's coefficient can be determined from a measurement of $k_S$, given the values of $r_{evp}$ and $n_{0,l}$.  This is quite easy to do, because $k_F$ is so much larger than $k_S$ that after a short time the exponential term in $k_F$ will go to zero, and the impurity concentration as a function of time is then entirely described by the exponential term in $k_S$ (see Eq.~\eqref{eq:two_eq_solution}).  Then $k_S$ can be identified as the fractional rate of change with time of the impurity concentration in the liquid
\begin{equation}
 k_S = - \frac {1} {c_{i,l}} \frac {d c_{i,l}(t)} {dt} = - \frac {d \ log(c_{i,l}(t)} {dt},
\end{equation}
which we will therefore also refer to as the ``cleaning" time constant for LAr.

The full expression for Henry's coefficient can be obtained by solving the second equation in Eq.~\eqref{eq:rate_constants} for $H_{xx}$: 
\begin{equation}\label{eq:Henry_full}
  H_{xx} = \frac {k_S^2 \cdot n_{0,g} \cdot n_{0,l} + r_{evp}^2  -k_{dis} \cdot k_S \cdot n_{0,g} \cdot n_{0,l} - k_S \cdot (n_{0,g} + n_{0,l}) \cdot r_{evp}} {k_{dis} \cdot n_{0,g} \cdot (k_S n_{0,g} - r_{evp})}.
\end{equation}
Using the fact that the first term in Eq.~\eqref{eq:kSexpansion} dominates, an approximation to Henry's coefficient can be written as
\begin{equation}\label{eq:Henry_approx}
 H_{xx}\sim h\equiv k_S \cdot n_{0,l}/r_{evp}, 
 \end{equation}
 and in terms of $h$ we can obtain an expansion of Eq.~\eqref{eq:Henry_full} in the small quantities $\delta \mathcal{E}$ and $\delta n$ as an infinite series:

\begin{equation}\label{eq:Henry_series}
  H_{xx} = h  +(1-h)  \frac {\delta \mathcal{E}} {\delta n}  +  h \sum_{j=1}^{\infty} \left ( h \, \delta n \right )^j.
  \end{equation}

For $h=0.9$ the error made by taking $h$ to be $H_{xx}$ is dominated by the second term in this expansion, which is 0.002 for the values of $\delta n$ and $\delta \mathcal{E}$ stated above.

Henry's coefficient can of course be measured directly as the ratio of concentrations of the impurity in the gas to that in the liquid at equilibrium (Eq.~\eqref{eq:henry_law}), providing that convection is sufficient to thoroughly mix the gas and liquid and that the sampling rate approaches zero.  Samples of liquid and gas can be taken simultaneously and the impurity concentrations can be measured. Since instruments that measure impurity concentrations work only with gas mixtures, the liquid sample must be converted to gas before measurement.  However this is done, the sample must be isolated from the bulk of the liquid before it is completely evaporated to gas.  Otherwise, equilibration of the gas and liquid will occur (at a rate given by $k_F$), and the impurity concentration in the evaporated liquid sample will be identical to that in the gas: the deduced Henry's coefficient would then reduce to one.  Therefore, the sampling is usually done through a capillary tube so that the flow rate in the tube is much greater than the diffusion and convection rates in the liquid~\cite{preston3}.  The procedure to be described here, measuring $k_S$ and using Eq.~\eqref{eq:Henry_series} to obtain $H_{xx}$, has the advantage that since $k_S$ is the logarithmic rate of change of impurity concentration, the sampling method only needs to be proportional to concentration with a time independent proportionality constant.

%% file: measurement_CET.tex
\section{Model Application: Oxygen in LAr}\label{sec:measurement}

\subsection{BNL 20-L LAr Test Stand} \label{sec:setup}

\begin{figure}[ht]
\centering
\includegraphics[width=0.65\textwidth]{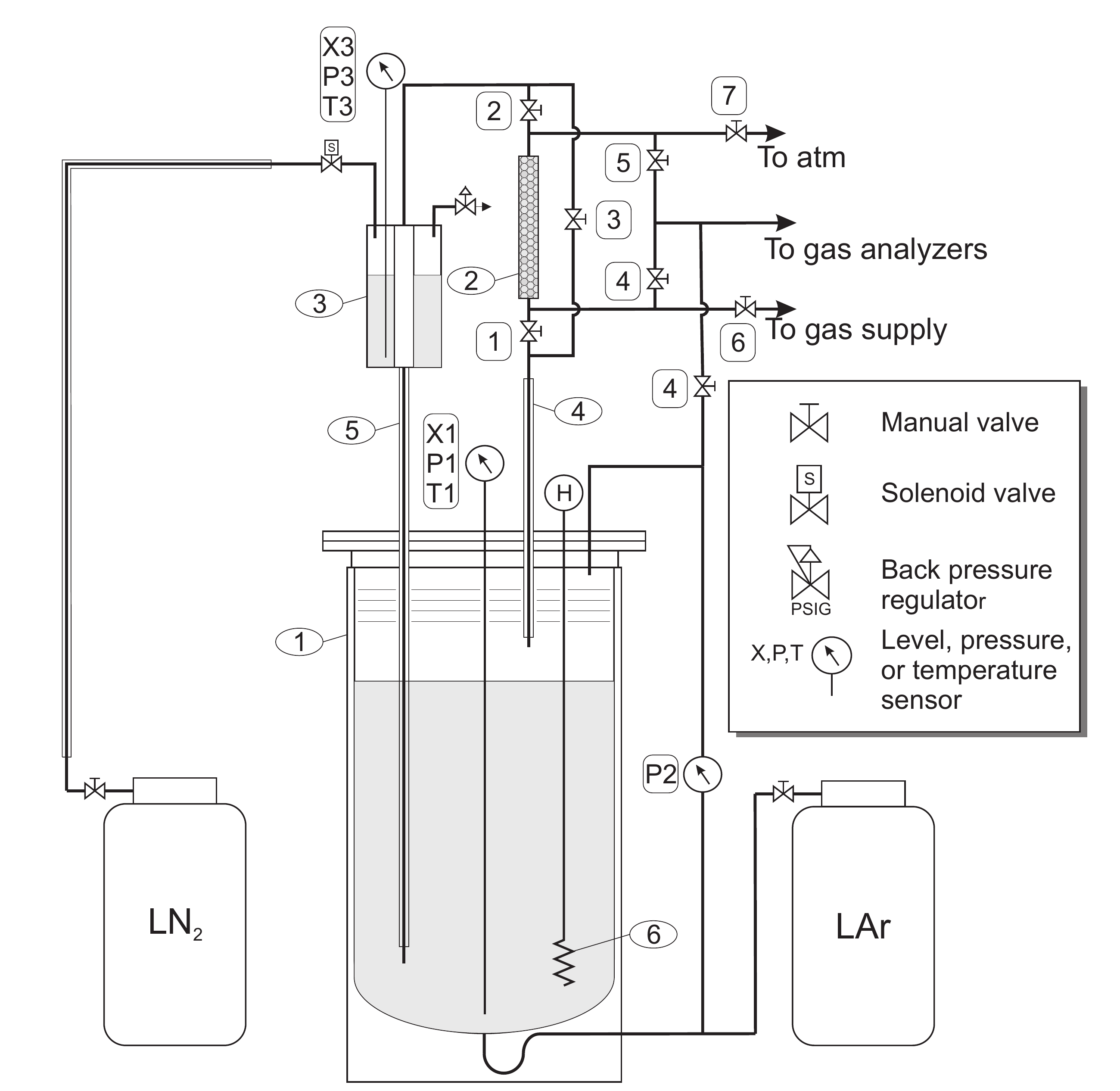}
\caption{The cryostat of the BNL 20-L test stand.  A gas purifier and condenser are mounted on the top flange to remove impurities from the gas, condense the gas, and return the liquid to the bottom of the cryostat. A detailed description of the system can be found in Ref.~\cite{jinst_Li}.} 
\label{fig_teststand}
\end{figure}

The measurements were made using the 20-L, multi-purpose LAr test stand at BNL. The details of the system setup and operation are reported in Ref.~\cite{jinst_Li}.
A diagram of the essential parts of the system is shown in Fig.~\ref{fig_teststand}, and the specifications that are relevant to determining the parameters of the model are listed in Table~\ref{table:test_stand}.
The cylindrical cryostat has internal dimensions of 24~inches in depth by 9.46~inches in diameter with an ellipsoidal bottom.
A 12-inch diameter conflat flange, containing 15 feedthroughs, is used to seal the cryostat at the top. These feedthroughs provide connections for the gas outlet and purified liquid inlet as well as penetrations for all the test instruments including high voltage and power supply cables, signal cables, and temperature and pressure sensors.
Five thin stainless steel plates, spaced at 1~inch intervals, are installed just below the top flange to provide thermal insulation and reduce convective flow. 
This set of  heat shield plates significantly reduces the heat transfer from the top flange due to radiation and gas convection. Their effectiveness is confirmed by temperature measurements of the outside surface of the top flange with LAr filled just below the bottom plate. The outside surface remains near room temperature, preventing the condensation of water on the flange, which occurs without the heat shield. Gas to be purified is withdrawn from just below these plates.
A tube passing through the insulating vacuum space at the bottom of the cryostat is used to fill and drain LAr and provides a connection to the bottom of the LAr to measure the liquid head pressure.

The principle of system operation is briefly described as follows.
LAr, together with dissolved impurities, evaporates to gas that flows upward to an outlet tube just below the lowest heat shield plate.  The GAr is then conducted to a purifier cylinder containing 13X molecular sieve and GetterMax 133T catalyst\footnote{\href{https://www.catalyst-central.com/product}{https://www.catalyst-central.com/product.}}.  The purifier can be bypassed by opening and closing appropriate valves (valves 1, 2, and 3 in Fig.~\ref{fig_teststand}). There is no direct LAr purification in the liquid.  The GAr, purified or not, flows into a stainless steel condenser, cooled by pressurized liquid nitrogen (LN2), and the resulting LAr flows back into the bottom of the LAr volume in the cryostat by gravity. 
The pressure in the LN2 condenser is maintained by a back pressure regulator\footnote{Equilibar EB1HF2-SS} set nominally at 41~psia by reference to a pressure maintained by a spring loaded diaphragm pressure reducing regulator\footnote{Equilibar model10} which is supplied by house compressed air. 
The reference pressure is not sensitive to atmospheric pressure, and has a manufacturer specified variation of $0.1$\% of the supply pressure variation. 
The house compressed air is probably stable to about $\pm2$ psi, so the net result is a variation of the back pressure reference of $\pm 2\times10^{-3}$ psi. 
Temperatures were measured with Omega PR-10E-2-100-1/4-2 RTD resistors (Class B accuracy, $\pm$1.2~K at 90~K) and Omega solid state absolute pressure transducers (model PX209-060A5V, $\pm$0.15~psi accuracy for nitrogen and PX209-030A5V, $\pm$0.075~psi accuracy for argon). 
The evaporation rate of the LAr in the cryostat is proportional to the heat input to the liquid, which is introduced by: (i) conduction down the cryostat inner wall and along instrumentation penetrating the top flange and entering the liquid, (ii) convection in the gas from the top flange to the liquid surface, (iii) heat leakage through the insulating vacuum of the cryostat wall, and (iv) a resistive heater immersed in LAr, which can produce up to 150~W. 
The first three sources together are treated as one effective heat leak, $P_0$, which can be deduced from the data as we will show. 

The temperature of the bulk LAr is determined both by an RTD installed at about 39.2~cm below the bottom heat-shield, and also by converting the pressure measured in the ullage to temperature on the LAr boiling line~\cite{Tegeler1999}. The measured values are not consistent: the temperature at the liquid surface is systematically 0.25~K colder than that near the RTD.
This temperature inversion does not seem likely, especially considering that the cold LAr from the condenser enters at the bottom of the cryostat.  Therefore, we believe that either the measurement of pressure, or temperature, or both are systematically in error.  We have shifted each individually and both together to obtain best fits to the vapor pressure. The shifts found are well within the manufacturers quoted accuracy bands for the two devices. 
The best fit to  our data gives an average temperature of 89.0$\pm$0.7~K and an average pressure of 0.121$\pm$0.009~MPa. 
These uncertainties are caused principally by changes in heater power and LN2 level during the system operation. 

These temperature and pressure fluctuations are the result of several imperfections in our system.  In an ideal implementation of our system, the temperature of the LAr in the cryostat would equal the temperature of the LN2 in the condenser, which in turn would equal the temperature at the LN2 vapor pressure set by the back pressure regulator.  
There are many realities of our implementation that reduce these two equalities to approximations.  
First, because the condenser wall separating the LN2 and condensing GAr has a finite thermal resistance, the temperature of the condensed LAr increases as the heat content of the GAr increases.  The heat content of the GAr is equal to the heat input to the LAr plus the heat input to the GAr as it flows through the purifier and tubing to the condenser.  This section of the apparatus is wrapped in six inches of Cryo-Lite insulation\footnote{Lydall, Cryo-Lite cryogenic insulation, \href{http://www.lydallpm.com/products/low-temperature-insulation/cryolite-cryogenic-insulation/overview/}{http://www.lydallpm.com/products/low-temperature-insulation/\newline cryolite-cryogenic-insulation/overview/}.}, so the heat input depends on the flow rate of the gas and the ambient temperature.  In addition, the LN2 is not continuously replenished, but rather is added periodically to keep the LN2 level between 55\% and 85\% of the maximum LN2 capacity.  The result is that, as the level of the LN2 falls and rises with the consumption and refill of LN2, the contact area between the LN2 and the GAr decreases and increases, the LAr temperature in the cryostat varies by about 0.4~K peak-to-peak with a period between 0.36 and 1.4 hours depending on the heat input to the LAr.  

In addition, the back pressure is not held truly constant: the LN2 pressure changes slightly as the LN2 level and the power into the GAr and the LN2 (through the Cryo-Lite insulation) change, causing the LN2 evaporation rate to change.  This is only a small effect because the back pressure regulator was chosen to have a small change in pressure with change in flow (in control parlance, a large compliance).  However, when the condenser is filled with LN2 the large flow of N2 gas produced by cooling the initially hot fill line and the top of the condenser volume (which is vented through the back pressure regulator) is sufficient to cause the back pressure to rise momentarily, leading to semi-periodic spikes of up to 1.5~K in the condenser temperature.
As a result of all these fluctuations, the average pressure and temperature of LN2 are 0.281$\pm$0.012~MPa and 87.2$\pm$0.6~K, respectively.  These are consistent with the literature value for LN2 vapor pressure~\cite{Span2000}.

Finally, the flow of evaporation gas through the pneumatic resistance of the purifier causes the pressure in the cryostat to be higher than that in the condenser.  This pressure difference increases with the evaporation rate, which is proportional to the heat input to the LAr.  The consequence of this is that the temperature of the LAr in the cryostat is higher than the temperature of the condensed LAr that returns, with a temperature difference proportional to the evaporation rate (about 0.012 K/W).  Also, as a consequence, the LAr level rises in the return tube to create a head pressure to balance the pressure drop across the purifier.  We have made a thermodynamic calculation of the system, based on the considerations described above, that quantitatively reproduces the observed pressure and temperature time variations in our system. This calculation shows that the combinations of the thermal resistance of the condenser and the pneumatic resistance of the purifier with the heat capacity of the LAr cause the linear time dependence of the LN2 level to be integrated to give the quadratic dependence of LAr temperature between LN2 fills that can be observed by comparing the top and bottom panels of Fig.~6 in Ref.~\cite{jinst_Li}. 

During operation, LAr is sampled by a tube immersed in the liquid with a small orifice at the end. The tube is in thermal contact with the top flange which helps evaporate the liquid rapidly into gas.  The orifice diameter is small (0.75~mm) and the liquid sampling rate is high enough that the gas produced flows away rapidly and has little time to come into equilibrium with the liquid.  Therefore we believe that the concentration measured by the analyzer is a good measure of the liquid concentration~\cite{preston3}.  Oxygen concentration is measured and recorded every 30 seconds by a Servomex DF-560e oxygen gas analyzer\footnote{\href{https://www.servomex.com/gas-analyzers}{https://www.servomex.com/gas-analyzers.}}.  The stated range of concentration is 0 to 20 ppm oxygen, the minimum detection level is 75 ppt, and the systematic calibration uncertainty is the greater of $\pm$3\% of reading or $\pm$0.1 ppb.  The signal averaging time of the analyzer is set to a value of 800~s, per the manufacturer’s instructions\footnote {This is the value associated with the instrument filtering settings of 200 for ``Weight" and ``Low Noise" for ``Response Type".}.  We have measured the response time of the oxygen impurity concentration for the entire system by injecting air into the gas volume.  The mean value for the 10\% to 90\% rise time is 1210$\pm$180 s.  This is significantly longer than the instrument response time, presumably due to diffusion during the transport of the gas to the oxygen analyzer.

The LAr level gradually decreases due to sampling.  It is measured by a differential pressure transducer\footnote{Alpha Instruments, Inc Model 175.}
connected between the bottom of the cryostat and the gas ullage, and by a capacitance level probe immersed in the liquid.  The capacitance probe does not have an absolute calibration; it was calibrated against the differential pressure measurement. 
The level measurements made by these two methods agree with each other over the entire range of LAr levels (14 cm to 42 cm) within $\pm$0.43 cm, which is $\pm$0.20 liters or $\pm$6.8 moles of LAr.

The O2 concentration reading is stated by the DF-560e manual to have a sensitivity of ``a few percent" to the gas flow rate within the range of 0.24-1.9 L/min, converted to argon flow rate.   About 40\% of the argon flow was periodically diverted to a water concentration measuring instrument. The decrease in the measured LAr level as a function of time, corrected for this diversion, implies an average GAr flow rate (at NPT) of 0.32 $\pm$ 0.09 L/min, with minimum and maximum values of 0.23 L/min and 0.44 L/min. This is consistent with the flow observed on the rotameter in the DF-560e. 
Therefore the flow rate was generally within the manufacturer’s recommended range. However, it is at the lower limit of the range, and the flow fluctuations are probably a significant contributor to the observed noise in the concentration measurements that is discussed in Sec.~\ref{sec:measurement}.  Atmospheric pressure variations\footnote{The atmospheric pressure recorded by the Meteorology Group at BNL has a RMS deviation of $\pm0.77$\% over a period of one year (2018).} have a negligible effect on the flow compared to the measured pressure variation ($\pm7$\%) in the cryostat. 
The system operating parameters - such as LN2 and LAr temperatures, pressures, and levels - were recorded with a slow control system~\cite{jinst_Li}, which recorded data every 60 seconds.

Before filling the cryostat with commercial LAr, it was evacuated with a Pfeiffer Model TSH 071 turbo-molecular pump. We have estimated the leak rate of oxygen into the cryostat to be $1 \times 10^{-9}$~mol/s from the measured pressure ($4 \times 10^{-5}$~torr, using the nominal pumping speed of the turbo-molecular pump (33 L/s), the calculated conductance of the piping between the pump and the gauge (2 L/s), and the fraction of oxygen in air (21\%). The cryostat was not baked and no special effort was made to clean the system.

\begin{table}[!htbp]
\caption{\label{table:test_stand}The properties of the BNL 20-L test stand (some of the uncertainties are only estimates).}
\vspace{-0.2cm}
\begin{center}
\begin{tabular}{|c|c|c|c|}
\hline
Property & Value & Uncertainty & Units \\\hline
Internal diameter & 24.03 & 0.10 & cm \\\hline
Internal volume of cryostat & 27.9 & 0.5 & L \\\hline
Maximum volume of LAr & 22.1 & 0.5 & L \\\hline
Typical heat leaking into LAr & 23 & 7 & W \\\hline
Operating temperature & 88.9 & 0.7 & K \\\hline
Nominal leak rate & $1\times10^{-9}$ & $5\times10^{-10}$ & mol/s \\\hline
Gas/Liquid contact area & 0.046 & 0.002 & m$^2$\\\hline
Gas/Surface contact area & 2.2 & 0.4 & m$^2$ \\\hline
Liquid/Surface contact area & 0.36 & 0.04 & m$^2$ \\\hline
\end{tabular}
\end{center}
\end{table}

\subsection{Measurements}  

The results reported in this paper are from the analysis of four data sets. 
Each data set is a continuous measurement lasting 240 to 560 hours and containing up to 100,000 concentration measurements. 
Data set \#1 was taken in February 2016. During these measurements the insulating vacuum was poor and water condensed on the outside of the cryostat. 
The insulation volume of the cryostat was evacuated, and data set \#2 was then recorded in June 2016. 
Data sets \#3 and \#4 were taken in December 2018 and April 2019, respectively.  The vacuum insulation volume was again re-evacuated between these two data sets. 
Each data set therefore has a different value of heat leakage, $P_{0}$ (see Eq.~\eqref{eq:evaporation_rate}).
Before each set of measurements, the cryostat volume was evacuated for a few days to remove air and adsorbed surface impurities. The base pressure achieved was never greater than $5 \times 10^{-5}$~torr.

For convenience in the analysis, each of the four sets are divided into a series of periods, numbered sequentially starting at 1. 
During each period the system operating parameters and the coefficients to be used in the differential equations are considered constant. 
Various operations were performed only between periods, such as changing the heater power, engaging or disengaging the purifier, or adding LAr or air to the system.

%% file: numerical_calc.tex
\subsection{Comparison of the Model to Measurements}\label{sec:model_calc}  

In this section, we compare the data to numerical calculations made with the model using, when available, the best values for the relevant physical parameters taken from the literature. The remaining parameters are adjusted to obtain a best fit to the data. 
The results show that the model accurately describes the behavior of oxygen in argon. 
These calculations also determine Henry's coefficient, the leak rates for oxygen under various conditions, and the upper limits of the sorption parameters, as well as providing estimates of systematic errors.
In addition, we show that the sorption of oxygen is generally a small effect compared to the leak rate, and its effects can only be observed if the leak rate and the oxygen concentration are both very small.

\begin{figure*}[htp]
       \centering
       \includegraphics[width=0.86\textwidth]{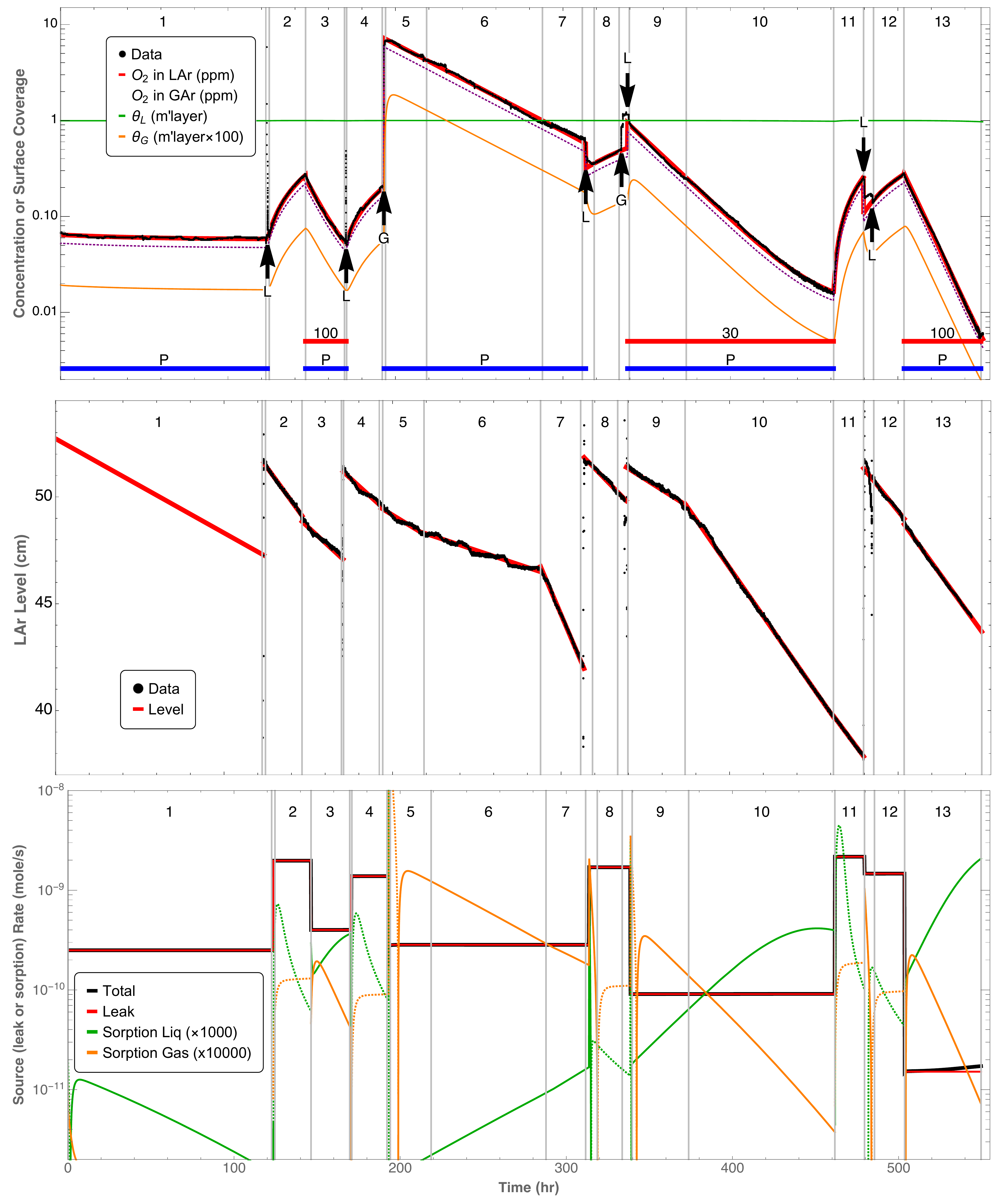}
       \caption{The oxygen concentrations (top), LAr level (middle), and oxygen source rates (bottom) as a function of time. The black points are the measurements. 
       The data are analysed in the 13 periods numbered sequentially across the top.
       The solid red lines are either outputs from (LAr and surface concentrations and impurity leak rates), or inputs to (LAr level), the model calculations. 
       In the top panel the dashed line is the calculated oxygen concentration in GAr; the green and orange lines are the surface coverages in liquid and gas, respectively; the thick blue horizontal lines with label ``P" indicate the gas was purified during those periods; the thick horizontal red lines indicate that the heater was on at the power in $W$ above the line; and the labels ``G" or ``L" indicates air or commercial LAr was injected into the cryostat.
       In the bottom panel the black line is the effective leak rate, the green and orange lines are the rates of surface sorption in LAr and GAr, respectively (solid indicates a positive rate and dashed indicates a negative rate), and the red line is the total of these three rates.
       }
       \label{fig:Feb2016DataSet}
 \end{figure*}

Fig.~\ref{fig:Feb2016DataSet} displays the complete first data set, containing 65673 measurements of the oxygen concentration in LAr over a duration of 560 hours.
The black points in the top (middle) panel are the measured oxygen concentration (the LAr level).  
This full data set is divided into 13 continuous periods, during each of which the parameters of the model (i.e., the coefficients in the differential equations) remain constant.
When the measurements began the oxygen concentration was very close to the steady state value and remained constant for $\sim$123~hours (see period 1 in Fig.~\ref{fig:Feb2016DataSet}).  The purifier was then bypassed and at the same time LAr was injected directly from a commercial LAr dewar through the fill tube at the bottom of the cryostat; this is the start of period 2.  The oxygen concentration before and after the injection did not change significantly; this is because the LAr in the supply dewar had approximately the same concentration as the LAr in the cryostat. However, the momentary pressure increase from the gas injected with the liquid momentarily upset the system and the oxygen analyzer, causing the concentration reading to spike.  During this period without purification, air leaking into the ullage accumulated and the oxygen concentration increased linearly (in the approximation of low concentration). 
At $\sim$146~hours (period 3) the purifier was reconnected and the heater was turned on at 100~W.  During this time the gas was purified and the oxygen concentration decreased exponentially at a cleaning rate characteristic of 100~W additional heat input. 
Starting at $\sim$171~hours (period 4) the heater was turned off and the purifier was bypassed, and the concentration again increased linearly as air leaked into the cryostat.
At $\sim$193~hours (the beginning of period 5) a small amount of air was injected into the system between the purifier and condenser, so that the oxygen concentration increased sharply, and the purifier was again reconnected so that the concentration in the liquid began decreasing at the cleaning rate for 0~W of additional heater power.  Note that at the end of period 7 when LAr from the supply dewar was again injected into the system, the concentration decreased, since the LAr in the cryostat was then less pure than that in the supply dewar.
Subsequently, these various operations were repeated a few times.  During some periods the liquid sampling rate was increased by flowing gas to a water analyzer in parallel with the oxygen analyzer (the increased loss rate of LAr is particularly obvious in periods 7 and 10 through 13).

The various lines in Fig.~\ref{fig:Feb2016DataSet} compare the values obtained by solving the four coupled differential equations of Eq.~\eqref{eq:dcigdt} to the data.  The red line in the top panel is the oxygen concentration in the liquid, $c_{i,l}(t)$, in ppm. The dashed line is the oxygen concentration in the gas, $c_{i,g}(t)$, also in ppm, and the green and orange lines are the oxygen concentrations $\theta_l(t)$ and $\theta_g(t)$, in molecular monolayers, on the surfaces in the liquid and the gas, respectively. 
In the middle panel the red line is a best fit to the LAr level data, which is used to obtain the set of sampling rates, $r_{sam}$, in each period, which are inputs to the calculation.  Note that the recording of the LAr level was not started until the beginning of period 2; before that the red line is simply a reasonable guess of the LAr level, based on the subsequent values.

The red line in the bottom panel is the value of the oxygen leak rate ($r_{lek}$) determined by fitting the model to the data; the rate has been adjusted for each period to best fit to the oxygen concentration data in the liquid. 
The net sorption rates of impurities leaving the cryostat surfaces and entering the LAr and GAr are computed from Eqs.~\eqref{eq:outgassing_1} and ~\eqref{eq:sorption_in_liquid}, using the solutions obtained for $\theta_l(t)$, $\theta_g(t)$, $c_{i,l}(t)$, and $c_{i,g}(t)$ with the sorption parameters in Table ~\ref{tab:sorption_params}. Written out in full these rates are
\begin{equation}
r_{srp,p}(t) \equiv \left ( \frac {dn_{i,p}} {dt} \right )_6 = n_{i,sp}^{sat}  \left(k_{des,p} \cdot \theta_p(t) - K_{ad,p} \cdot k_{des,p} \cdot c_{i,p}(t) (1-\theta_p(t))\right),
\label{eq:desorption_rate}
\end{equation}
where $p$ is either $g$ or $l$ for the gas or liquid phase. The sign of $r_{srp,p}(t)$ is determined by which of the two terms in parentheses in this expression is larger, desorption or adsorption. 
The absolute values of these two rates are shown by the green line for the surface in LAr ($|r_{srp,l}(t)|$) and by the orange line for surface in GAr ($|r_{srp,g}(t)|$). 
The sign is indicated by a solid line for a positive rate (net desorption, i.e. more impurity going from the surface into the bulk phase) and a dotted line for a negative rate (net adsorption, i.e. more impurity leaving the the bulk phase and going onto the surface).  Note that they are two to three orders of magnitude smaller than the leak rate for this data set (note the changes of scales), and therefore make a negligible contribution to the total impurity source rate.  This is not surprising: one monolayer adsorbed on the surface area in contact with the liquid is $4\times10^{-6}$ mol of oxygen, which when mixed with the $\sim$700 moles of LAr gives a incremental concentration increase of only 6 ppb.  Or, if the initial desorption rate were equal to the observed total leak rate, in 3 hours the desorption rate would be reduced to 1\% of this.

The model parameters are determined by adjusting them to produce a best fit of the calculated oxygen concentration to the measured oxygen concentration as a function of time. We measure the ``goodness of fit" by the function
\begin{equation}
    \begin{aligned}
    \phi^2_n &= \sum_{j=1}^{N_{obs,n}} \left( \frac {c_{obs,n,j} - c_{calc}(t_{n,j}; \,H_{xx}, \,P_{0}, \,c_{0,n}, \,r_{lek,n})} {f \cdot c_{obs,n,j}} \right)^2, \\
    \phi^2 &= \sum_{n=1}^{N_p} \phi^2_n, \\
    N_{obs} &= \sum_{n=1}^{N_p} N_{obs,n},\\
    \end{aligned}
    \label{eq:goodness_of_fit}
    \end{equation}
where the index $n$ indicates each period in the data set, $j$ indicates each discrete observation time within the period, $c_{obs,n,j}$ is the measured oxygen concentration in LAr at the time $t_{n,j}$, $f$ is the fractional uncertainty in the measurement, and $c_{calc}(t_{n,j};\,...)$ is the corresponding calculated concentration. The number of measurements in period $n$ is $N_{obs,n}$, the number of periods is $N_p$, and the total number of measurements in the set is $N_{obs}$.  The quantity $\phi^2$ is the sum of squares of the fractional deviation of the calculation from the data at each point for the set compared to the measurement uncertainty, so $\sqrt{\phi^2 / N_{obs}}$ is easily understood as the weighted root-mean-square (RMS) fractional deviation of the model from the data. The smaller $\phi^2$, the higher is the quality of the representation of the data by the model. Notice that since $f$ is a constant, it is merely an overall factor that scales $\phi^2$ and it has no effect on the minimization procedure.   

The parameters to be adjusted to minimize $\phi^2$ are $H_{xx}$ and $P_{0}$, which apply only to the periods in the data set with purification on; and the set of initial oxygen concentrations ($c_{0,n}$) and the oxygen leak rates ($r_{lek,n}$), one pair for each period in the set. 
The other parameters ($k_{dis}$ and the sorption parameters) are fixed for all data sets at the values given in Sec.~\ref{subsec:process_1} and in Table~\ref{tab:sorption_params}. 
For sorption on the liquid-surface (gas-surface) interface, the values of set S1 (S3) were used, since they represent the maximal physically reasonable sorption rates.  
As stated above, the sampling rate for each period is fixed at the value determined by fitting the LAr level data for that period with a straight line. 
Given these parameters, one can numerically solve the four differential equations in Eq.~\eqref{eq:dcigdt} and calculate $\phi^{2}$. 

Because $H_{xx}$ and $P_{0}$ are common to all periods, and $c_{0,n}$ and $r_{lek,n}$ are specific to each individual period in the set, the minimization of $\phi^2$ is implemented in two steps of iterations. 
First, with $H_{xx}$ and $P_0$ fixed, the values of $c_{0,n}$ and $r_{lek,n}$ that minimize $\phi_n^2$ are found for each of the $N_P$ periods individually.  In the second step, these values are fixed and the values of $H_{xx}$ and $P_{0}$ are found that minimize $\phi^2$ for all periods with purification combined.  This minimization procedure is iterated until there is no significant decrease in $\phi^2$ or changes in the values of the parameters. At most eight iterations were required to achieve this for each set. 

The calculated oxygen concentrations in the liquid that best fit the data are shown by the red line in the top panel, and the leak rates that produce this fit are shown by the red line in the bottom panel of  Fig.~\ref{fig:Feb2016DataSet}.  The values found for Henry's coefficient and the heat leakage into the cryostat that best represent the data and the minimum value of the RMS fractional deviation per point ($\sqrt{\phi^2 / N_{obs}}$) of the concentration from the best fit are given in Table~\ref{tab:best_fit_params} for all data sets.

\begin{table}[!htbp]
\vspace{-0 cm}
\caption{Best Values for $H_{xx}$ and $P_0$ for each data set. The mean value is 0.861, standard deviation is 0.045, and maximum deviation from the mean is 0.055. }
\begin{center}
\vspace{-0.5cm}
\begin{tabular}{|c|c|c|c|}
\hline
Set  &  $H_{xx}$ & $P_0 (W)$ & $\sqrt{\phi^2 / N_{obs}}$ \\\hline 
  1   &  0.810   & 33.8      & 0.032 \\\hline
  2   &  0.863   & 19.0      & 0.042 \\\hline
  3   &  0.853   & 16.7      & 0.047 \\\hline
  4   &  0.919   & 7.7       & 0.029\\\hline
\end{tabular}
\label{tab:best_fit_params}
\end{center}
\end{table}

With the choice of $k_{dis}=$100~s$^{-1}$, the equilibration of the liquid and gas is extremely rapid (see Eq.~\eqref{eq:kFexpansion}), so the calculated oxygen concentration in the GAr, as shown by the dotted line, always lies just below the oxygen concentration in the LAr, and is always approximately $H_{xx}$ times the concentration in LAr. Since the concentration in the gas was not measured, there is no data to verify this. 

The leak rates for all periods with no purification (2, 4, 8, 11, and 12) are found to be relatively constant, with a mean value of $(1.7\pm0.15) \times 10^{-9}$~mol/s. The standard deviation of these five values is $\pm0.33\times10^{-9}$ mol/s and the average uncertainty in the values obtained from the fitting procedure is only $\pm0.04 \times 10^{-9}$~mol/s, almost 8 times smaller. 
Therefore we believe there is a real systematic, slow variation of the leak rate.  This variation can be attributed to the fact that the pressure relief valve was operated too near to the limiting pressure at which it seals.  To ensure ``leak tight" operation, the manufacturer specifies a maximum ``sealing" pressure of 50\% of the set relief pressure, which was 8 psig. For this data set, the operating pressure was above the ``sealing" pressure 9\% of the time. 

This explanation for the large and variable leak rate is supported by the fact that leak rate determined in the initial evacuation of the system, $1\times10^{-9}$ mol/s (see Sec.~\ref{sec:setup}), is about the same as that with the cryostat filled with LAr. 
It would be expected that the leak rate under vacuum would be much larger than that with an internal pressure above atmospheric pressure.  In the first case, the leak rate would be largely due to viscous flow inward, and in the second case to the much smaller inward diffusion against the outward flow of argon.  It would, however, be true if the dominant leak were pressure dependent, as would be the case for the failure of the relief valve to properly seal.  We have also observed increases in the leak rate after large pressure excursions that opened the relief valve to exhaust argon.

The leak rates for periods with purification are observed to be significantly lower than for those without purification, and the leak rate is observed to decrease with increasing heater power.  
The heat shield plates at the top of the ullage do not alter the rate of impurity entering the cryostat, nor the amount entering the ullage space between the bottom plate and the liquid surface, since the only outlet for gas is below the bottom plate.  The total impurity leak rate, delivered to this space, mixes with the gas produced by evaporation. It is this mixed gas that is removed from the volume for purification.  However, the mixed gas is not, itself, in contact with the surface. At the surface, and for some (perhaps short) distance above it, there is a layer of rising evaporated LAr, which (for non-zero leak rates) is purer than the mixed ullage gas.  Any impurity, in order to dissolve in the liquid, must pass through this layer.  This can only occur by diffusion of the impurity against the gas flow~\cite{Hertz1923}.  However, with the purifier bypassed, this is all irrelevant, since all of the impurity entering the cryostat by leaks is condensed with the argon and returned to the liquid.  Thus we must distinguish between the actual leak rate into the cryostat and the fraction of the total leak rate which arrives at the LAr surface, which we call the effective leak rate.  This will be discussed in more detail in Sec.~\ref{sec:baffle}.

As we pointed out earlier, the sorption rates are expected to be very small compared to the effective leak rate. 
This is only true if the effective leak rate is large and the oxygen concentrations are high, as is the case for all of sets \#1, \#2, and \#3.
However, for one period of set \#4, the effective leak rate was low and for the last thirty hours of data the oxygen concentration was low enough to observe the effects of desorption of oxygen on the liquid concentration.  The model can adequately fit this data (at roughly to 90\% confidence level), with either 1) an effective leak rate less than $\sim6\times 10^{-12}$~mol/s, or 2) a desorption rate constant from the surface in the liquid of between $\sim10^{-3}$ s$^{-1}$ and $\sim10^{-6}$ s$^{-1}$ and 3) a surface saturation of less than $\sim6$ monolayers, or 4) an appropriately chosen combination of these. Therefore sorption parameter sets S1 and S2 are both compatible with our data, as is no sorption at all.

\begin{figure}[htb]
\centering
\vspace{-0.1 cm}
\begin{minipage}[]{0.48\textwidth}
 \includegraphics[width=1.0\textwidth]{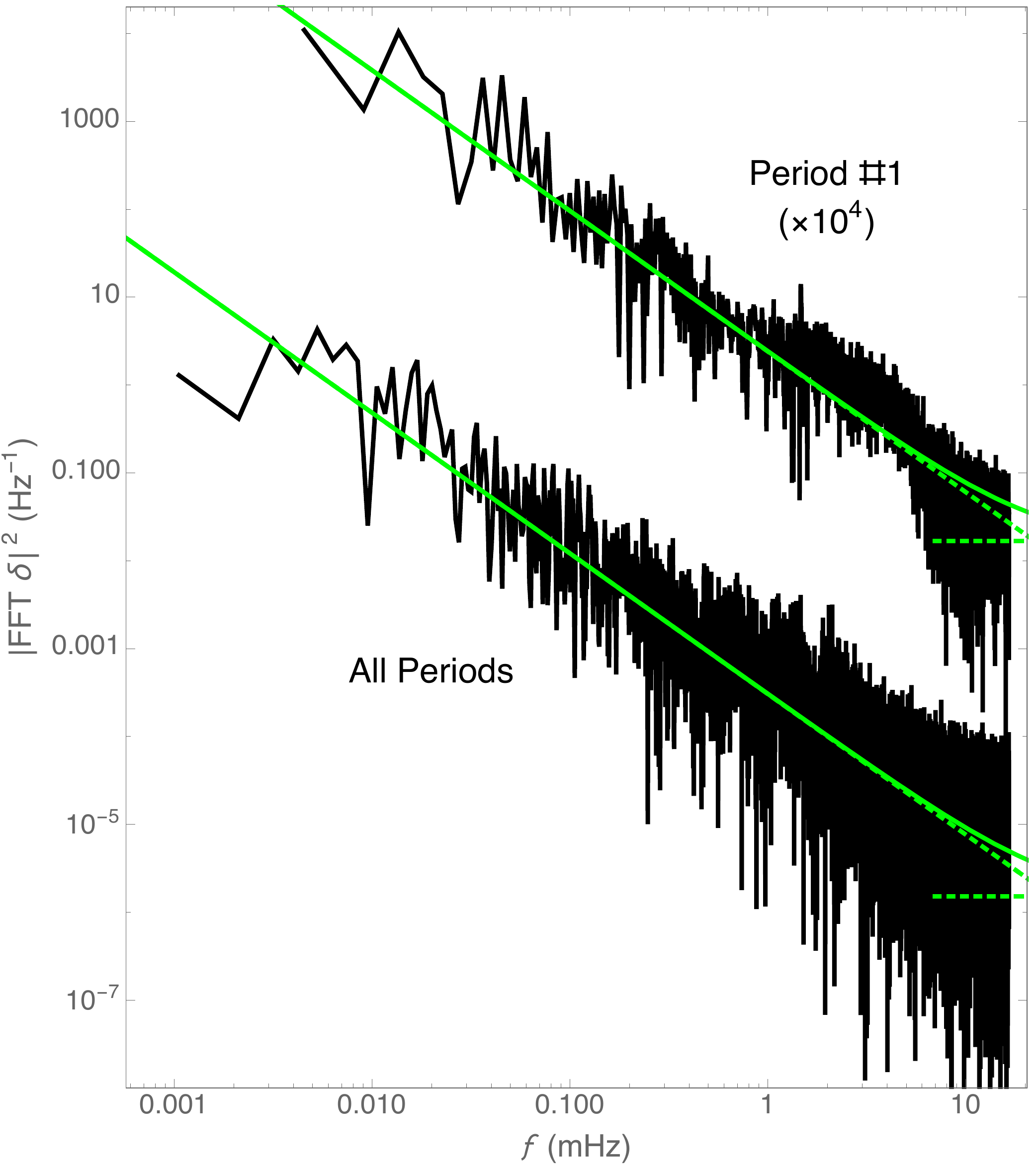}
  \end{minipage}
 \begin{minipage}[]{0.48\textwidth} 
\includegraphics[width=1.0\textwidth]{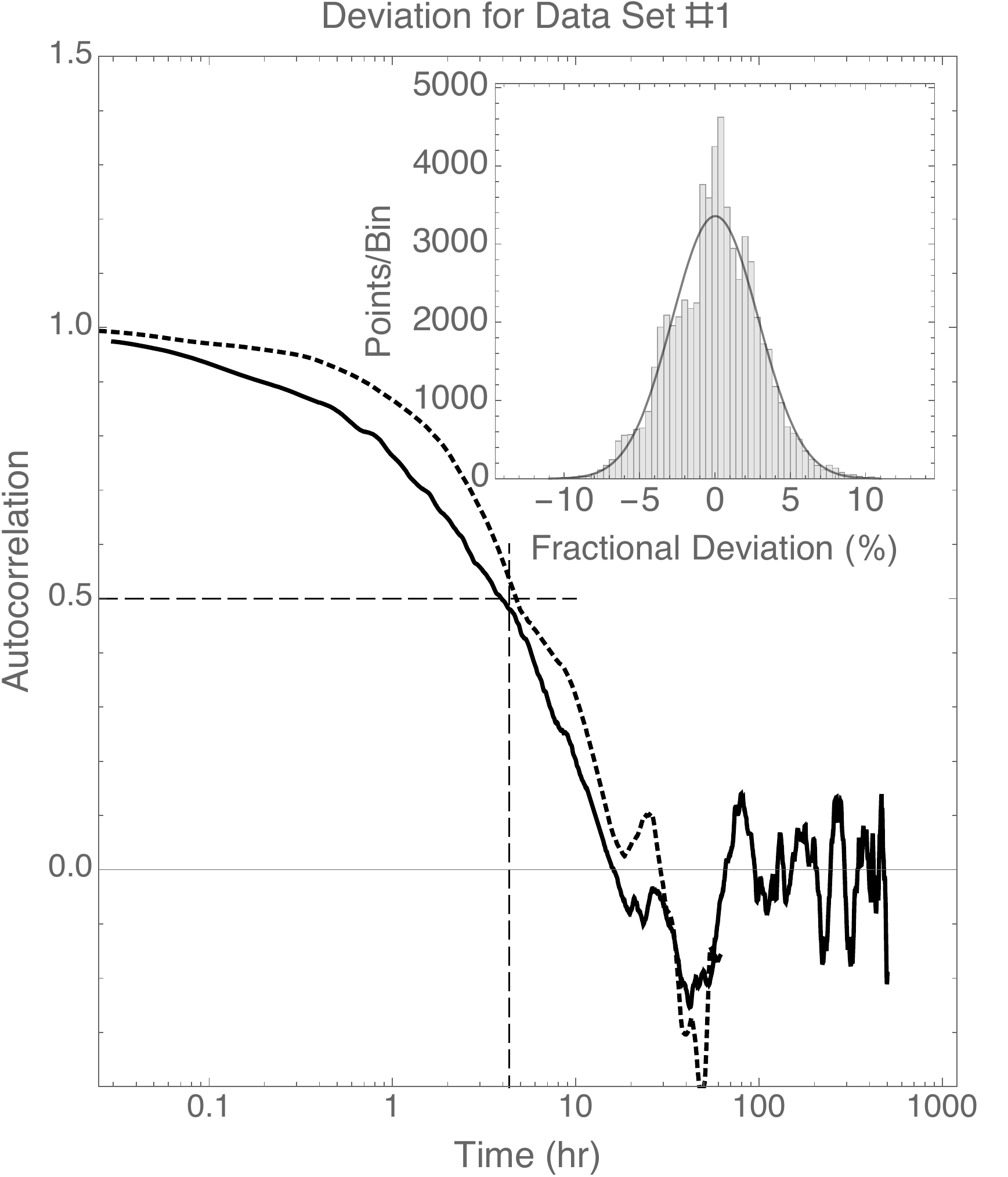}
  \end{minipage}
 \vspace{-0.1 cm}
       \caption{Left: The square of the magnitude of the Fourier transform of the fractional deviation of the measured oxygen concentration from the model calculation for the data shown in Fig.~\ref{fig:Feb2016DataSet} as a function of frequency.  This is computed for both the whole data set and for the first period of the data set only (labeled Period \#1, multiplied by $10^4$). Both lines are a fit with a slope of 1.85.  Right: The autocorrelation function for the same data.  The solid line is for the entire set and the dotted line is for the first Period only.  The inset is the histogram of the values of fractional deviation for each point in the entire set.}
\label{fig:Noise_Spectrum}
\end{figure}

The ``noise'' in the oxygen concentration data, defined as the fluctuation of the data from an ``average" baseline, can be measured as the fractional deviation of the data from the model calculation at each data point. 
It is found to have a roughly Gaussian distribution centered at zero (see the the right panel of Fig.~\ref{fig:Noise_Spectrum}). 
However, the noise does not have a uniform frequency distribution and so it is not ``white'' noise.  The square of the magnitude of the Fourier transform of this noise (the power spectral density, PSD) is shown in the left panel of Fig.~\ref{fig:Noise_Spectrum}. 
The PSD of the noise is shown both for the entire data set \#1 and for just the first period, during which time the concentration was at the steady state and did not change. 
The result is essentially identical in both cases - the noise density increases with decreasing frequency, $\nu$, as $1/\nu^\alpha$, where $\alpha=1.85$, over the entire frequency range, as indicated by the straight lines in the figure. Therefore, as the duration of a measurement increases, the low frequency limit decreases, and the total noise increases.

Some of the low-frequency excess noise can be attributed to the long signal averaging time (800~s) of the oxygen analyzer. Noise with a $1/\nu^\alpha$ spectrum is generated by the fractional integration with order $\alpha/2$ of white noise~\cite{Radeka1969}.  Therefore, a filter implemented as a simple Riemann-Liouville integral (fractional order 1) evaluated at this integration time would account for a $1/\nu^2$ power spectral dependence only at frequencies above $\sim$0.6~mHz.

The excess noise at frequencies below this is presumably accounted for by other integrating processes within the system (such as the RC combinations of the thermal resistance of the condenser and the pneumatic resistance of the purifier with the heat capacity of the LAr), and by fluctuations in the various operating conditions of the system, such as heat input, leak rate, argon pressure and temperature (caused by the cyclical filling of liquid nitrogen in the condenser, and convective mixing of the impurity within the gas and liquid. 
Diffusion, which occurs in many parts of the system, also generates a $1/\nu^2$ power spectrum. Diffusion and convection could be especially significant in the liquid since the purified liquid inflow, the heater, and the LAr sampling point are all well separated within the LAr volume. Finally, as discussed above, a significant source of variation in the total leak rate is believed to be due to the imperfect sealing of the pressure relief valve.  Some of these processes are quasi-periodic while others are chaotic; there is little evidence in the PSD for any strictly periodic disturbances. 
The measured concentration therefore fluctuates significantly from the average at varying frequencies.  This excess low frequency noise leads to problems, as we demonstrate in the next section, in comparing means and variances between samples of data computed for short periods over long times. 

The Wiener-Khinchin theorem states that, under appropriate conditions the Fourier transform of the PSD is identical to the autocorrelation function (ACF)~\cite{Radeka1969}.  A consequence is that noise with a $1/\nu^\alpha$ PSD has an ACF that is large over some initial time range.  This means that successive measurements within the interval of time are correlated, and each measurement is not independent of its neighbors. The right panel of Fig.~\ref{fig:Noise_Spectrum} displays the autocorrelation function of the noise, which is positive for times less than about 18 hours, and is greater than 1/2 for times less than about 4 hours.  This implies that only a small fraction of the measurements in each set are independent.

If each data point were an independent measurement, the individual deviations were normally distributed, and the true uncertainty in each measurement were a constant fraction $f$ of $c_{obs,n}$, then $\phi^2/f^{2}$ would be distributed as $\chi^2$ and we could deduce the statistical significance of the fit and the uncertainties in the parameters from the least-square minimization procedure. Unfortunately none of these conditions are strictly met, and this makes it difficult to assign statistically justified confidence levels to the parameters obtained from the minimization procedure.

\begin{figure}[!htb]
\vspace{-0 cm}
       \centering
       \includegraphics[width=0.6\textwidth]{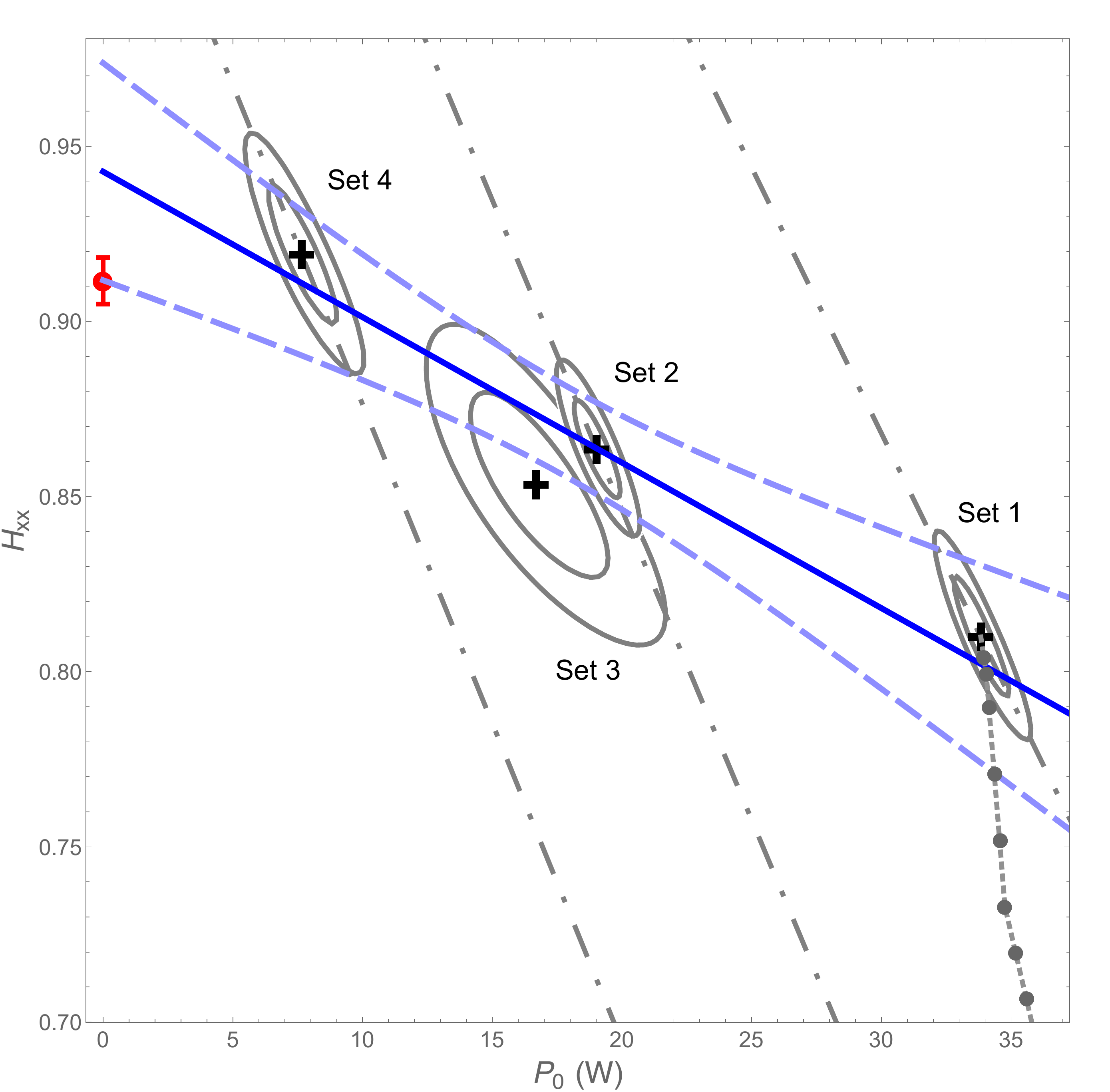}
       \caption{The best fit values and contours at an estimated confidence level 0.95 and 0.9999 for all data sets assuming $k_{dis}=100$~$s^{-1}$. The dot-dashed lines show the anti-correlations between H$_{xx}$ and P$_{0}$ for each set.  The solid blue line is the fit to the four best (H$_{xx}$, P$_{0}$) pairs and the dashed blue lines are the uncertainty in the fit. The dotted gray line and solid points associated with Set 1 indicate the trajectory of the best fit values for $k_{dis}$ values set to 0.5, 0.1, 0.09, 0.07, 0.05, 0.03, 0.028, and 0.026~$s^{-1}$, reading from top to bottom. The red point at $P_0 = 0$ is the literature value for Henry's coefficient at $89.0\pm0.7$ K.}
       \label{fig:Hxx_Pin_Contours}
 \end{figure}

The main problem in determining the statistical significance of these parameters is that the correlated nature of the noise in the measurements makes it difficult to know the true number of degrees of freedom in each data set.  
However, if we make the simple assumption that the noise is perfectly correlated for times less than the time at which the autocorrelation function has the value of 0.5 and totally independent for times longer than this (see the right panel in Fig.~\ref{fig:Noise_Spectrum}), then we can estimate the number of independent measurements in a data set as the total measurement time (in hours) divided by this correlation time. 
For Set \#1, the correlation time is about 4.3 hours, which implies 99 independent points or 97 degrees of freedom if two parameters are fitted.  The upper limit of integration of the chi-squared function per degree of freedom, with 97 degrees of freedom, required to include a probability less than 0.683 (1 sigma) is 1.063. 
Choosing the value of the uncertainty in each point ($f$ in Eq.~\eqref{eq:goodness_of_fit}) to be 0.0314, the value of $\phi^2$ per point becomes 1.063, the expected value for $\chi^2/n_{dof}$. 
The values of the integral of the chi-squared function with 97 degrees of freedom that include probabilities less than 0.95 and 0.9999 are 1.255 and 1.624, respectively. 
The regions in $H_{xx}$ and $P_0$ that have  $\phi^2/n_{dof}$ values less than these two values are indicated by the pair of contours in Fig.~\ref{fig:Hxx_Pin_Contours} for Set \#1.  The figure shows the $\phi^2/n_{dof}$ contours we find in the same manner for sets \#2, \#3, and \#4.
For values of $H_{xx}$ and $P_0$ lying inside the 0.05 probability contour, the quality of the representation of the data by calculation is indistinguishable from that shown in Fig.~\ref{fig:Feb2016DataSet}. 
For values on the 0.0001 probability contour it is noticeably worse. 
We can take the maximum and minimum values of $H_{xx}$ and $P_0$ on the 0.05 probability contour to represent acceptable uncertainties in these parameters. Although they have only a tenuous statistical significance, they should at least indicate relative uncertainties among the four data sets.

Within each set (except perhaps set \#3) the values of $H_{xx}$ and $P_{0}$ are anti-correlated along the parallel lines $H_{xx}=H_{xx,best}-0.0208 \left(P_0-P_{0,best}\right)$, indicated by the dash-dotted lines in the figure. However, the best fit values fall along the blue line, with much lower slope. 
Unless the estimated uncertainties represented by the ellipses were vastly larger than what we have estimated, the four determinations are not statistically compatible. 
Therefore it seems likely that the data indicates a systematic process that is not included in the model, and which causes the deduced $H_{xx}$ to depend on $P_0$. 
A possible candidate for such an effect is boiling at the surface of the electrical heater.  The bubbles that reach the surface without thermal equilibration with the liquid would carry some of the heater power into the gas. This would reduce the actual evaporation rate over that obtained by computing the evaporation rate assuming that all the heater power goes into the LAr as we have done.  Increasing the heat input from the walls would presumably allow more bubbles containing hotter gas to reach the surface.  Since this process is not fundamental to the typical operation of LAr detectors, we will not consider it further, but simply suggest that future measurements should avoid the problem by ensuring that the power density at the surface of the heater is kept below a value that causes boiling.

We can ignore  this systematic effect and estimate of the value of $H_{xx}$ by simply taking the average and standard deviation of the four determinations in Table~\ref{tab:best_fit_params}, to obtain a value  for $H_{xx}$ of 0.86$\pm$0.04. Alternately, we can try to eliminate the systematic dependence on $P_0$ by extrapolating $H_{xx}$ to $P_0=0$ assuming a linear dependence.  This gives a best estimate of 0.938$\pm$0.024. Both these values are consistent with the literature value.  Of course, the use of a linear extrapolation has no physical justification. In fact, if boiling is the cause, one might expect a function that approaches a constant value as the heater power approaches zero.  And we have found other two-parameter functions that produce a similar or better quality of fit with extrapolated values of Henry's coefficient ranging from about 0.90 to 1.04. 

We have also examined, using data set \#1, the systematic uncertainty introduced by our lack of knowledge of the value of $k_{dis}$ by varying its value and repeating the least-squares procedure. It is found that for $k_{dis}$ greater than 0.7~s$^{-1}$, there is no change in the best values for $H_{xx}$ and $P_{0}$. As $k_{dis}$ is decreased below this value, however, the best $H_{xx}$ ($P_{0}$) value decreases (increases), as indicated by the dashed line and solid points in Fig.~\ref{fig:Hxx_Pin_Contours}. The MS fractional deviation increases slowly along this line as $k_{dis}$ decreases, from 0.032 at $H_{xx}$=0.80 to 0.039 at $H_{xx}$=0.70. 
Therefore, the data shows little preference for any value of $k_{dis}$ greater than about 0.09~s$^{-1}$, and it is not possible to define a ``best" value of $k_{dis}$, except to observe that as the value decreases below 0.01~s$^{-1}$ the value obtained for Henry's coefficient goes to zero.
Therefore we conclude that the dissolution rate constant must be larger than $\sim$0.1~$s^{-1}$. Further measurements are needed to define the dissolution rate constant.  We discuss the implications of other systematic uncertainties in Sec.~\ref{subsec:systemtics}.

%% file: Henry_Coeff.tex
\subsection{Henry's Coefficient for O$_2$ in LAr from Reduced Model} \label{sec:henryo2}

While the full model considers all the physical effects, it is difficult to use it in the error analysis of the extracted parameters because of the low frequency divergent noise in our data. Therefore, in this section, we describe the determination of Henry's coefficient using the same data with the reduced model obtained in Sec.~\ref{sec:time}, with an improved error analysis. 
The reduced model allows quick determinations of Henry's coefficient and can provide a cross check of the full calculation, when the appropriate conditions are met (valid in the 20-L system). 

\begin{figure*}[htbp]
       \centering
       \vspace{-2cm}
       \includegraphics[width=0.85\textwidth]{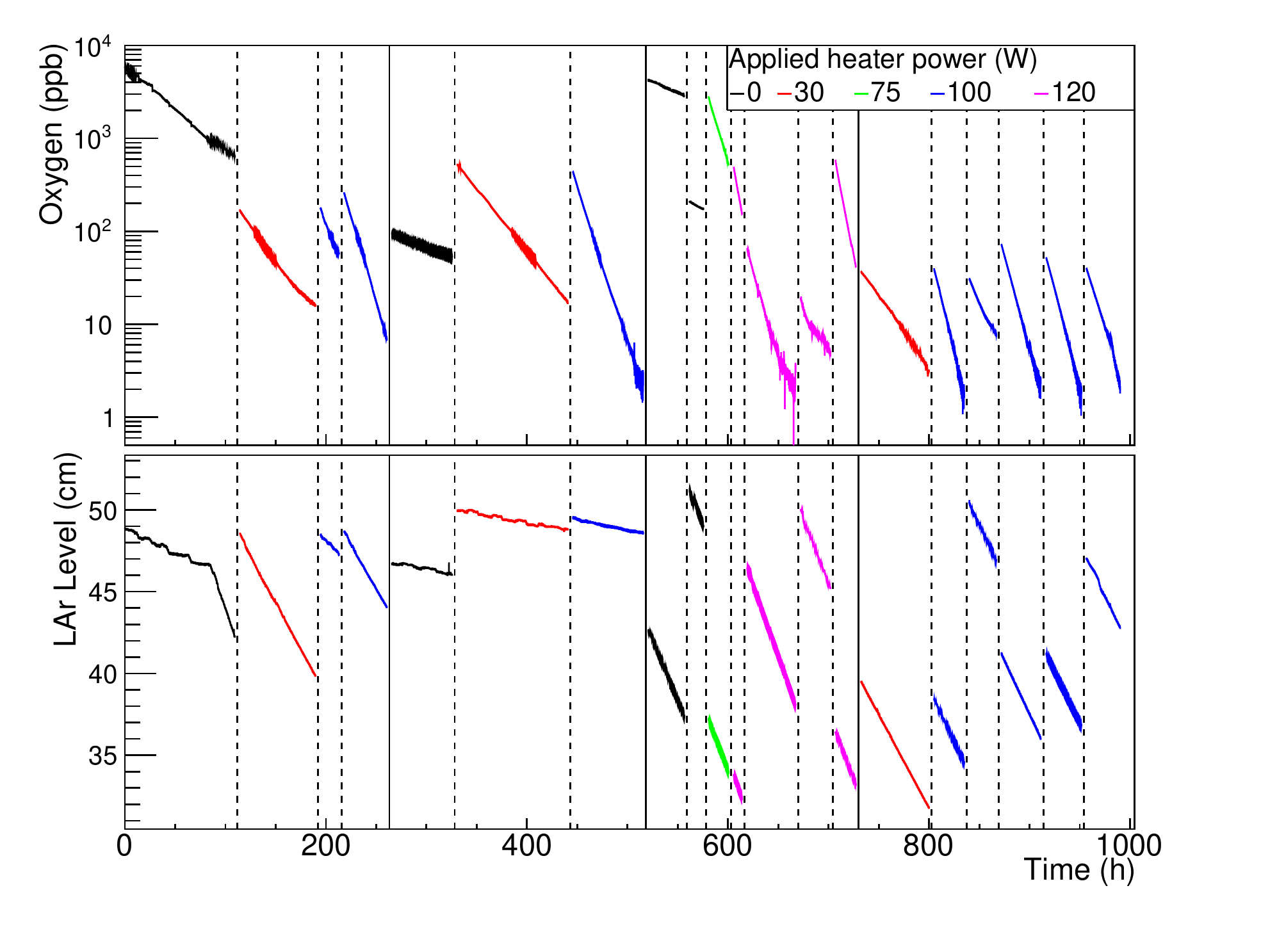}
       \vspace{-0.5cm}
       \caption{The oxygen concentration data selected from all data sets at various heater powers (top) and the corresponding liquid argon level in the detector (bottom) to be used in the analysis in Sec.~\ref{sec:henryo2}. The data shown are for sets \#1 to \#4 from left to right separated by the three vertical solid lines, and the applied heater powers are indicated by different colors. The time scale is a running total.}
       \label{fig:AllOxygenCleaningCurves}
 \end{figure*}

The oxygen concentration data selected in the analysis from all four data sets are shown in the top panel of Fig.~\ref{fig:AllOxygenCleaningCurves}.
The required conditions for data selection include 1) well behaved  logarithmic curves of oxygen concentration within each period, 2) LAr level decreasing at a constant rate (shown in the bottom panel of Fig.~\ref{fig:AllOxygenCleaningCurves}), and 3) a constant heater temperature.

We begin with Eq.~\eqref{eq:Henry_approx} and define a new rate (also referred as cleaning rate) as 
$r_S \equiv  \langle k_{S} \rangle _{Fit} \cdot \langle n_l \rangle_{Est}$, so that the approximation to Henry's coefficient is just this new rate divided by the evaporation rate:
\begin{equation}
\begin{aligned}
    H_{xx} & \sim h \equiv  \frac{r_S}{r_{evp}}, or \\
    r_{S} & \sim H_{xx} \cdot (P_{in,H}+P_0)/\Delta H_{evp}.
\end{aligned}
    \label{eq:Henry_appro4}
\end{equation}
In the $r_{S}$ definition, the slow rate constant $k_{S}$ is replaced by $\langle k_{S} \rangle _{Fit}$, which is obtained from fitting a single exponential function to the data\footnote{Note that the $k_{F}$ term in the impurity concentration solution (Eq.~\eqref{eq:two_eq_solution}) can be ignored as discussed in Sec.~\ref{discuss_kSkF}.} and the initial LAr amount $n_{0,l}$ in Eq.~\eqref{eq:Henry_approx} is replaced by $\langle n_l \rangle_{Est}$, an amount to be determined from the LAr level data. 
The evaporation rate is determined by the total power input ($P_{in}$) into the LAr (see Eq.~\ref{eq:evaporation_rate}), which is the sum of the heater power ($P_{in,H}$) and the heat leakage ($P_{0}$) into the LAr, with the assumption that $P_{0}$ is a constant throughout each data set.
Eq.~\eqref{eq:Henry_appro4} shows that Henry's coefficient is just the slope of a linear function between $r_{S}$ and the evaporation rate caused by the heater, $r_{evp,H} = P_{in,H}/\Delta H_{evp}$, with an x-axis intercept of $-P_0\cdot H_{xx}/\Delta H_{evp}$.

\begin{figure*}[htp]
       \centering
       \vspace{-0.5cm}
       \includegraphics[width=0.95\textwidth]{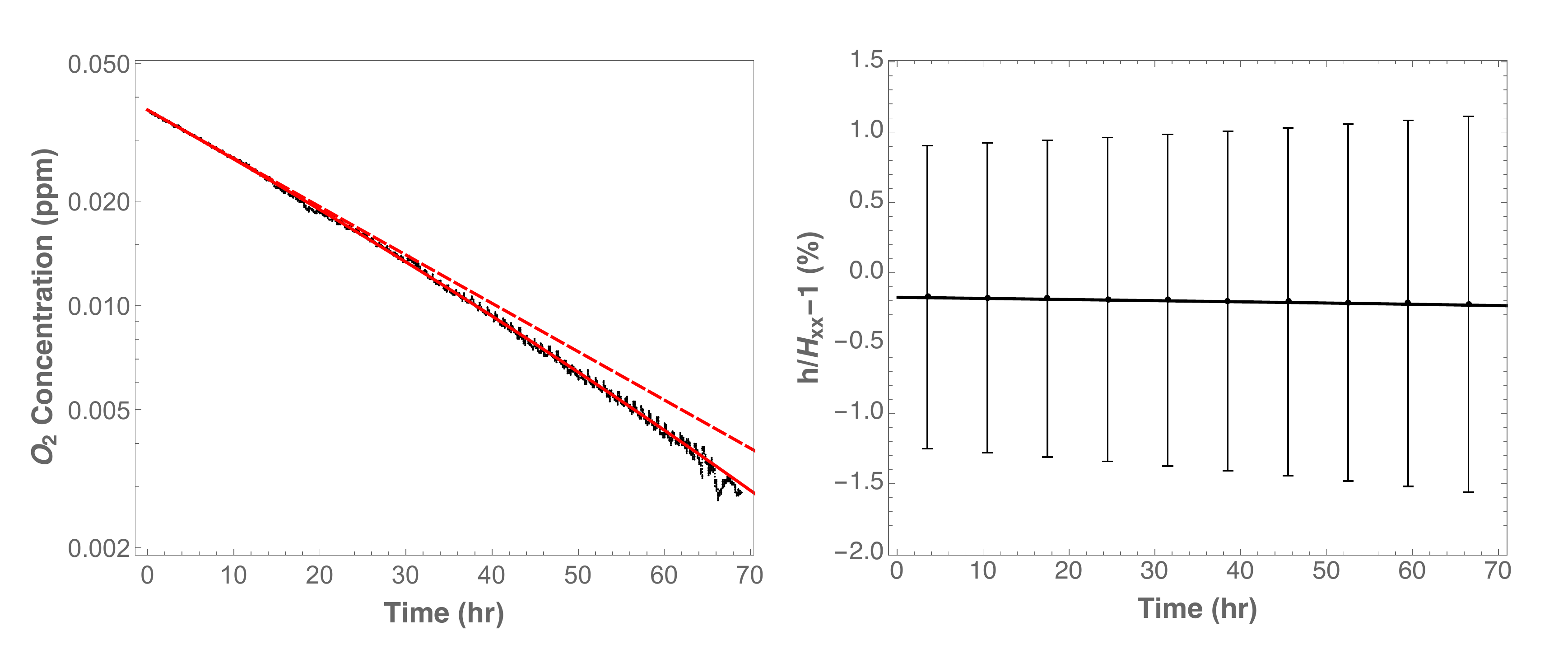}
       \caption{At left: The effect of sampling of the liquid on the cleaning rate of oxygen. The data are the black points and the two calculations are the red curves with all parameters identical except that for the dashed the sampling rate is zero, and for the solid it is the measured value of 1.81 mol/h. At right: the calculated error in the value of Henry's coefficient deduced from calculation with a sampling rate of 1.81~mol/h using the approximation of Eq.~\eqref{eq:Henry_approx} instead of the full calculation: the solid line is the error when $n_{0,l}$ is replaced with the actual LAr amount at each time, i.e. $n_l(t)$, and the seven points are the error when both the instantaneous cleaning constant and amount of LAr are averaged over 7-hour intervals.  The vertical bars indicate the limiting error when the amount of LAr is taken at the beginning or end of the 7-hour interval.}
        \label{fig:Sampling_Effect}
\end{figure*}

Since the amount of liquid decreases with time due to sampling, a single exponential function is not appropriate to describe each curve in the top panel of Fig.~\ref{fig:AllOxygenCleaningCurves}. 
In the analysis, we divide each curve into sub-regions of 4$\sim$7~hours\footnote{$1/k_S$ is estimated to be a few hours in Sec.~\ref{sec:time}.} and use the LAr level in the middle of each sub-region for calculating $\langle n_{l} \rangle _{Est}$. 
This is justified by the following discussion, where we compare $h$ obtained from Eq.~\eqref{eq:Henry_appro4} with $H_{xx}$ obtained in the model calculation. 
We choose the period of the data with the largest fractional change in the amount of LAr (taken from set \#4), which has a large sampling rate (1.81~mol/h) over a long period (68 hours) so it has the largest error that can result. 
The measured oxygen concentration as a function of time is shown in the left panel of Fig.~\ref{fig:Sampling_Effect}. 
Also shown are 1) the best fit to the full model, using the measured sampling rate and $H_{xx}=0.919$, and 2) the calculation with identical parameters except for the sampling rate being set to zero. 
The fractional deviation of $h$ as determined by Eq.~\eqref{eq:Henry_approx} from the $H_{xx}$ (i.e., the value used in the calculation) is shown in the right panel of Fig.~\ref{fig:Sampling_Effect}. The fractional deviation is shown for three choices of $\langle n_l \rangle_{Est}$: the initial value $n_{0,l}$, the instantaneous value $n_l(t)$, and the average value over ten hour intervals $\langle n_l(t) \rangle_{\Delta t=10}$. 
The best choice is the average value, and the systematic error in Henry's coefficient introduced by approximating the true value with $h$ is less than 0.3\% over a total measurement period of $\sim$70 hours.  
Even if the LAr amount at the beginning or end of the analysis interval is used to compute the Henry's coefficient the error is only $\pm$1\%. This is negligible compared to other systematic uncertainties (which will be further discussed in Sec.~\ref{subsec:systemtics}).

After the sub-regions are decided, each starting at its own $t_{0}$ is fitted with a single exponential function of time
\begin{equation}\label{eq:single_exp}
c_{i,l}(t)=A \cdot e^{-\frac{t-t_{0}}{\tau}}+c_{0},
\end{equation}
with $c_{i,l}(t)$ being the concentration of the impurity at time $t$, $c_{0}$ being the ultimate concentration that the system can reach,
$A$ being a coefficient depending on $c_{0}$ and $c_{i,l}(t_{0})$. 
Note that $c_{0}$ depends on the impurity leakage and total input heating power as discussed in Eq.~\eqref{eq:steadystate_liq}.
Here we have $\tau= \langle k_S \rangle _{Fit} ^{-1}$ and therefore the rate for cleaning of LAr is computed as $r_S= \langle n_{l} \rangle _{Est}/\tau$. 

Each curve in the top panel of Fig.~\ref{fig:AllOxygenCleaningCurves} at a fixed heater power is first fitted as a whole so that an overall $c_{0}$ is determined, which is then fixed when fitting each sub-region in that curve. The range of $c_{0}$ varies from $\sim$0 to 10~ppb for different heating powers and impurity leak rates, the impact of different values of $c_{0}$ (varied by a few ppb) is small to the extraction of $r_S$ within each sub-region.

\begin{figure}[t] 
\centering
   \includegraphics[width=0.6\textwidth]{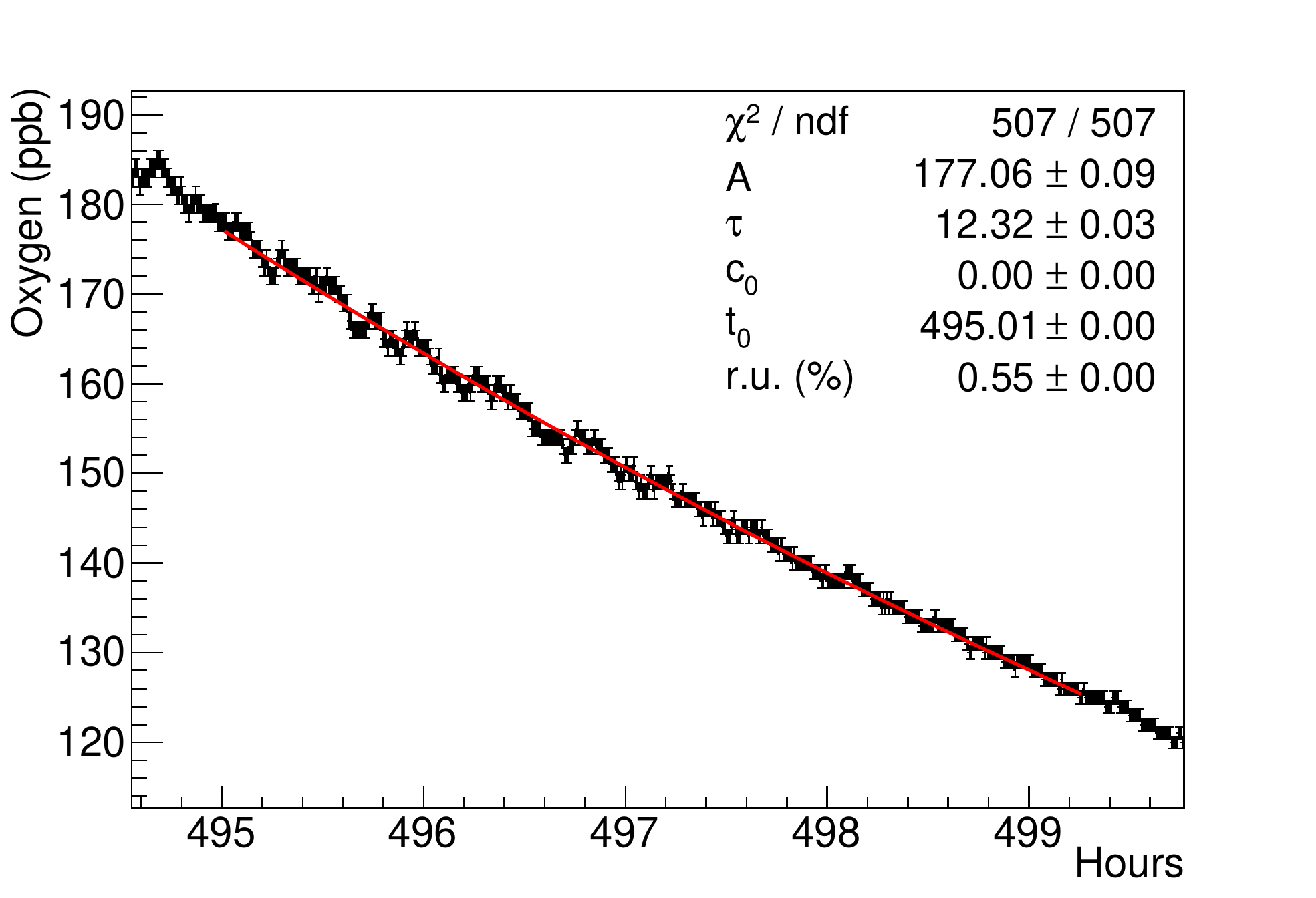}
   \caption{An example showing the exponential fit to oxygen concentration data for a subregion (from 495 to 499.25 hours in Fig.~\ref{fig:AllOxygenCleaningCurves}) in data set \#1 at 100~W heater power. $c_{0}$ and $t_0$ are fixed. About 0.55\% relative uncertainty (r.u.) is added to the data points to ensure $\chi^{2}/ndf=1$. The time constant in this fit is $\tau=12.3$~hours, the $r_S$ extracted from this fit is $0.016\pm0.00004$~mol/s.}
   \label{fig:subregion_fit_example}
\end{figure}

As has been discussed in the previous section, not all the data points are independent because of the nature of the noise. To account for this, additional relative uncertainties are added to the data points until the $\chi^{2}/ndf$ of the exponential fit to be close to unity.
For most sub-regions these relative uncertainties are below 1\%, a $\sim$10\% relative uncertainty is needed when the absolute concentrations in the sub-regions are below 10~ppb, where the fluctuation becomes relatively sizeable.
Fig.~\ref{fig:subregion_fit_example} shows an example fit to one sub-region from a curve in data set \#1 at 100~W heater power, from which $r_S$ is determined to be $0.016\pm0.00004$~mol/s. In this fit, a 0.55\% relative uncertainty is added to the data points to allow for $\chi^{2}/ndf$ being unity.

\begin{figure}[t] 
\centering
\vspace{-2cm}
    \begin{minipage}[]{0.48\textwidth}
       \includegraphics[width=1.1\textwidth]{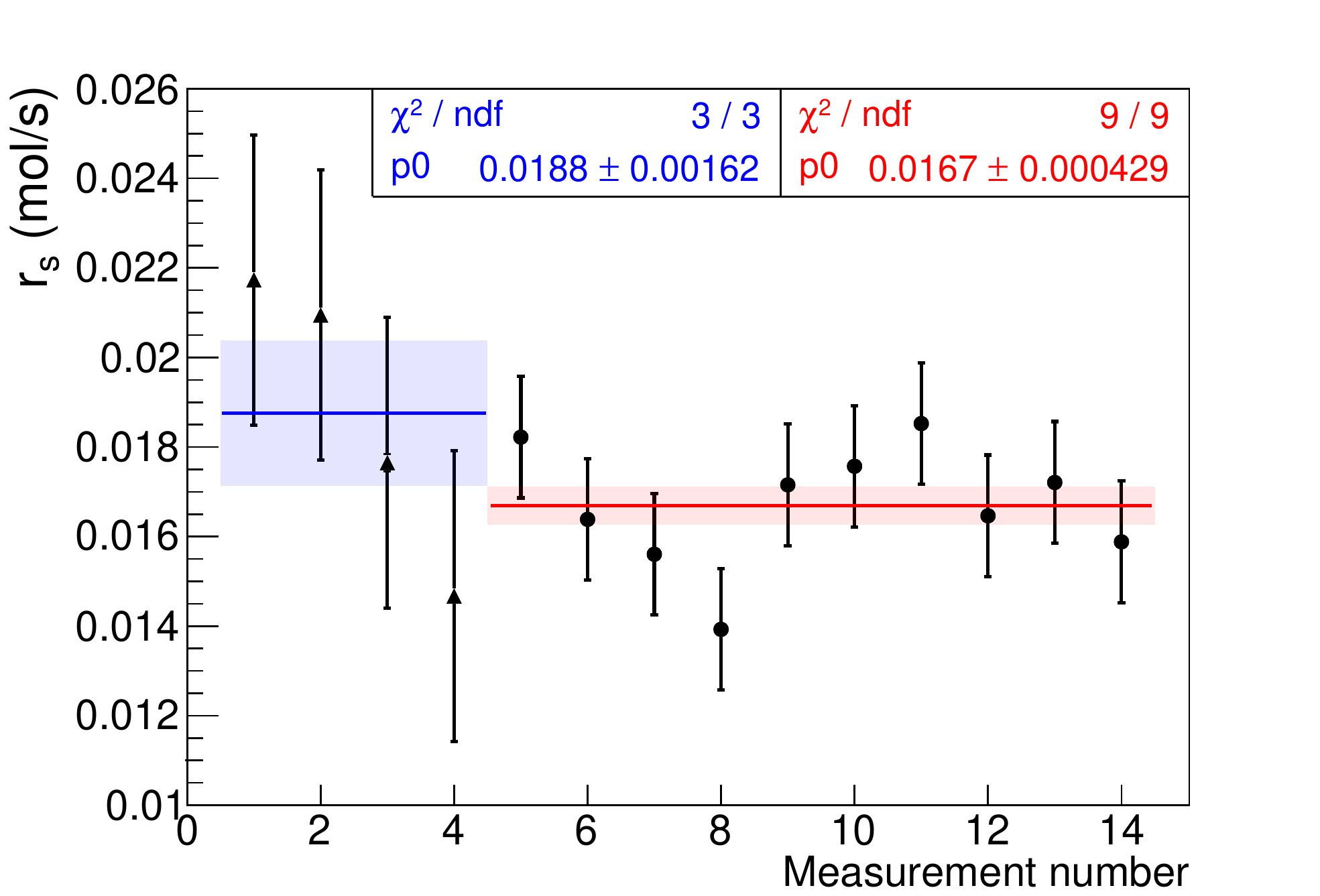}
    \end{minipage}
    \begin{minipage}[]{0.48\textwidth}
       \includegraphics[width=1.1\textwidth]{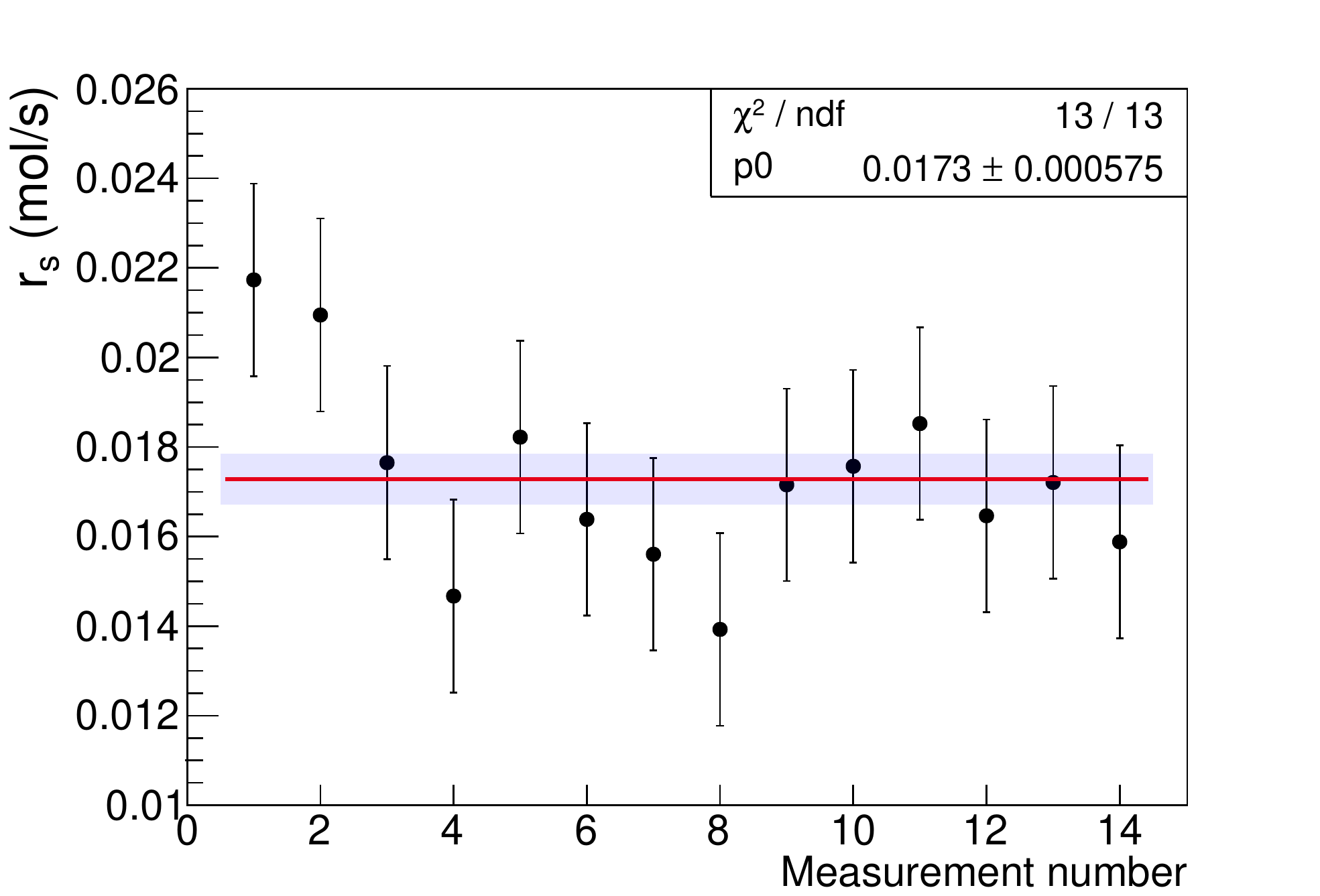}
    \end{minipage}
\caption{The $r_S$ values obtained from the sub-regions in the two 100 W heater regions in Set \#1. The first four data points are from four sub-regions from 28.5 to 47 hours, and the rest are from 10 sub-regions from 382 to 424.5 hours. The two groups of data points are assigned additional systematic uncertainties of 0.00324 and 0.001356~mol/s, to make $\chi^{2}/ndf$$\sim$1, and yield $r_{S}$ values of $0.0188\pm0.0016$ and $0.0167\pm0.0004$~mol/s, respectively (left plot). The blue and red lines and the shadows indicate the central values and their associated uncertainties. These two values are consistent and therefore all the data points are used in a linear fit with additional uncertainties of 0.00215~mol/s to produce a measurement of $r_{S}=0.0173\pm0.0006$~mol/s, indicated by the shadow and its central line (right plot).}
\label{fig:Feb2016_two_100W}
\end{figure}

After all sub-regions in the same curve are processed, we obtain a few measurements for $r_S$ at the fixed heater power. The best-fit $r_S$ values come out with very small statistical uncertainties (e.g.,$\pm$ 0.00004~mol/s in Fig.~\ref{fig:subregion_fit_example}). The difference among $r_S$ values from different sub-regions cannot be explained by merely the statistical uncertainties. 
Some sources of systematic uncertainties are briefly discussed in the following: 
the time variations of heating power leakage and air leakage into the cryostat could affect the cleaning rate; 
the constant concentration term being time dependent could be another factor whereas it is treated as fixed in each data set; 
also sampling can affect the cleaning rate as the amount of LAr decreases.
To estimate the systematic uncertainties, we compare the measured $r_S$ at different sub-regions. Similarly to the previous exponential fit, an additional uncertainty is added to each data point until a constant fit of results from all sub-regions giving $\chi^{2}/ndf\sim1$.
For example, Fig.~\ref{fig:Feb2016_two_100W} (left panel) shows the $r_S$ values obtained in the sub-regions of the two curves with 100~$W$ heater power in data set \#1. It is found that, by assigning additional uncertainties of 0.00324 and 0.00136~mol/s to the two curves independently, we obtain two measurements of $r_S$ values: $0.0188\pm0.0016$ and $0.0167\pm0.0004$~mol/s, respectively. These two measurements are consistent within one standard deviation. We can then combine all the data points and obtain one measurement of $r_S$ for the 100~W heater power data in data set \#1, which is $0.0173\pm0.0006$~mol/s as also shown in Fig.~\ref{fig:Feb2016_two_100W} (right panel).

\begin{figure}[!htbp] 
\centering
\vspace{-0.4cm}
   \includegraphics[width=0.6\textwidth]{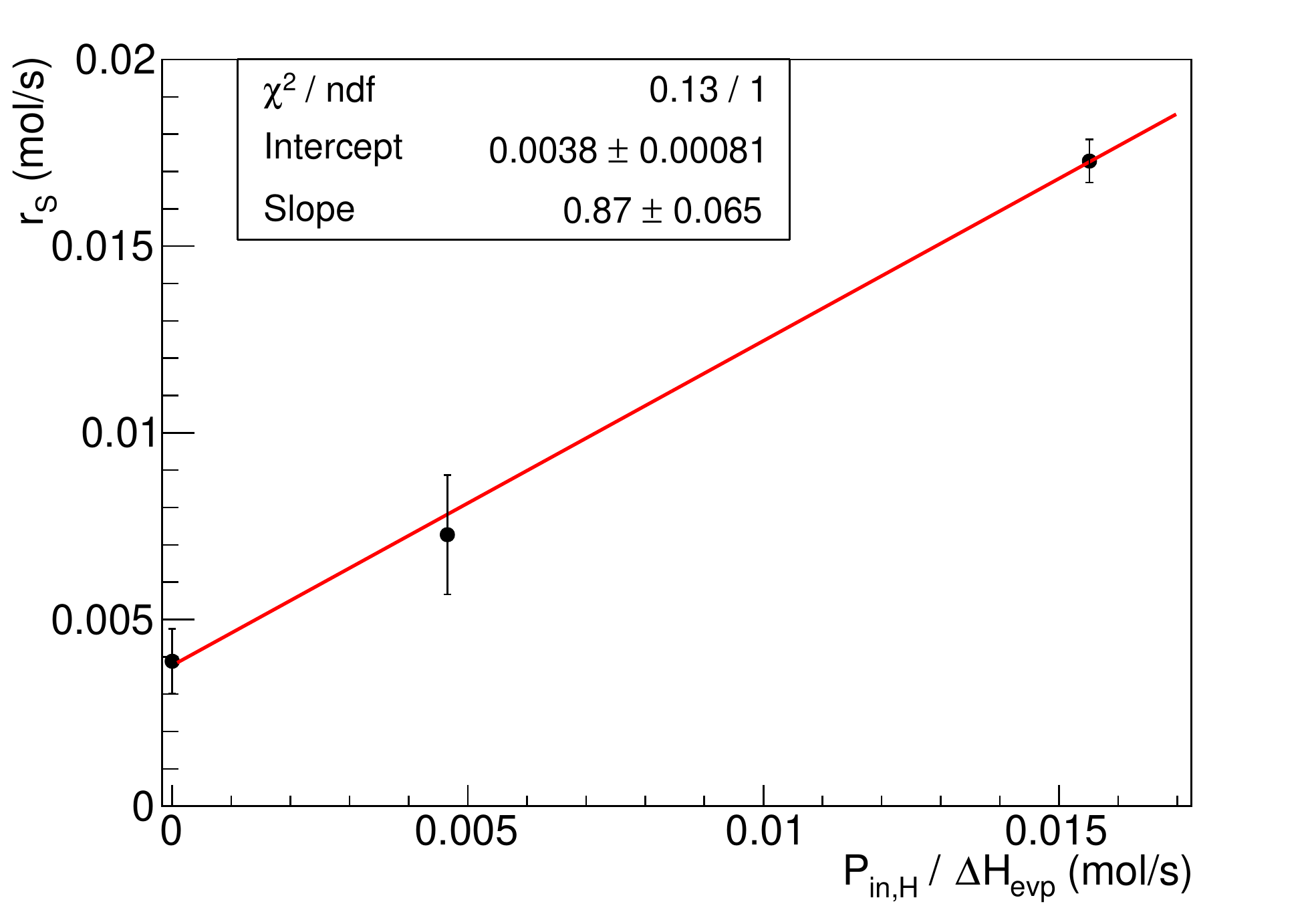}
   \caption{Cleaning rate for oxygen as a function of evaporation rate introduced by the heater, measured in data set \#1.}
   \label{fig:clnRate_vs_evpRate_Feb2016}
\end{figure}

The systematic uncertainty obtained above is considered relative as it is heating power dependent. With the above numbers, we obtain a relative systematic uncertainty of 12.4\% from the two 100~W heater power curves in data set \#1.
All the data sets are processed following the same procedures, an average relative systematic uncertainty is estimated to be 22\%, which is then added to the statistical uncertainties of $r_S$ when only one measurement exists (e.g., there is only one measurement for 0~W and 30~W heater powers in data set \#1, see Fig.~\ref{fig:Feb2016DataSet}).
Finally, $r_S$ as a function of $P_{in,H}/\Delta H_{evp}$ is plotted, e.g., Fig.~\ref{fig:clnRate_vs_evpRate_Feb2016} is such a plot for data set \#1.
A linear fit to the data points in Fig.~\ref{fig:clnRate_vs_evpRate_Feb2016} gives a slope of $0.87\pm0.07$ and an intercept of $0.0038\pm0.0008$. The slope measures the Henry's coefficient for oxygen in argon, and the heat leakage in the system from this data set is obtained as $27.9\pm6.4$~W when extrapolating the linear fit to $r_S=0$ (Eq.~\eqref{eq:Henry_appro4}).

\begin{table}[!htbp]
\vspace{-0.1 cm}
\caption{Best Values for $H_{xx}$ and $P_0$ for each data set from the reduced model analysis. }
\begin{center}
\vspace{-0.5cm}
\begin{tabular}{|c|c|c|}
\hline
Set  &  $H_{xx}$ & $P_0$~(W)  \\\hline 
  1   &  0.87$\pm$0.07   & 27.9$\pm$6.4       \\\hline
  2   &  0.95$\pm$0.19   & 12.7$\pm$3.8      \\\hline
  3   &  0.78$\pm$0.06   & 18.6$\pm$2.6      \\\hline
  4   &  0.92$\pm$0.12   & 9.6$\pm$12.7      \\\hline
\end{tabular}
\label{tab:best_fit_params2}
\vspace{-0.5cm}
\end{center}
\end{table}

We analyzed the four data sets independently with the above method and obtained values for $H_{xx}$ and $P_{0}$ for the data sets as shown in Table~\ref{tab:best_fit_params2}, the weighted average for $H_{xx}$ is 0.84$\pm$0.04. 
The results are consistent with those obtained in Sec.\ref{sec:model_calc} (Table~\ref{tab:best_fit_params}).

%% file: uncertainty_discuss.tex
\subsection{Further Discussion of Systematic Uncertainties}  \label{subsec:systemtics}

In this section, we discuss other sources of systematic uncertainties on $H_{xx}$ that are not covered in the above analyses. 
These additional systematic uncertainties are introduced by (1) our imprecise knowledge of the amount of liquid and gas in the system, (2) temperature variation of the liquid with time, (3) variation of the heat leakage into the LAr with LAr level, and (4) uncertainties in the heater power input to the LAr.

The amount of liquid is computed from the LAr level as measured by a differential pressure gauge. The accuracy of this device, as stated by the manufacturer, is $\pm 0.8\%$ of full scale (25 inches of water), which translates to $\pm0.37$~cm of LAr. Given the nominal geometry of the cryostat, this is $\pm0.9\%$ of the average amount of LAr. The uncertainty in the nominal geometry, which comes principally from the uncertainty in the domed shape of the cryostat bottom, is estimated to be $\pm1.5\%$.  The total uncertainty, added in quadrature, is about $\pm2\%$.

The change in $H_{xx}$ with temperature per degree, computed from Eq.~\eqref{eq:H_T_relation}, is 0.0025~K$^{-1}$. The observed maximum range of the temperature of the LAr over all measurements is 2.5~K. this change in temperature would therefore result in a maximum change of Henry's coefficient of 0.006.  Even this maximum value is negligible compared to the other uncertainties.

The heat leakage into the liquid, $P_{0}$, occurs because the contact area of the liquid with the walls, the heat conducted down the thin inner wall of the cryostat to the liquid, and the heat convected and radiated from the top flange through the gas into the liquid. We estimate the latter two to be much smaller than the first.  The contact area decreases with decreasing LAr level, and this effect would make the deduced $H_{xx}$ systematically smaller. We have looked for such an effect in the decrease in the cleaning rate constant with decrease in LAr level, but have been unable to detect any coherent effect, presumably because it is smaller than the noise in the data.  

The power output of the heater immersed in the LAr is known by measuring the voltage and current supplied to it.  However, boiling of the LAr probably occurs at the surface of the heater~\cite{Bewilogua} at higher powers.  If this occurs, then bubbles of GAr may reach the surface, carrying some of the heat directly into the gas, rather than dissipating it in the LAr, as we assume in computing $r_{evp}$ values for the model calculations. As the insulation leakage heat increases, the amount of vapor from boiling reaching the surface will presumably increase, and this effect will become larger. This could explain the  systematic dependence of $H_{xx}$ on $P_0$ observed in Fig.~\ref{fig:Hxx_Pin_Contours} and Table~\ref{tab:best_fit_params2}. 

It is difficult to estimate the magnitude of this effect without quantitative information about the amount of boiling, but this process would systematically make the true values of $P_{in,H}$ in the LAr smaller than calculated from the electrical input values, thus any boiling will cause the measured $H_{xx}$ value to be systematically too small.  Therefore we consider that this uncertainty can be estimated as the difference between the value of $H_{xx}$ found by extrapolation to zero heater power in Fig.~\ref{fig:Hxx_Pin_Contours} and the mean of the four determinations of $H_{xx}$ given in Table~\ref{tab:best_fit_params}, which is $+7.8$\%.

The systematic uncertainties are summarized in Table.~\ref{tab:summary_uncertainty}.  The total uncertainty is then the sum in quadrature of these uncertainties and the statistical uncertainty of 0.04, which gives a best estimate for Henry's coefficient, as determined from our data, of $0.84^{+0.09}_{-0.05}$.

\begin{table}[!htbp]
\vspace{-0 cm}
\caption{Summary of the systematic uncertainties contributed to the measured $H_{xx}$. }
\begin{center}
\vspace{-0.5cm}
\begin{tabular}{|c|c|}
\hline
Source  &  Contribution  \\\hline 
  LAr amount   &  $\pm2$\%    \\\hline
  Temperature variation   &  $\pm0.6$\%    \\\hline
  Total heat power   &  +7.8\%   \\\hline
  
\end{tabular}
\label{tab:summary_uncertainty}
\end{center}
\end{table}

%% file: baffle_CET.tex
\subsection{Reduction of Impurities by Adding a Baffle in a LAr Detector}\label{sec:baffle}

The observation in Sec.~\ref{sec:model_calc} that the oxygen effective leak rate decreases as the evaporation rate increases can be explained semi-analytically by a ``back-diffusion" model. After entering the cryostat at the top flange, the oxygen impurity moves by convection and diffusion past the heat shields into the gas ullage space below the bottom heat shield. In this region the impure gas mixes turbulently with the flow of evaporation gas, and this mixture is carried out to the purifier by a tube that starts just below the bottom heat shield. To reach the liquid surface and dissolve, the impurity in this mixture must move against the flow of  evaporated gas rising from from the liquid. As the surface of the liquid is approached the flow becomes quite laminar, since the evaporation occurs uniformly over the entire surface of the liquid; and, because the evaporation rate is low and the liquid surface area is large. the Reynolds number is low. 
The back-diffusion through this laminar layer is responsible for reducing the rate of impurity  reaching the liquid surface to a value less than the total leak rate entering at the top flange of the cryostat. 
Back-diffusion was first studied in Ref.~\cite{Hertz1923} and is further discussed in Ref.~\cite{reus} as a method for measuring diffusion constants in gas.  It is the basis of modern techniques for separating mixtures of isotopes.

\begin{figure}[htb]
       \centering
       \includegraphics[width=0.65\textwidth]{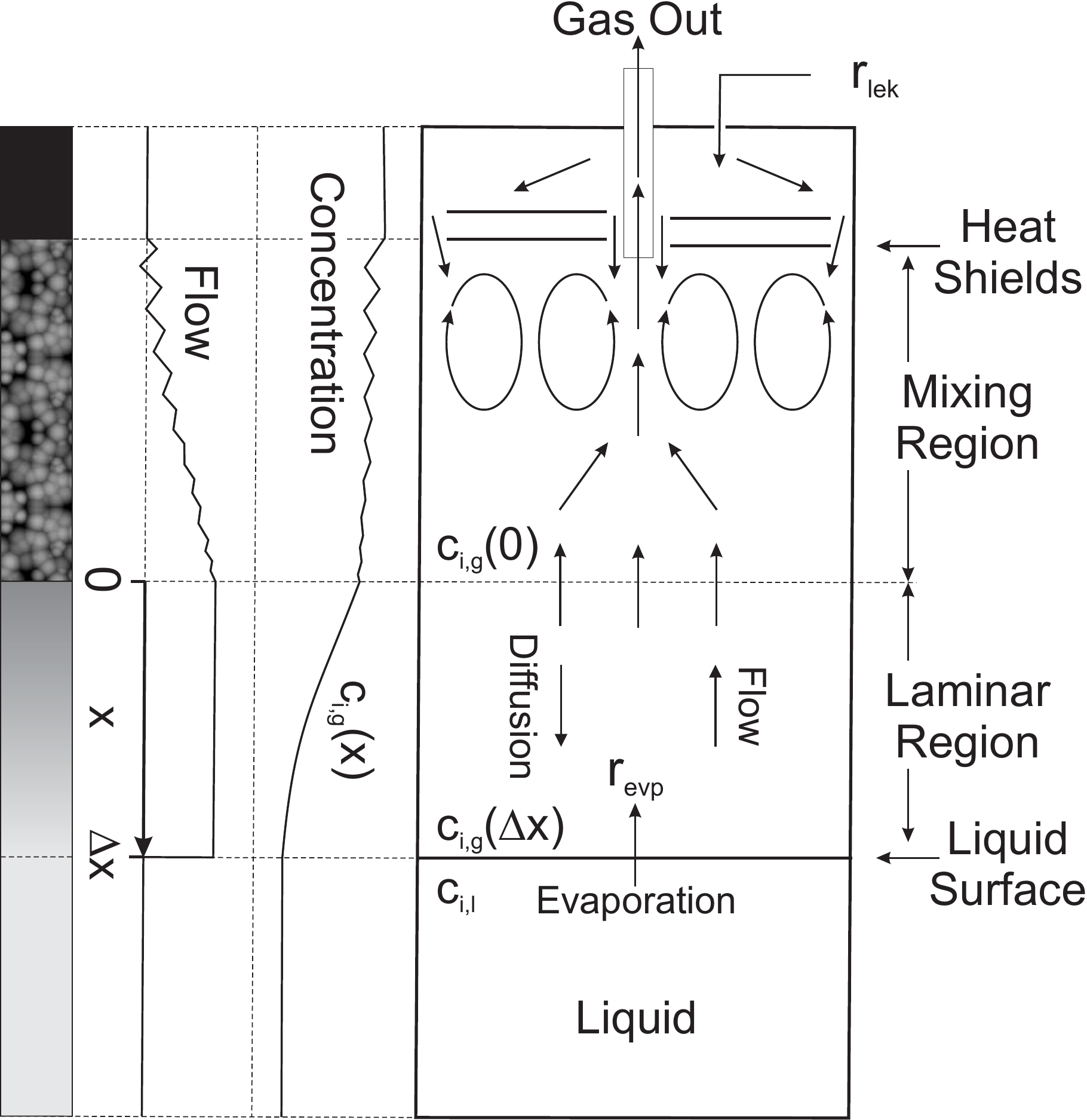}
       \caption{A schematic diagram of the conditions in the cryostat appropriate for the calculation of diffusion against flow as described in the text.}
       \label{fig:Flow_Against_Diffusion}
 \end{figure}

A one-dimensional model describing the application of this process to our cryostat is suggested in Fig.~\ref{fig:Flow_Against_Diffusion}.
As the surface of the liquid is approached the flow becomes more nearly laminar, since the evaporation occurs uniformly over the surface of the liquid.  

Below some height in the ullage, indicated by the dotted line across the cryostat in the middle of the Fig.~\ref{fig:Flow_Against_Diffusion}, we assume the mixing is essentially complete and a constant impurity concentration of $c_0$ is achieved.
The GAr linear flow velocity $v$ and the diffusion coefficient $D$ of oxygen in GAr are constant throughout this volume.
When the system comes to a steady state, the net flow of oxygen across the cross sectional area, $A_c$, of the cryostat must be zero.
Therefore we write the differential equation for the concentration in the laminar flow region:
\begin{equation}\label{eq:mass_balance} 
v \cdot c_{i,g}(x) + D\cdot \frac{dc_{i,g}(x)}{dx}=0,
\end{equation}
where $x$ is the distance from the start of the laminar flow boundary (pointing downward in Fig.~\ref{fig:Flow_Against_Diffusion}), $v$ is the GAr linear flow velocity (upward), and $c_{i,g}(x)$ is the impurity concentration.
The solution of this equation gives the concentration as a function of distance
\begin{equation}
c_{i,g}(x)=c_{i,g}(0) \cdot e^{-\frac{\textit{v} \cdot  x}{D}}.
\end{equation}
The GAr linear flow velocity is related to the evaporation rate of LAr by
\begin{equation}
\begin{aligned}
v=\frac{r_{evp}\cdot V_m}{A_{c}},
\end{aligned}
\end{equation}
where $V_m$ is the molar volume of GAr and we have the downward direction as positive.
Combining these last two expressions we obtain the concentration in the laminar region in terms of the evaporation rate
\begin{equation}\label{eq:conc_attenuation} 
c_{i,g}(x)=  c_{i,g}(0) \cdot e^{- \frac{r_{evp}\cdot V_m \cdot x}{D\cdot A_c}},
\end{equation}
where $c_{i,g}(0)=\frac{r_{lek}}{r_{evp}}$ can be considered as the concentration in the top region assuming complete mixing of the impurity into GAr.


An effective leak rate of oxygen, $r_{eff,lek}$, can be defined as 
\begin{equation}\label{eq:effective_attenuation_conc} 
r_{eff,lek} \equiv r_{evp} \cdot c_{i,g}(\Delta x)=r_{lek} \cdot F,
\end{equation}
where $\Delta x$ is the thickness of the laminar flow region and $F$ is an attenuation factor defined as
\begin{equation}\label{eq:gammaFunction}
    F = e^{-\gamma \cdot r_{evp}},
\end{equation}
and $\gamma=V_m \cdot \Delta x/(D\cdot A_c)$ being the decay constant.

\begin{figure*}[!htb]
       \centering
       \includegraphics[width=0.6\textwidth]{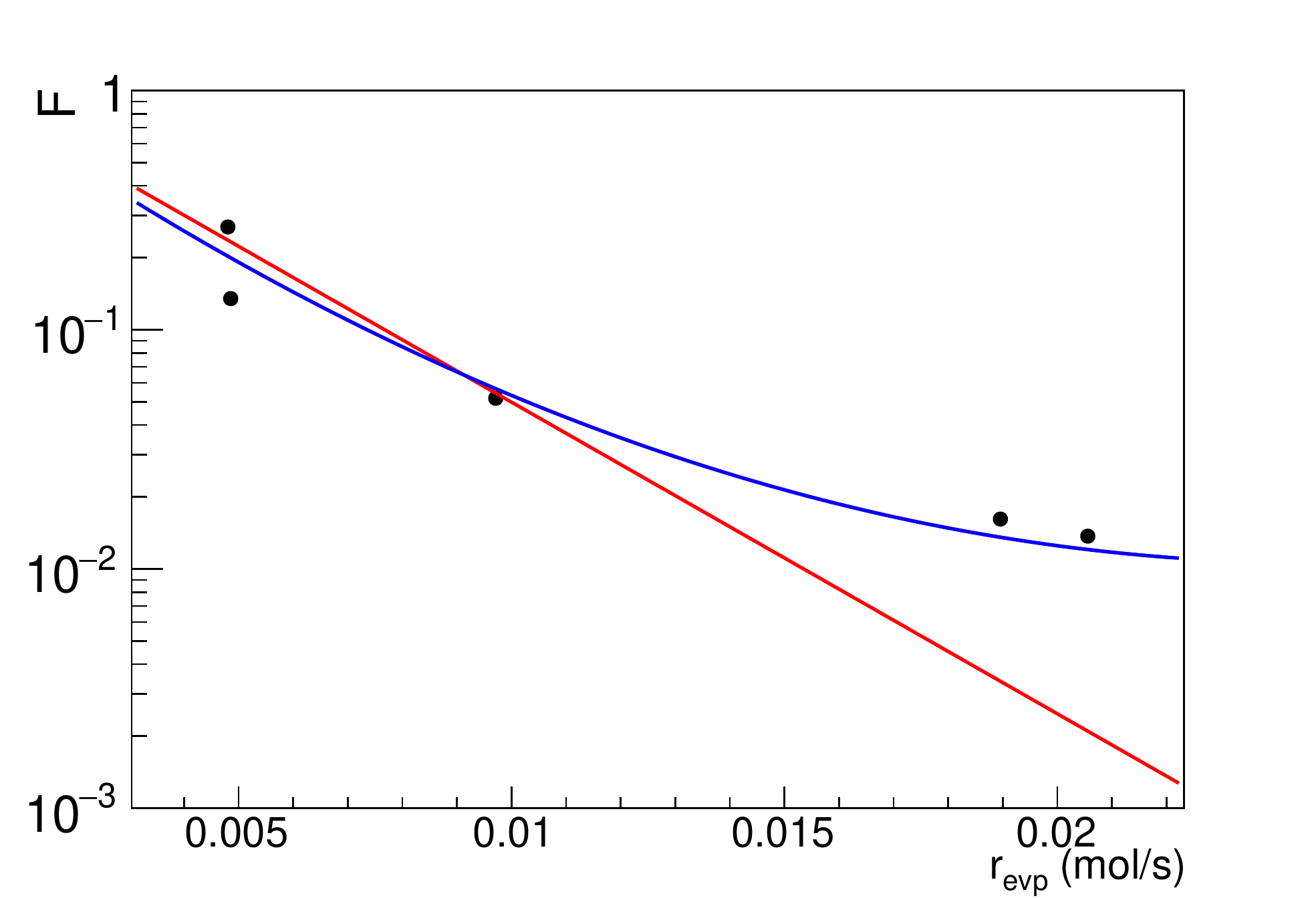}
       \caption{The leak rate attenuation factor deduced from fitting the model to the concentration data in Fig.~\ref{fig:Feb2016DataSet} as a function of the evaporation rate. Another data point comes from data set 4 at 100~$W$ heater power.
       The red line is fitted with Eq.~\eqref{eq:gammaFunction}, while the dashed line is calculated by adding a quadratic term of $r_{evp}$ to $\gamma$.
       }
       \label{fig:Leak_Attenuation}
 \end{figure*}

As discussed in Sec.~\ref{sec:model_calc}, the effective leak rate (the rate of oxygen from the gas arriving at the liquid-gas surface) can be determined by comparing the model to the data at a given evaporation rate, while the total oxygen leak rate is determined from the data without purification. The attenuation factor is computed as the ratio of leak rates with and without purification; the values determined at a several evaporation rates are shown by the data points in Fig.~\ref{fig:Leak_Attenuation}. Fitting the data points with Eq.~\eqref{eq:gammaFunction} yields $\gamma\sim328$~s/mol, however, it is found that adding a quadratic term of $r_{evp}$ in the fitting function will describe the trend better. This can be understood as a second order effect, which we will not quantitatively discuss further.

\begin{figure*}[!htb]
       \centering
       \includegraphics[width=0.5\textwidth]{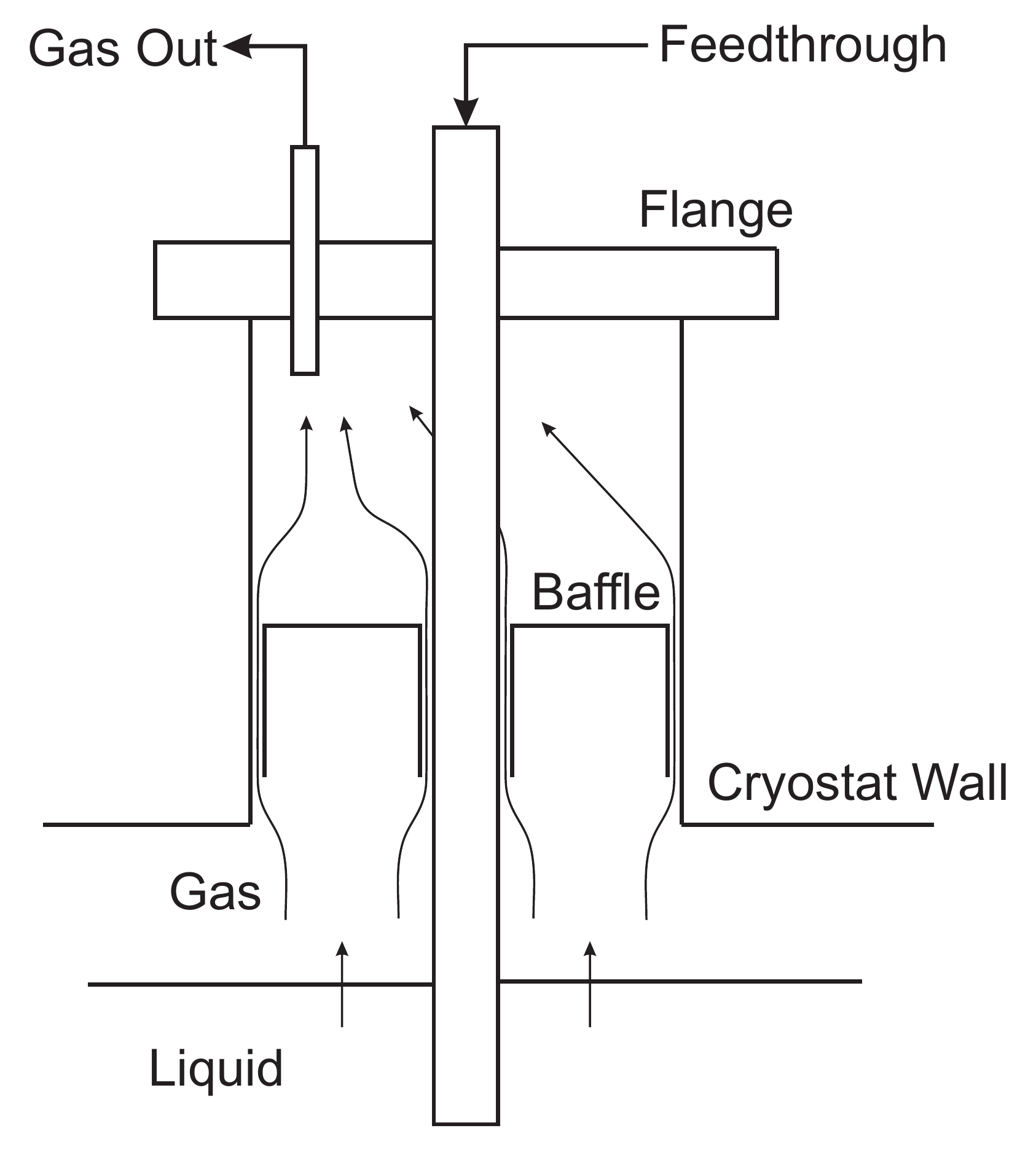}
       \caption{A schematic cross section of a flange with a feedthrough at the top of a cryostat with a baffle installed.}
       \label{fig:Baffle_Implementation}
 \end{figure*} 
 
As it appears from these measurements, that back-diffusion near the surface reduces the ultimate concentration at the liquid-gas interface and hence the concentration in the liquid, then it should be possible to exploit the effect to obtain even high ultimate purity.
One way to do this is to deliberately create a high velocity laminar flow region well above the surface by placing a barrier, covering most of the cross sectional area in the ullage, designed so as to create narrow long channels separating the leakage source from the liquid, through which evaporation gas must flow (Fig.~\ref{fig:Baffle_Implementation}).
The goal of a good design is to create only laminar flow channels with large values of $\Delta x / A_c$, and therefore a large attenuation of any leak rate present above the baffle.  As the velocity in the narrow long channels is the ratio of the total area to the restricted area the local gas velocity can be greatly increased and therefore the rate of back diffusion decreased.
As a result, the ultimate impurity concentration in LAr can be reduced by a factor of $F$ as in Eq.~\eqref{eq:gammaFunction}. 
We refer to this barrier as a ``baffle".
In summary, such a baffle should reduce impurity concentrations in the liquid by increasing the flow velocity of the evaporation gas, decreasing the open area available for gas flow, and increasing the length of flow channel(s) formed by the ``rims" on baffle plate.
For future large experiments, a baffle designed with these characteristics could be highly effective if installed in the ``chimneys" used to mount flanges for feedthroughs, as shown schematically in Fig.~\ref{fig:Baffle_Implementation}. 

Such a baffle should not interfere with the initial purging of air from the cryostat. If the “piston” part of the purge process is done slowly from the bottom of the cryostat, and the gas is withdrawn above the baffle, it should not alter the effectiveness of the purge even in the region above it. In practice, if the baffle causes more atmospheric impurities to be left in the small volume at the top of the cryostat, the continued gas purging after the piston phase and during the gas cool down should efficiently remove even this small additional burden.

%% file: summary.tex
\section{Summary and Conclusions}~\label{sec:summary}
In this paper, we present a mathematical model describing the time evolution of impurity distributions in both LAr and GAr in a typical LAr detector.  This model is implemented as a fourth order, nonlinear differential equation with coefficients that depend on the operating parameters and a small set of well defined physical parameters, some with known and others with uncertain or unknown values. 
Under assumptions that are typically valid for LAr detectors, this model can be reduced to a simpler, approximate model, which can be solved in closed form expressions for the impurity concentrations.
To validate the model, we operated a 20-L LAr test stand to obtain large sets of oxygen concentration data under various conditions.
We demonstrate that the data can be accurately described by the model with known, or reasonable approximations of, the physical parameters, and show that it is capable of determining leak rates, sorption rates, and the Henry’s coefficient for oxygen in LAr.
Further, we show that using the closed form approximate solution we can achieve a consistent measurement of Henry's coefficient for oxygen in LAr.  
The comparison of the calculations to the data indicates that the sorption process of oxygen on the inner cryostat surfaces is negligible for surfaces in the gas, and for surfaces in the liquid it makes an observable contribution only at the lowest oxygen concentrations (below $\sim$10~ppb).  Under these conditions, the model distinguishes between the two different sorption isotherms presented in the literature, and prefers the one with a saturation adsorption significantly less than one monolayer. 

After consideration of the uncertainties introduced by the system, we believe many improvements could be made in future investigations. Among these are: 
1) a mass flow meter should be installed between the gas purifier and condenser to determine the evaporation rate, independent of assumptions on the power input to the LAr; 
2) the pressure relief valve in our 20-L system, installed as a safety feature, should be replaced with a burst disk to eliminate the problem of inadequate sealing of a poppet valve;
3) the elimination of leaks in the top flange and the addition of a controlled leak valve would provide a known and stable leak rate;
4) the need to introduce the sampling process should be eliminated by returning the exhaust of the gas from the gas analyzers, after passing it through a purifier, to the system.

Extending the results from the model to a typical LAr detector, we suggest that a configuration with properly designed baffle in the gas volume can be helpful in reducing the impurity concentrations in the liquid.
The baffle configuration will presumably become cost-efficient for building and operating large LAr detectors for its capability of effectively limiting the impurity leak from air, because otherwise great efforts need to be made on welding joints in the cryostat to eliminate leakage. It should be noted that this also requires an efficient gas purifier, or purification of the condensed liquid, so that any impurity entering the GAr can be removed before it is returned to the bulk LAr~\cite{andrews}.  The effectiveness of the baffle concept needs further experimental verification and will be a subject of future experimental studies.

Finally we comment that, since the present model provides a good description of oxygen as an impurity, the obvious next step would be to apply it to describe water in LAr. 
Unfortunately, almost nothing is known about the properties of water in LAr.  Henry's coefficient, sorption properties, and even the solubility of water in LAr are all unknown.  The vapor pressure of ice at 90~K is reported to be $1.4\times10^{-22}$ bar~\cite{Fray2009}. For a mixture of ideal gasses, this would give a concentration of $1.0\times10^{-13}$ ppb of water in argon gas.  Naively, then, there is essentially no water in argon at 90~K.  Obviously, this is not true; there must be a strong interaction between water and argon that creates a highly ''non-ideal" solution of the two. In fact, a value of $H_{xx}=3.4\times10^{-9}$ for water in LAr at 90~K of is given by the NIST fluid properties program, REFPROP ~\cite{Kunz:REFPROP}.  If this were true, gas purification would have no effect on water concentrations in LAr. Yet we have observed that water can be removed by gas purification about as rapidly and thoroughly as oxygen, so the solution of water and LAr must be even more favorable than this. Sorption, on the other hand, should be a dominant effect. Since water is a solid below 273~K, the number of monolayers on a surface in LAr is in principle unlimited.  We have in fact observed that adsorption does dominate over purification when LAr is first introduced to the cryostat.   However,  since the  solubility  of  water  in  LAr  is known to be limited, presumably between 10 ppb and one ppm~\cite{rest}, a fifth phase and a fifth differential equation would be required  in  our  model to  adequately  describe  water.   Designing experiments to further investigate these topics would be relatively straightforward.

%% file: bibliography.tex
\renewcommand\bibname{References}